\documentclass[preprint]{aastex}
\usepackage{graphicx}
\usepackage{natbib}
\begin{document}

%%%%%%%%%%%%%%%%%%%%%%%%%%%%%%%%%%
\title{Interstellar and Ejecta Dust in the Cas A Supernova Remnant}
\shorttitle{Interstellar and Ejecta Dust in Cas A}
\shortauthors{ARENDT ET AL.}
\author{Richard G. Arendt\altaffilmark{1,2}, 
Eli Dwek\altaffilmark{2}, 
Gladys Kober\altaffilmark{2,3}, 
Jeonghee Rho\altaffilmark{4,5},
and Una Hwang\altaffilmark{6,7}}
\altaffiltext{1}{CRESST, University of Maryland -- Baltimore County,
Baltimore, MD 21250, USA; Richard.G.Arendt@nasa.gov}
\altaffiltext{2}{NASA Goddard Space Flight Center, Code 665, Greenbelt, MD 
20771, USA}
\altaffiltext{3}{Department of Physics, IACS, Catholic University of America, 
Washington DC 20064, USA}
\altaffiltext{4}{SETI Institute, 189 Bernardo Ave, Mountain View, CA 94043}
\altaffiltext{5}{SOFIA Science Center,  NASA Ames Research Center, MS 211-3, 
Moffett Field, CA 94035}
\altaffiltext{6}{NASA Goddard Space Flight Center, Code 662, Greenbelt, MD 
20771, USA}
\altaffiltext{7}{The Henry A. Rowland Department of Physics and Astronomy, Johns 
Hopkins University, 3400 N. Charles Street, Baltimore, MD 21218, USA}

\setcounter{footnote}{6}
%%%%%%%%%%%%%%%%%%%%%%%%%%%%%%%%%%
\begin{abstract}
Infrared continuum observations provide a means of investigating the physical 
composition of the dust in the ejecta and swept up medium of the Cas A 
supernova remnant.
Using low resolution {\it Spitzer} IRS spectra (5--35~$\micron$), 
and broad-band {\it Herschel} PACS imaging (70, 100, and 160~$\micron$), 
we identify characteristic dust spectra, associated with 
ejecta layers that underwent distinct nuclear burning histories.
The most luminous spectrum exhibits strong emission
features at $\sim9$ and 21 $\micron$ and is closely associated 
with ejecta knots with strong Ar emission lines.
The dust features can be reproduced by magnesium silicate grains with 
relatively low Mg to Si ratios.
Another dust spectrum is associated with ejecta having strong 
Ne emission lines. It has no indication of any silicate 
features, and is best fit by Al$_2$O$_3$ dust.
A third characteristic dust spectrum shows
features that are best matched by magnesium silicates with a relatively 
high Mg to Si ratio. 
This dust is primarily associated with the X-ray 
emitting shocked ejecta, but it is also evident in regions where 
shocked interstellar or circumstellar material is expected.
However, the identification of dust composition is not unique, and each 
spectrum includes an additional featureless dust component 
of unknown composition.
Colder dust of indeterminate composition is associated with 
emission from the interior of the SNR, where the reverse shock has 
not yet swept up and heated the ejecta. Most of 
the dust mass in Cas A is associated with this unidentified cold 
component, which is $\lesssim0.1$  $M_{\sun}$.
The mass of warmer dust is only $\sim 0.04$~$M_{\sun}$. 
\end{abstract}

%%%%%%%%%%%%%%%%%%%%%%%%%%%%%%%%%%
\section{INTRODUCTION}

Interstellar dust models that fit the average interstellar extinction curve, the diffuse infrared emission and scattering, polarization, and abundance constraints employ a very limited variety of dust compositions, consisting primarily of polycyclic aromatic hydrocarbons (PAHs), graphite or amorphous carbon, and astronomical silicates \citep{Li:2001,Zubko:2004,Brandt:2012,Siebenmorgen:2014}. 
Yet observations of the primary sources of interstellar medium (ISM) dust 
(or at least the metals therein), asymptotic giant branch (AGB) stars and supernovae (SNe), 
reveal a significantly richer variety of dust compositions.
For example, magnesium sulfide (MgS) is inferred from spectral features of pre-planetary nebulae \citep{Omont:1995}
with similar features in carbon-rich AGB stars and PNe \citep{Forrest:1981, Hony:2002}. \cite{Cherchneff:2012} 
contains a detailed model of the formation of a wide variety of molecules and dust in an AGB star.
More directly, presolar grains of supernova or stellar origin, such as silicate carbide (SiC), silicon 
nitride (Si$_3$N$_4$), and aluminum-, calcium- and titanium-oxides are found as 
meteoritic inclusions \citep[e.g.][] {Zinner:2008}. 
The absence of a wide variety of specific compositions in interstellar dust models indicates that 
these compositions are not required for fitting various manifestations of dust in the general 
ISM, either because of their low abundance relative to silicates and carbonaceous dust, 
or due to the fact that they may have been processed in the ISM.

Supernovae can be important sources of interstellar dust. They produce all the refractory elements needed for the formation of dust, and their ejecta largely retain the compositional inhomogeneity of the progenitor star. They may therefore be sources of dust with unusual chemical and isotopic compositions. Furthermore, SNe are drivers of the chemical evolution in galaxies, and therefore potentially the most important sources of interstellar dust. In young populations
before low mass stars have evolved off the main sequence, e.g. high redshift galaxies, SNe are the dominant source of thermally-condensed dust,
though additional grain growth by cold accretion within dense clouds may be
required to explain the inferred dust mass in these systems
\citep{Dwek:2011, Valiante:2011}. 
Determining the mass and composition of SN condensed dust is therefore key for understanding the origin, evolution, and processing of dust in galaxies.

The Cas~A remnant is an ideal object for studying the composition and abundance of dust that formed in the ejecta of a core collapse SN.
The SN was not definitively recorded at the time that it occurred. 
Studies of the expansion of the supernova remnant (SNR) estimate that the explosion was
in the year 1681$\pm$19 \citep{Fesen:2006}. Yet fortunately, light echoes of the Cas~A
SN have allowed studies of this old event with modern instruments. Such
observations have revealed that Cas~A was a Type IIb SN \citep{Krause:2008}
with distinct asymmetry in its explosion \citep{Rest:2011}. 
Dynamical and compositional asymmetries are still imprinted on the 
Cas~A SNR today, but the dominant structure of the Cas A SNR is characterized
by a clear distinction between the forward shock sweeping up the interstellar 
(and/or circumstellar) medium, and the reverse shock through which the 
SN ejecta is expanding. 

The ejecta consists of three main components: the first, containing most of the mass, is a low density 
phase that is heated by the reverse shock to X-ray emitting temperatures ($\gtrsim 10^6$ K). 
The second component consists of dense fast-moving knots (FMKs) that have gone through 
the reverse shock, and are radiatively cooling by line emission at optical and infrared (IR) wavelengths. 
A third component comprises ejecta that has not yet encountered the reverse shock, and is 
primarily heated by the ambient radiation within the SNR interior. 

In this paper we revisit the analysis of the mid- to far-IR spectra of the dust in Cas~A. 
IR emission can arise from dust in each of the ejecta components discussed above, as well as the 
circumstellar medium (CSM) or ISM that is shocked by the 
advancing SN blast wave, i.e. the forward shock.
Our main goal is to separate and identify different types of dust that associated with different 
ejecta (and ISM or CSM) components and to determine the spatial distribution of the different types of dust.
Our approach will provide important information on the physical processes that facilitate or inhibit the nucleation of dust in the different layers of the ejecta, the resulting dust composition, and the dust heating mechanisms that give rise to the IR emission. Our approach is different from previous ones \citep{Ennis:2006,Rho:2008} which only grouped the IR emission into distinct spectral components, with no relation to the nature of the ejecta from which they originated.

The outline of our analysis is: \\
(1) We identify a set of {\it spatial templates} that are used as the initial indicators of 
regions of different ejecta composition and/or physical conditions around the SNR. 
%The selection of these templates is described in Section 3.1 and illustrated in Figure 1.\\
These are illustrated in Figure \ref{fig:spatial_templates}, with the details of their 
derivation in the Appendix.\\
(2) For each spatial template, we identify {\it zones} (subregions) where that template is most 
prominent with respect to the other templates. These are described in Section 3.1 and shown 
in Figure \ref{fig:zones}.\\
(3) Within each of the zones and at each wavelength, we use the spatial correlation between 
the data and the template (see Eq. 1) to derive the {\it characteristic spectra} associated 
with each spatial template. These are presented in Section 3.1 and Figure \ref{fig:spectral_distributions}. 
Analysis of 
these characteristic spectra provides indications of the dust composition and 
temperature(s).\\
(4) Finally, at each spatial location across the entire SNR, the spectrum is decomposed 
as a linear combination of the characteristic spectra. The coefficients of these 
decompositions are mapped out to reveal images of the {\it spatial distributions} of the 
dust that gives rise to each characteristic spectrum. This is described in Section 3.2
and illustrated in Figure~\ref{fig:spatial_distributions}.

Section 2 of this paper describes the preparation of the {\it Spitzer} IRS data to create a spectral cube of the 
dust continuum emission of Cas A. Section 3 explains the data analysis steps described above, with the details
and results of modeling the characteristic spectra presented in Section 4. In Section 5 we discuss the results, 
including what conclusions can and cannot be drawn concerning the dust composition and mass. The work is summarized
in section 6.

%%%%%%%%%%%%%%%%%%%%%%%%%%%%%%%%%
\section{DATA PREPARATION}

Our work began with the low resolution IRS spectral data cubes used by \cite{Rho:2008}.
There are 4 cubes, generated from the 1st and 2nd order spectra in short
and long wavelength low resolution modules. At each wavelength, the 
data were convolved to the spatial resolution at the longest IRS wavelength, 38.33 $\micron$,
using kernels derived from Tiny Tim / Spitzer
(STINYTIM)\footnote{\url{http://irsa.ipac.caltech.edu/data/SPITZER/docs/dataanalysistools/tools/contributed/general/stinytim/}} 
PSFs according to \cite{Gordon:2008}. The three shorter wavelength cubes were
then reprojected to the same scale and orientation as the longest wavelength (SL1)
cube. The data are then combined into a single IRS low-resolution cube covering 
5 -- 38 $\micron$ with a fixed spatial resolution. Weighted averages were used 
at the wavelengths that overlap between the different spectral orders and modules.

For the purpose of analyzing the dust emission, we created a continuum spectral cube
by replacing all the emission lines with smooth polynomial fits to the continuum
on either side of each line. This procedure is performed independently for each 
spatial pixel of the cube and each line. It avoids any need to fit the lines themselves,
which can be found at various velocity components at each pixel
because of the high expansion velocities of Cas A. 
The lines that were thus removed are listed in the row and column headings of Table \ref{tab:cc}. 
Finally, to remove remaining artifacts and
improve the signal to noise, the continuum cube was smoothed in wavelength to a 
spectral resolution of $R = \lambda/\Delta\lambda \approx20$, and trimmed to a maximum
wavelength of 35 $\micron$. After this smoothing, there are only $\sim40$ truly 
independently-sampled wavelengths in the data cube.

Longer wavelength information was obtained by using {\it Herschel} PACS 
\citep{Pilbratt:2010,Poglitsch:2010} observations at 
70, 100, and 160 $\micron$, with the appropriate spatial convolution of the 
70 and 100 $\micron$ images whose native resolution is better than the IRS at 
38 $\micron$. (All {\it Herschel} data used here were Level 2 products 
generated from Standard Product Generation SPG 4.1.0.)
At 160 $\micron$ the SNR is substantially confused by ISM emission. 
This confusion was reduced by subtracting an empirically 
scaled version of the {\it Herschel} SPIRE \citep{Griffin:2010} 
250 $\micron$
image in which the synchrotron emission of the SNR was subtracted by extrapolation 
from the 350 and 500 $\micron$ bands. The subtraction is an improvement, but is still
imperfect, in part due to the lower spatial resolution of the longer 
wavelength SPIRE data. 

The synchrotron emission is removed from the final {\it Spitzer}
continuum cube by subtraction of the VLA radio template \citep{DeLaney:2004} extrapolated 
assuming a power law spectrum $S_\nu \sim \nu^{-0.71}$.

%%%%%%%%%%%%%%%%%%%%%%%%%%%%%%%%%%%%%%%%%%%%%%
\section{ANALYSIS}

\subsection{Derivation of Characteristic Spectra}
One of the primary goals of this analysis was to find the continuum spectra of dust that is found in 
regions of various physical parameters and compositions throughout the SNR. 
The regions to be investigated are identified via six distinct spatial templates.
These templates are shown in Figure \ref{fig:spatial_templates} and Table \ref{tab:purpose}, 
and are described in more detail in the Appendix.
To identify 
the spectra associated with each of the six emission templates, % shown in Fig. \ref{fig:spatial_templates},
we superimposed all the templates and for each one we identified the zones where its relative
emission was the dominant component. Because of the strong and widespread [\ion{Ar}{2}] emission, 
in some cases the zones necessarily included [\ion{Ar}{2}] emission as well.
The zones selected for each of the templates are outlined in Figure \ref{fig:zones}.

For all pixels, $i$, of each zone of the 6 selected emission templates, $T_j(i)$, 
the data $D(i,\lambda)$ can be represented by:
\begin{equation}
D(i,\lambda) = \sum_{j=1}^6{S_j(\lambda)T_j(i)}\ + \ C(\lambda).
\end{equation}
The parameters $C(\lambda)$, a constant term (which should be small) to account for errors in the background 
levels, and $S_j(\lambda)$, the spectra associated with each of the templates, can thus be derived via
a least-squares fit or linear regression between the data and the templates.
 
For the \ion{Ar}{2} zone, the sum is strongly dominated by this template, and 
$S_{ArII}(\lambda)$ is determined as the slope of a linear least squares fit between $D(i,\lambda)$ and 
$T_{ArII}(i)$. Some \ion{Ar}{2} emission is unavoidable in all other zones, therefore for the other zones 
the spectra $S_j(\lambda)$ were derived using a linear regression between $D(i,\lambda)$ and two templates: 
$T_{j}(i)$ and $T_{ArII}(i)$.
The characteristic spectra 
derived are shown in Figure \ref{fig:spectral_distributions}. 

When these derived spectra are multiplied by the spatial templates over the entire SNR (not
just within the selected zones) and the result is subtracted from the 
data cube, we found that there was one region of residual emission that was particularly strong and 
consistently positive at all wavelengths. This region is in the south central portion 
of the SNR. One additional zone was created to cover this region (the ``South Spot'' in 
Fig. \ref{fig:zones}), and the measured spectrum within this zone is included in Fig. 
\ref{fig:spectral_distributions}. 

The characteristic spectra are key data that allow detailed examination of the nature of the
dust in different regions of the SNR.
The modelling and analysis of the characteristic spectra are presented in Section 4.

\subsection{Spatial Distributions of the Characteristic Spectra}

In principle, the spatial templates, $T_j(i)$ and their characteristic spectra, $S_j(\lambda)$,
provide a complete description of the continuum data cube. However, because the characteristic
spectra were derived only from limited zones of the SNR, it is informative to revisit the data 
modelling given by Equation (1), but this time we use the seven $S_j(\lambda)$ characteristic spectra as 
known quantities, and solve for the spatial distributions, $I_j(i)$, of each spectum over the entire SNR. 
\begin{equation}
D(i,\lambda) = \sum_{j=1}^7{S_j(\lambda)I_j(i)}\ + \ C(i)
\end{equation}
In this fitting, a constant term, $C(i)$, is still present to account for background errors, but 
now this constant is a function of position rather than wavelength. 
If the characteristic spectra are uniquely and perfectly correlated with the spatial templates, then 
we should find that the spatial distributions are identical to the templates, $I_j(i) = T_j(i)$.
Differences between the initial templates and these derived spatial distributions 
would identify (a) regions where the template emission is present, but not 
accompanied by the characteristic spectrum, and (b) regions where emission matching a characteristic 
spectrum is present but without corresponding emission of the initial template.

The derived spatial distributions of the different characteristic spectra
are shown in Figure~\ref{fig:spatial_distributions}. Comparison with 
Figure \ref{fig:spatial_templates} shows that the derived spatial distribution of the
dust associated with \ion{Ar}{2} emission is very similar to the input template, 
$I_{ArII}(i) \approx T_{ArII}(i)$. Differences are more evident for dust 
associated with \ion{Ne}{2} and X-ray Fe emission. Yet in both cases the 
brightest portions of the derived spatial distributions do follow the original
templates, even in areas that were outside of the zones where the characteristic 
spectra were derived. The spatial distribution of the radio dust (i.e. dust associated with the radio template) is very different
from original template, as it tends to be concentrated in areas at the periphery of the SNR 
that are expected to lie between the forward shock and the reverse shock. Material here
should be dominated by swept up interstellar or circumstellar material. Interestingly, the 
spatial distribution identifies a large area on the east size of the SNR which matches
the characteristic spectrum, but had not been within the zone used to define that spectrum.
The agreement between the \ion{Si}{2} dust's spatial distribution and the initial template is 
very rough. This spatial distribution is dominated by the morphology of the emission at
wavelengths $\lambda \geq 70$ $\micron$.
The spatial distribution of dust matching the \ion{Ar}{3} characteristic
spectrum is confined to a few of the brightest peaks in the continuum and [\ion{Ar}{3}] 
images.

Guided by these results, we find that a traditional 3-color image can be 
constructed to emphasize the different components 
via the tint of the continuum emission, as shown in Figure 
\ref{fig:3color}. This image combines the emission at 11.8, 20.8, and 70 $\micron$
from the smoothed continuum data cube. These wavelengths are chosen by selecting 
those where the characteristic spectra (Fig. \ref{fig:spectral_distributions})
have the greatest differences from one another. 

\subsection{Total Emission}

\begin{figure}[h]
  \includegraphics[width=6in]{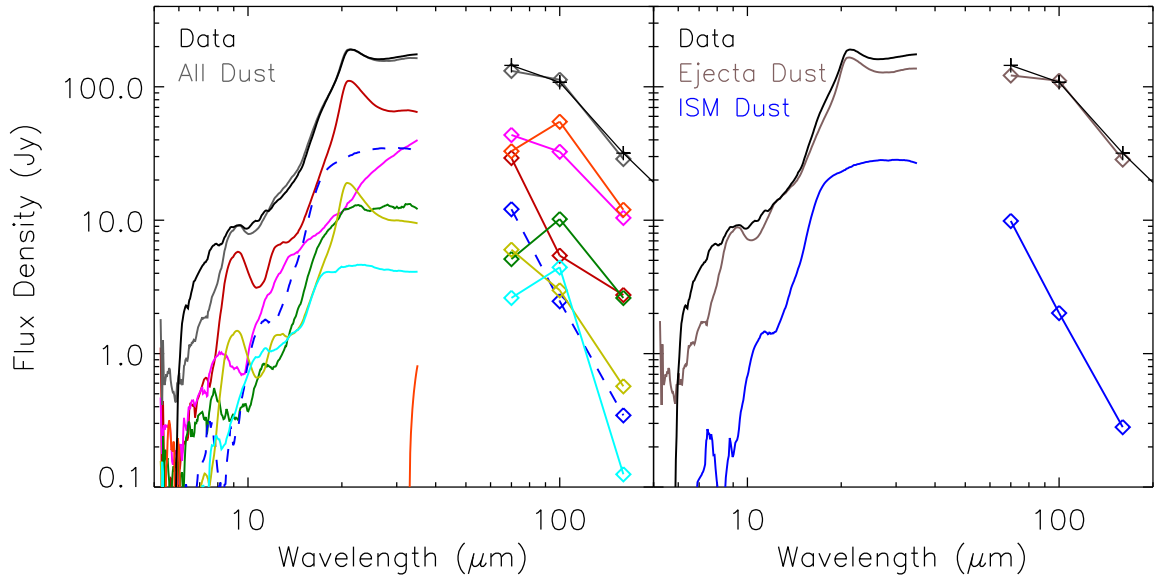}
   \caption{(left) Comparison of the relative intensities of each 
   of the spectral types with the total emission of Cas A. 
   The gray line is the sum of all the characteristic spectra.
   The black line at $\lambda \leq 35$ $\micron$ is the total of our {\it Spitzer} 
   IRS spectral cube.
   The black points at $\lambda \geq 70$ $\micron$ are the {\it Herschel} PACS 
   measurements with synchrotron emission and ISM background subtracted \citep{Barlow:2010}.
   (right) Comparison of the total Cas A emission with the sum of all the ejecta dust 
   components and the ISM dust component. The ISM dust component is that associated
   with the radio emission of the forward shock. 
   \label{fig:spectra}}
\end{figure}

Figure \ref{fig:spectra} shows the seven Cas A characteristic spectra with their actual 
intensities (i.e. without normalization). 
The total emission of each is calculated by integrating the characteristic
spectra, $S_j(\lambda)$ (Fig. \ref{fig:spectral_distributions}), over the regions of the corresponding
spatial distributions, $I_j(i)$ (Fig. \ref{fig:spatial_distributions}).
This comparison shows that in the ejecta, the \ion{Ar}{2} dust component is dominant
at wavelengths from 8 to 35 $\micron$. The dust associated with the radio emission is assumed
to be swept up ISM dust rather than ejecta, because of its spatial distribution turns out to be largely confined to 
the periphery of the SNR. The brightness of the ISM dust
is generally $<1/3$ of the brightness of the ejecta dust.
The sum of the all the spectra is also shown as a ``fit'' to the total Cas A spectrum 
as measured directly from {\it Spitzer} and {\it Herschel} data. 
The general agreement at 10 -- 35 $\micron$ 
indicates that there are no major missing components. At shorter wavelengths, as the SNR emission 
drops towards the level of the background and the noise, the comparison is not as good. 

%%%%%%%%%%%%%%%%%%%%%%%%%%%%%%%%%%
\section{DUST COMPOSITIONS FROM THE CHARACTERISTIC SPECTRA}
Six of the seven characteristic spectra (Figure \ref{fig:spectral_distributions}) 
show strong emission at wavelengths from 10 to 40 $\micron$. 
The {\it Spitzer} IRS spectra across these wavelengths 
reveal distinctive features in some of these spectra. The features are
 suggestive of silicates with peaks of varying sharpness 
at $\sim10$ and $\sim20$ $\micron$. In order to identify the particular type
of silicates and/or other grain compositions that produce each of the characteristic
spectra, we assembled a set of grain absorption efficiencies from published optical constants (Table \ref{tab:emiss}), 
and fitted each characteristic spectrum, $S_j{(\lambda})$, as the weighted, $M_{n,i,j}$,
sum of different grain mass absorption coefficients, $\kappa_n(\lambda)$, applied to blackbody 
emission at different temperatures, $B(T_i)$: 
\begin{equation}
S_j(\lambda) = \sum_{n}\sum_{T_i=20\rm{K}}^{2000\rm{K}} M_{n,i,j} \kappa_n(\lambda) B(T_i)\ 
\end{equation}
where $n$ and $i$ are indices for dust composition and temperature respectively.
The mass absorption coefficients, $\kappa_n(\lambda) \equiv 3Q_n(\lambda,a)/4\rho_n a$, or more directly the
absorption efficiencies, $Q(\lambda,a)$, are calculated
using Mie theory and assuming a relatively small grain size $a = 0.01$ $\micron$ \citep[e.g.][]{nozawa:2008,nozawa:2010}, although $Q(\lambda,a)/a$ is essentially independent
of $a$ for $a \lesssim 0.1$ $\micron$ at these wavelengths. ($\rho_n$ is the mass density of grains of compostion $n$.)

\begin{figure}[hp]
  \includegraphics[height=3.5in]{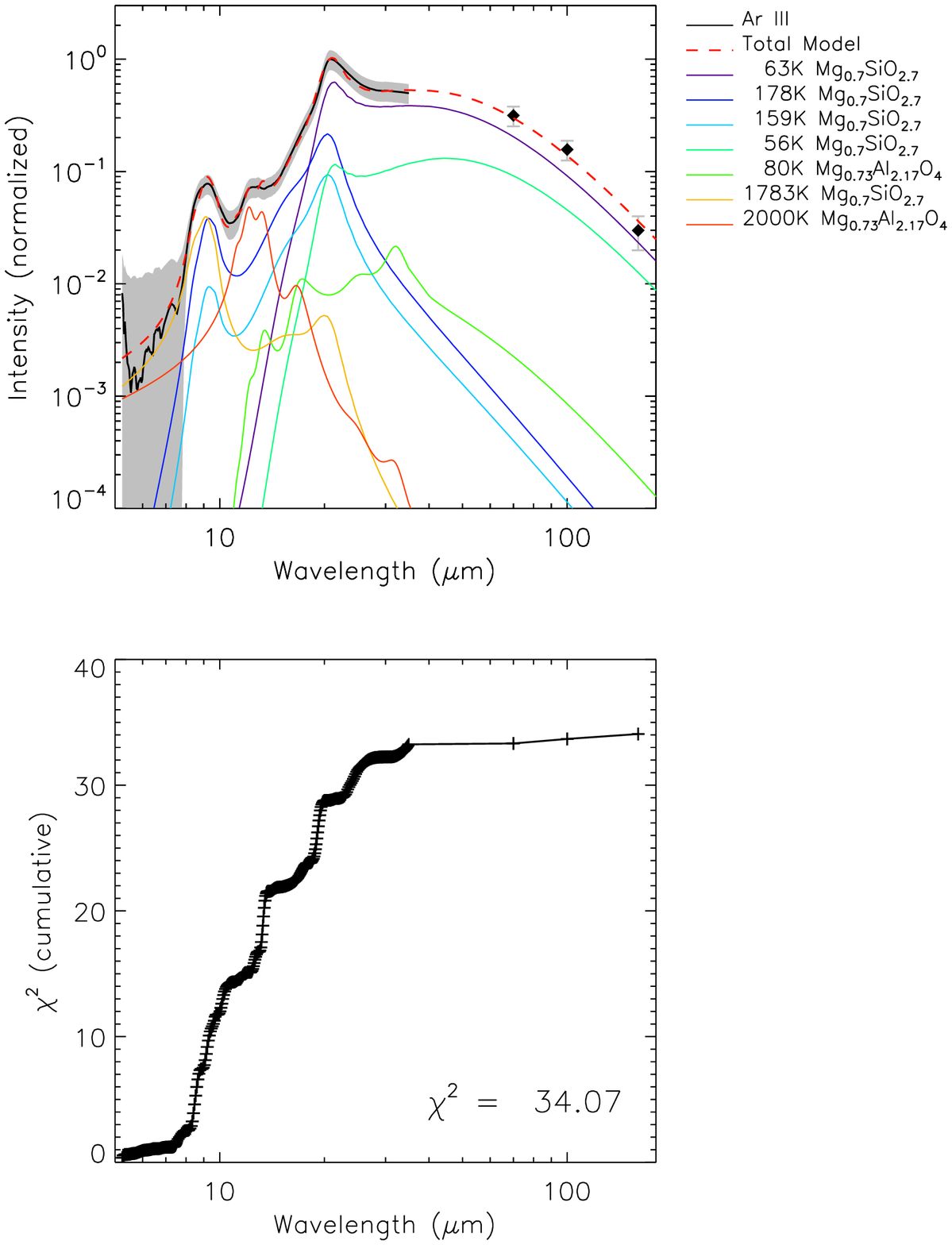}~~~~~~
  \includegraphics[height=3.5in]{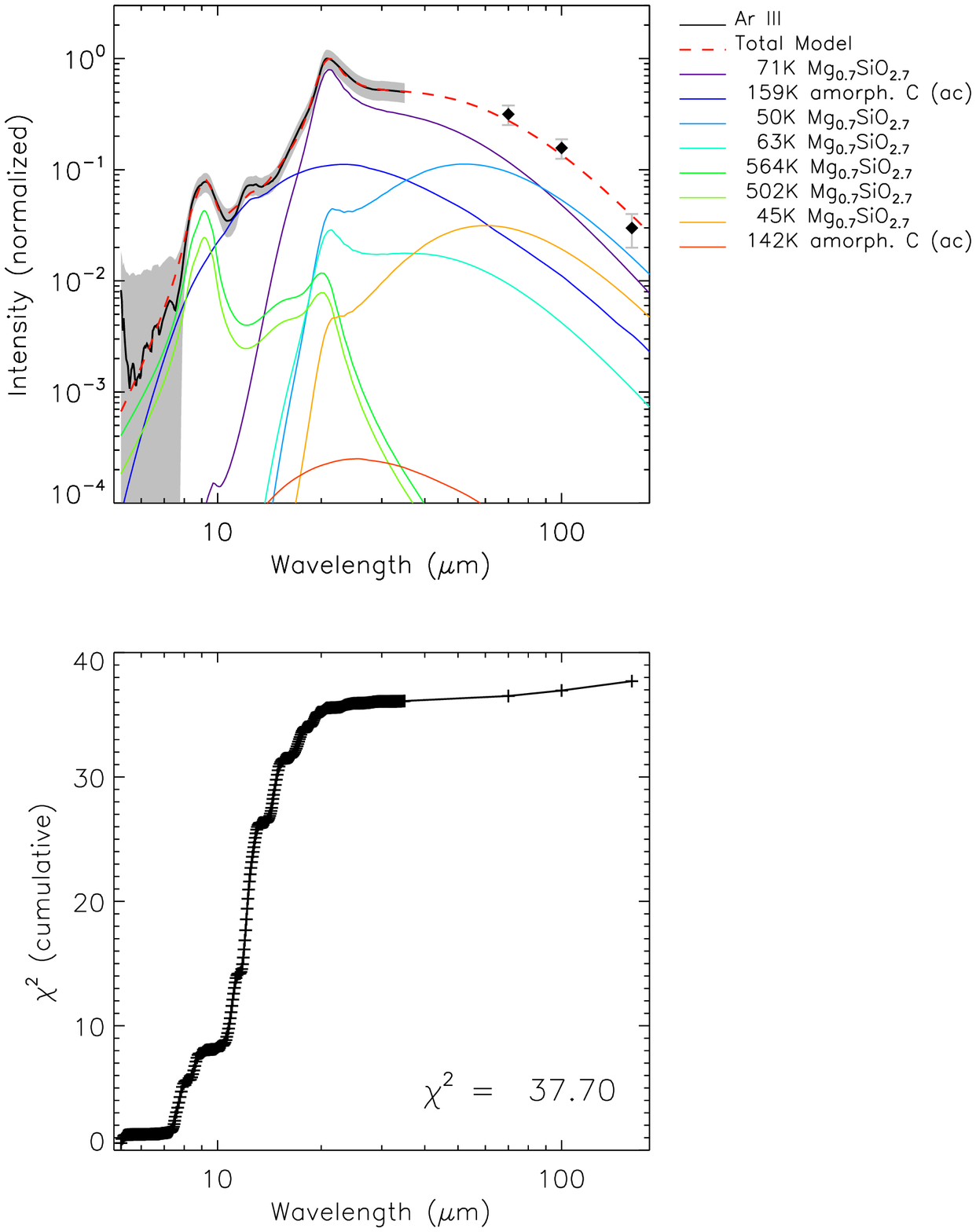}\\
  ~~\\
  \includegraphics[height=3.5in]{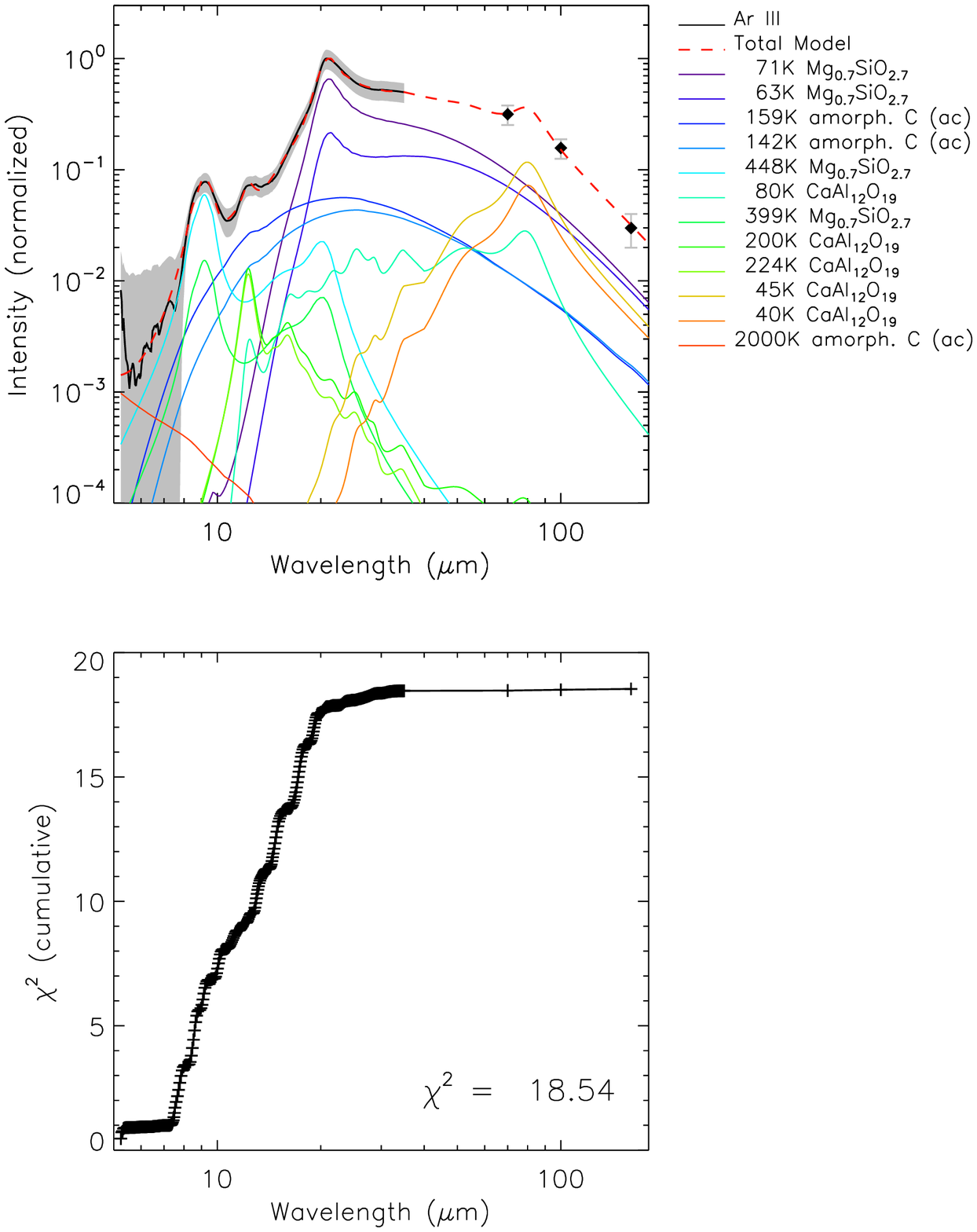}\\
   \caption{(upper left) The best 2-composition fit to the \ion{Ar}{3} spectrum makes use of the 
   12 $\micron$ feature of nonstoichiometric spinel.
   (upper right) Alternate 2-component fits to the spectrum using featureless
   compositions (e.g. amorphous C) fit nearly as well, but do not match the 12 $\micron$ peak.
   (lower left) The best 3-component fit uses CaAl$_{12}$O$_{19}$ 
   (hibonite) to fit the 12~$\micron$ peak.
   \label{fig:ariiidust}}
\end{figure}

\begin{figure}[hp]
  \includegraphics[height=3.5in]{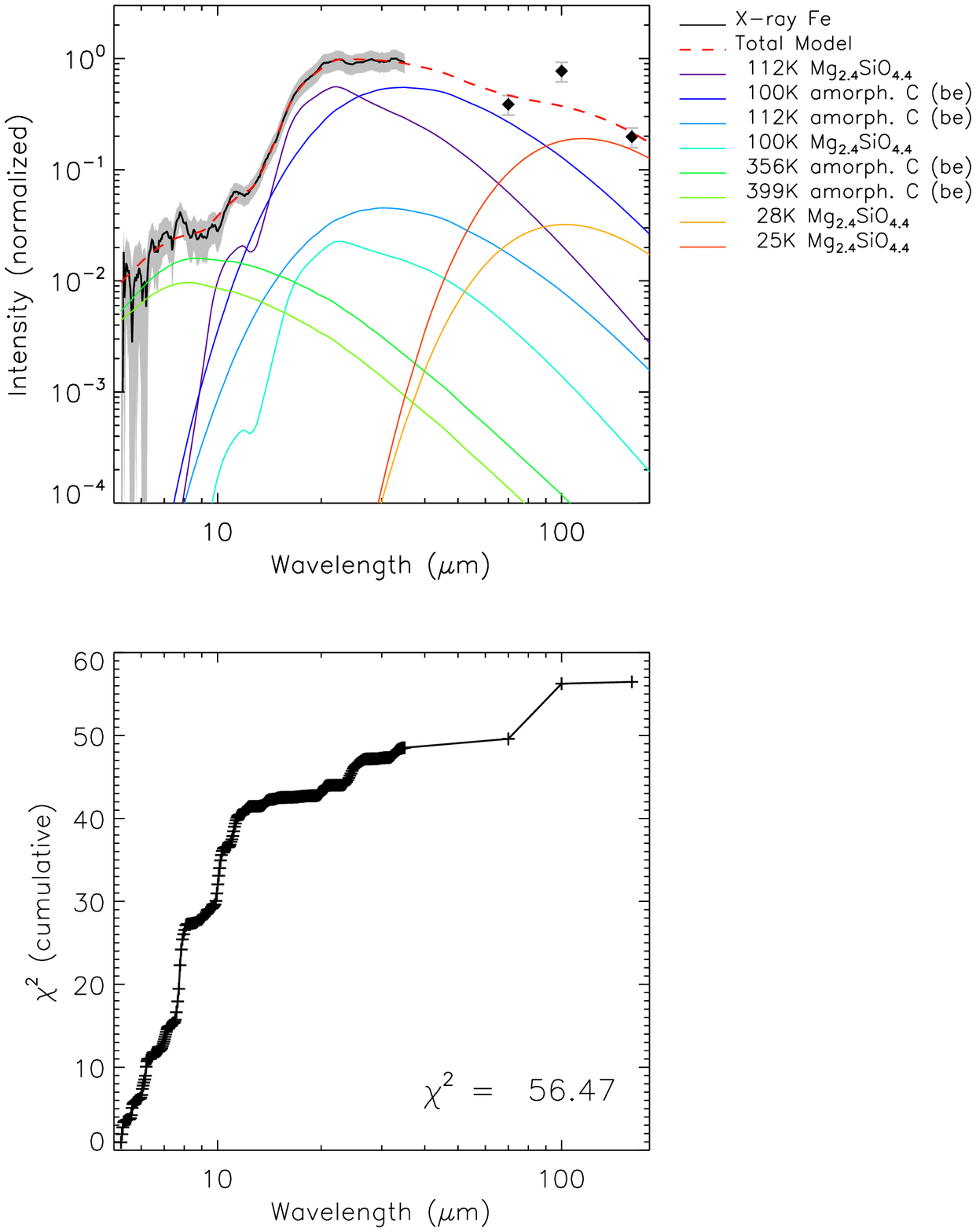}~~~~~~
  \includegraphics[height=3.5in]{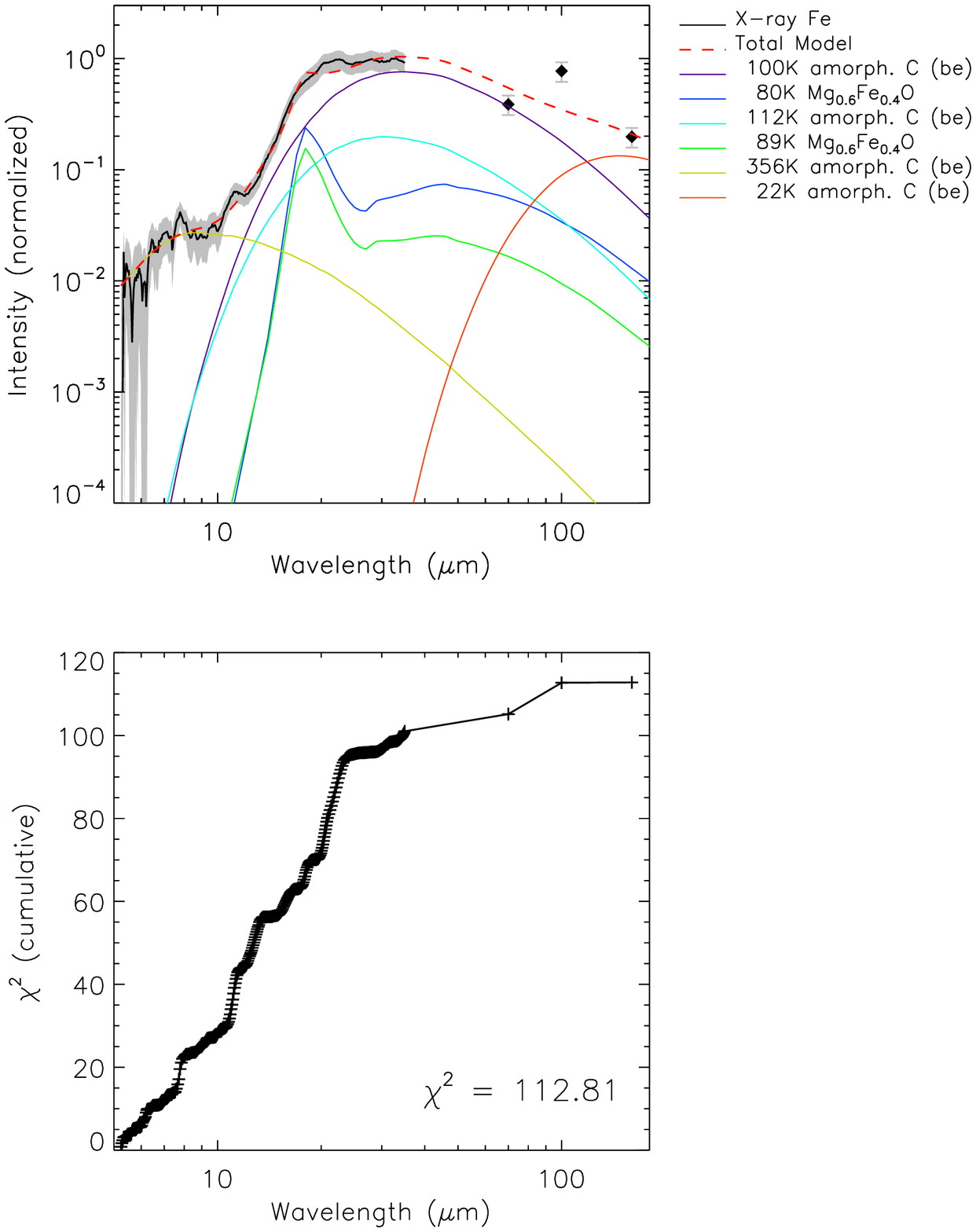}\\
  ~~\\
  \includegraphics[height=3.5in]{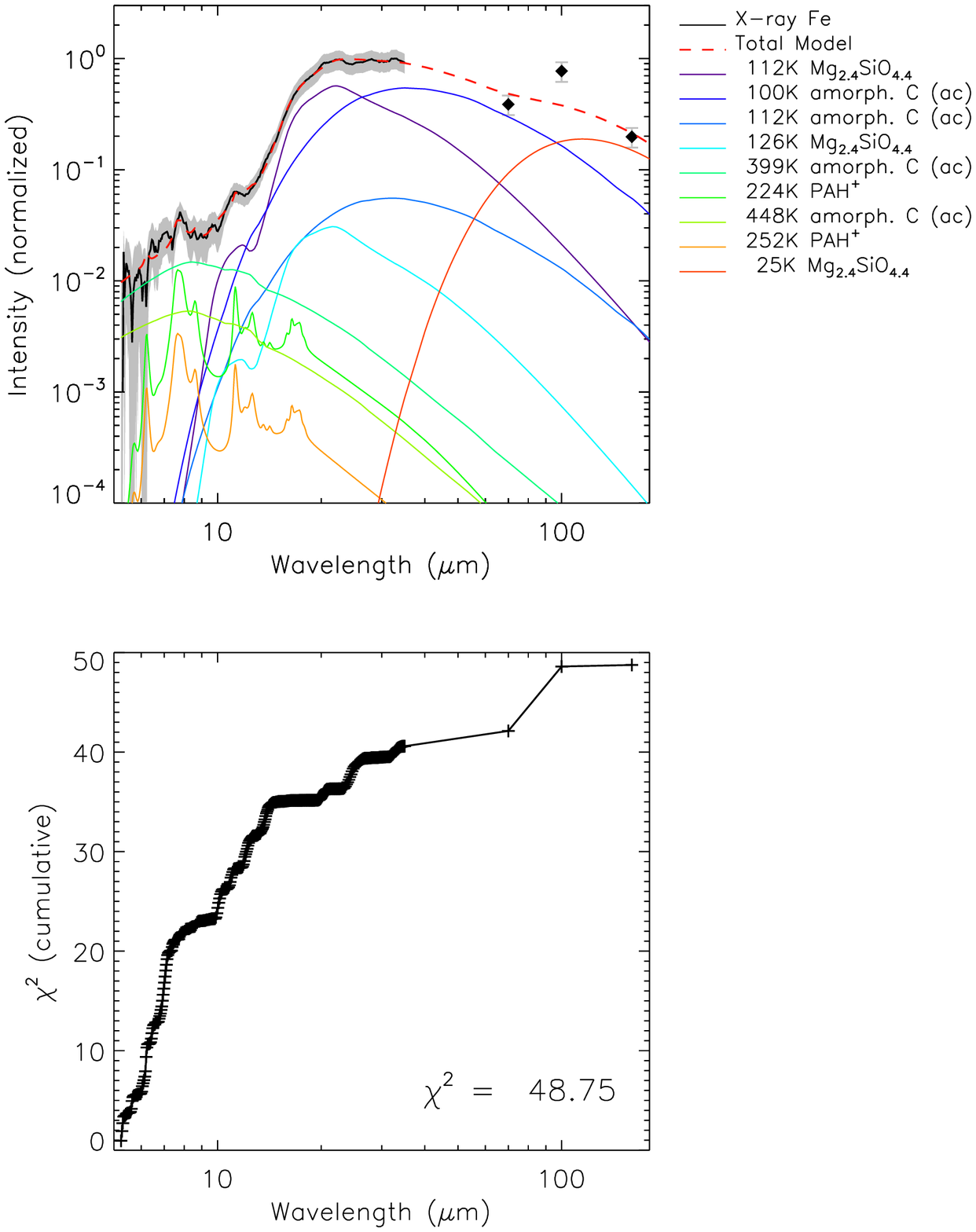}\\
   \caption{(upper left) The best 2-component fit to the X-ray Fe spectrum uses a Mg silicate in 
   combination with a featureless composition. 
   (upper right) This alternate 2-component fit (without silicates) 
   is about a factor of 2 worse than the best fit, and is thus 
   only marginally acceptable.
   (lower left) The best 3-component fit uses ionized PAHs to
   match the small (and possibly spurious) bump at 8 $\micron$. 
   \label{fig:xrayfedust}}
\end{figure}

\begin{figure}[hp]
  \includegraphics[height=3.5in]{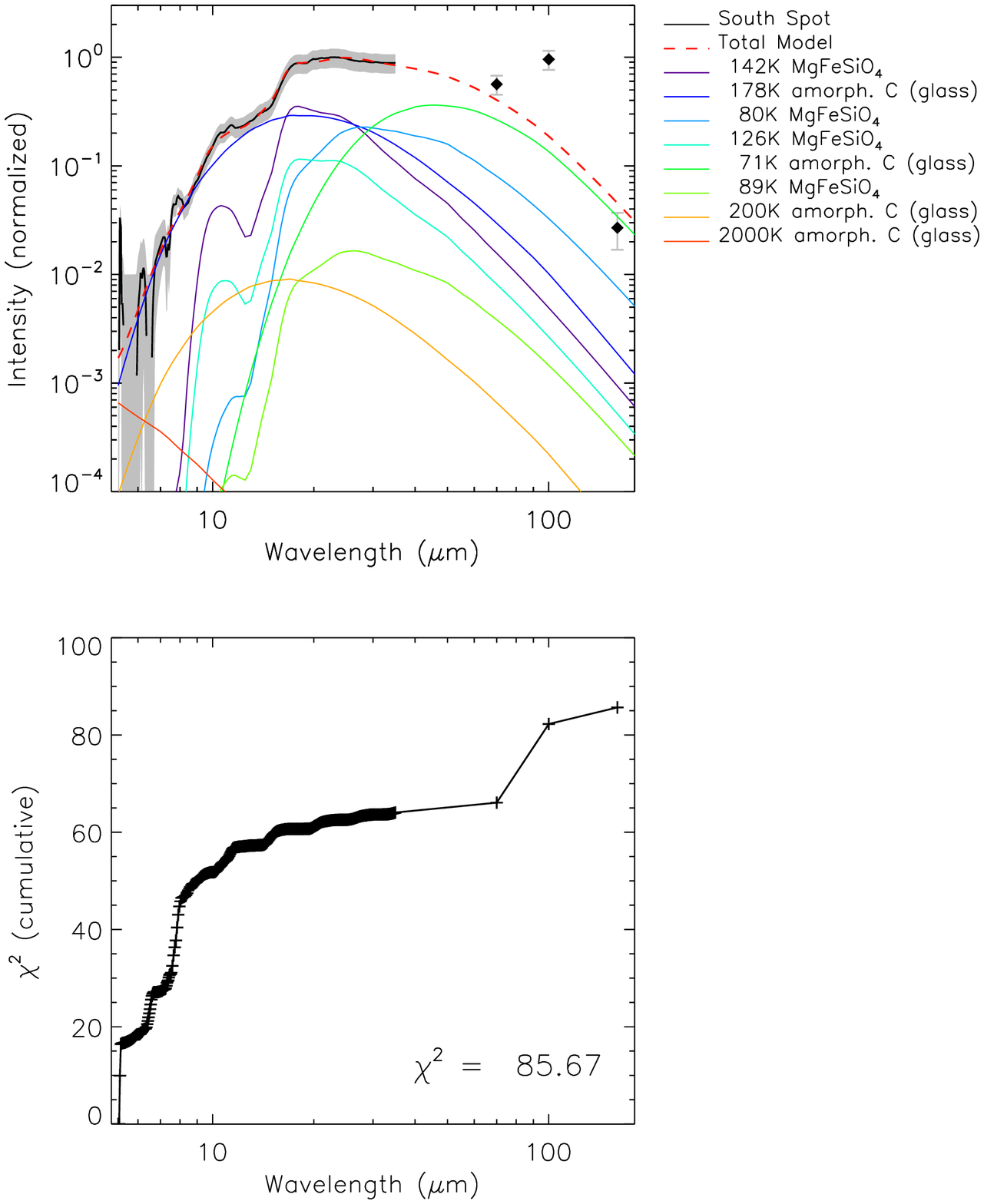}~~~~~~
  \includegraphics[height=3.5in]{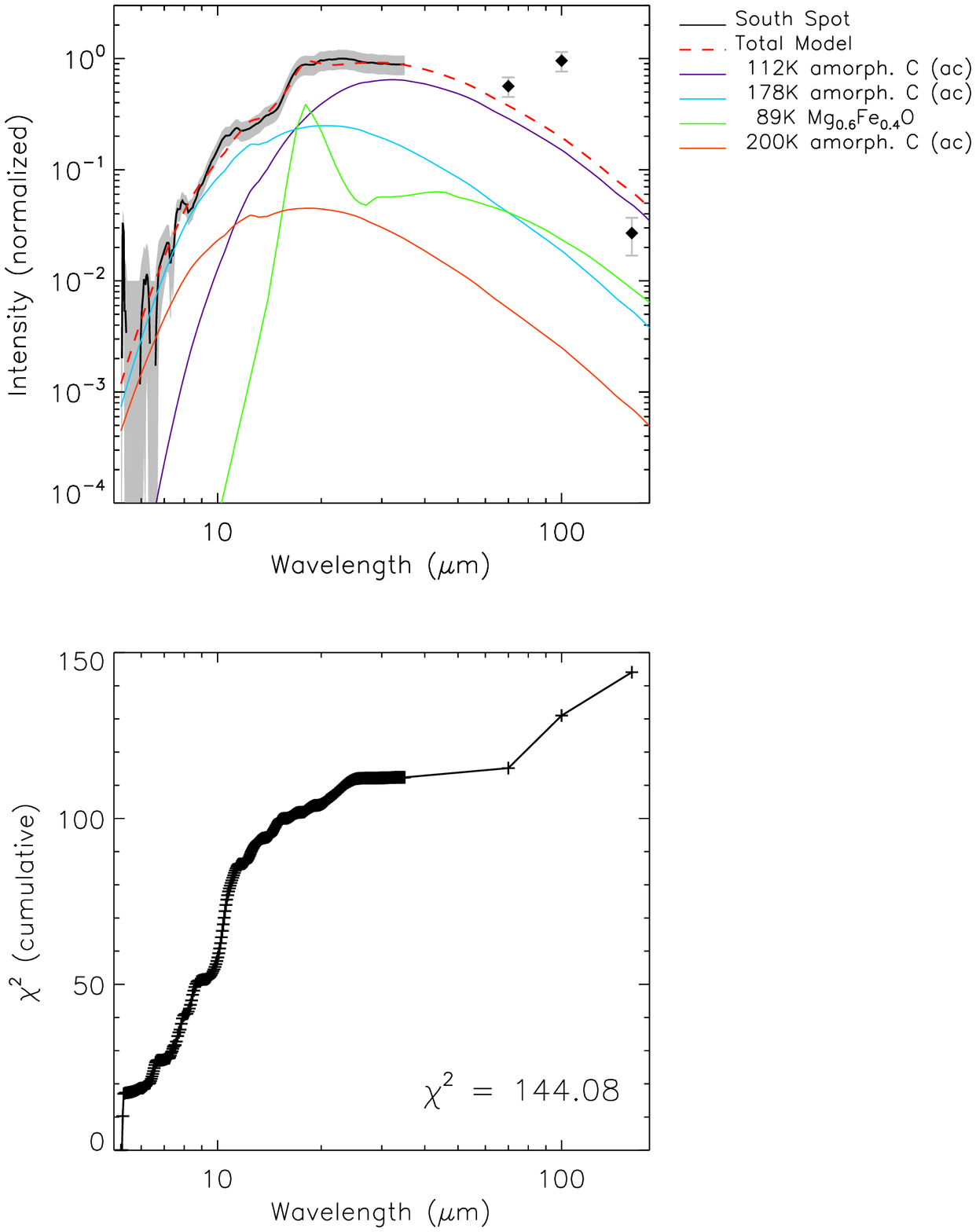}\\
  ~~\\
  \includegraphics[height=3.5in]{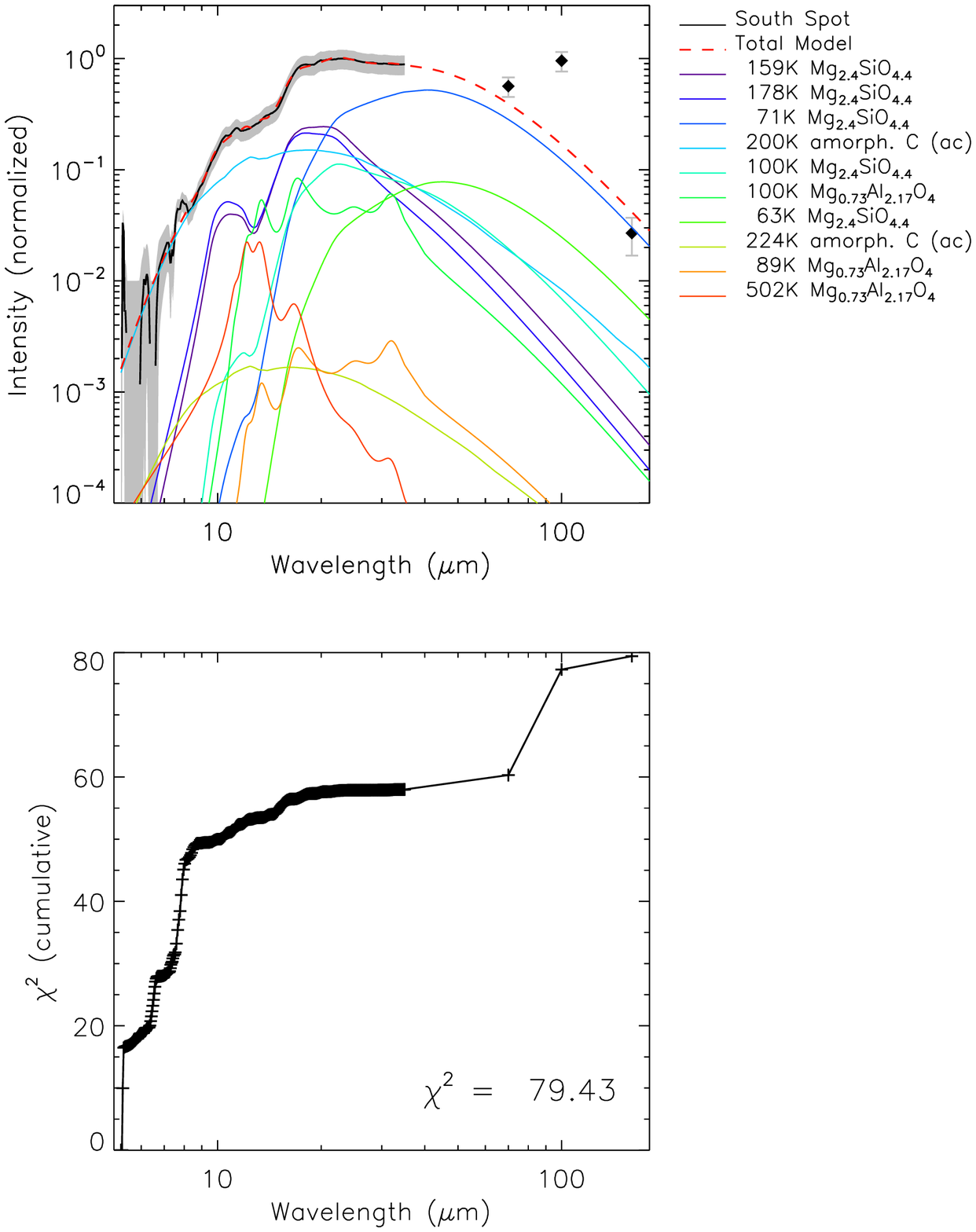}\\
   \caption{(upper left) The best 2-component fits to the South Spot spectrum are similar
   to those for the X-ray Fe spectrum, but involved higher dust temperatures
   for the dominant components.  
   (upper right) An alternate 2-component fit again shows 
   that reasonable (though worse) fits can also be obtained without the use 
   of silicates.
   (lower left) The best 3-component fit uses 
   nonstoichiometric spinel as a third component to make relatively minor 
   adjustments to the model.
   \label{fig:ssdust}}
\end{figure}

\begin{figure}[hp]
  \includegraphics[height=3.5in]{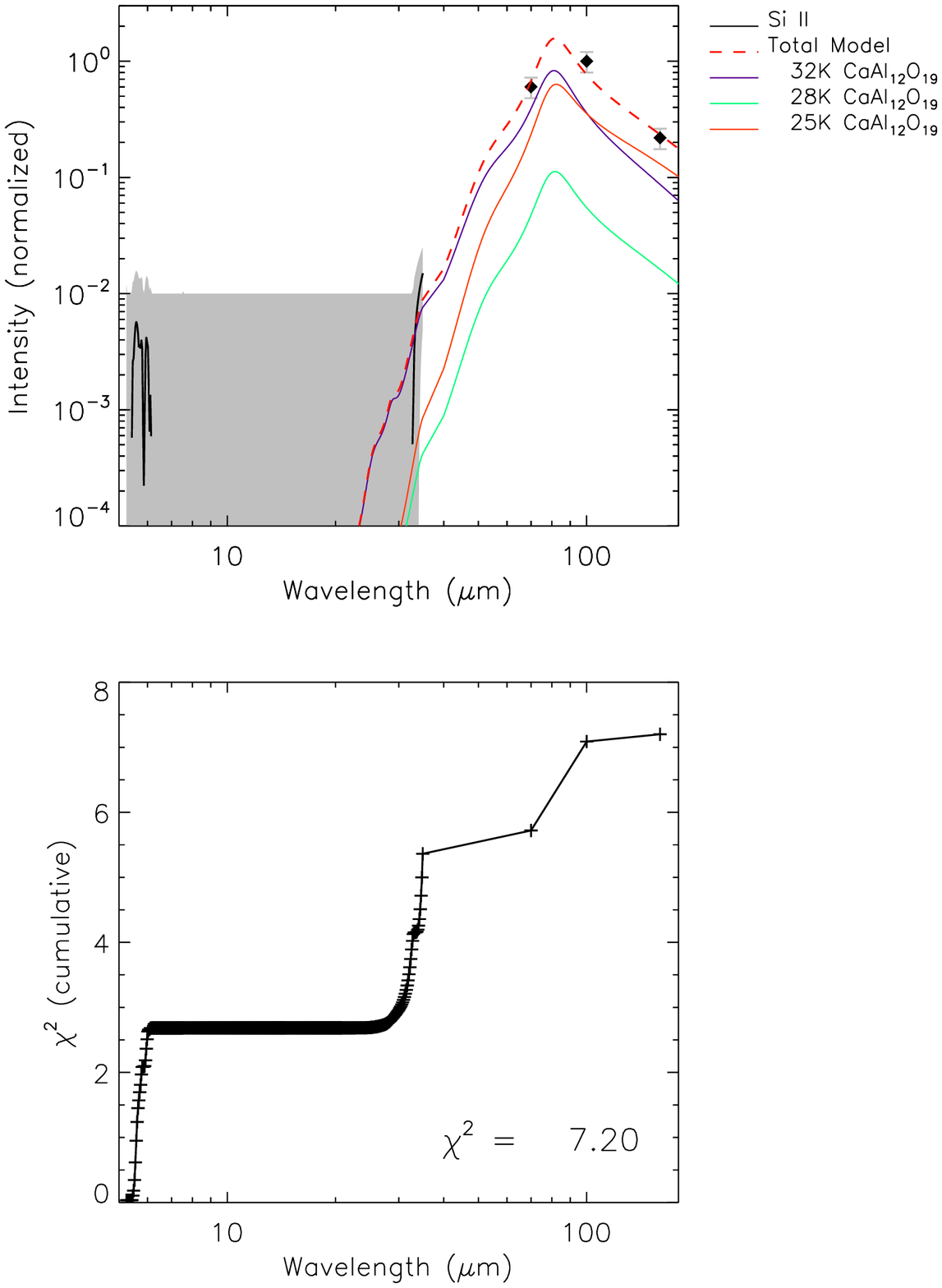}~~~~~~
  \includegraphics[height=3.5in]{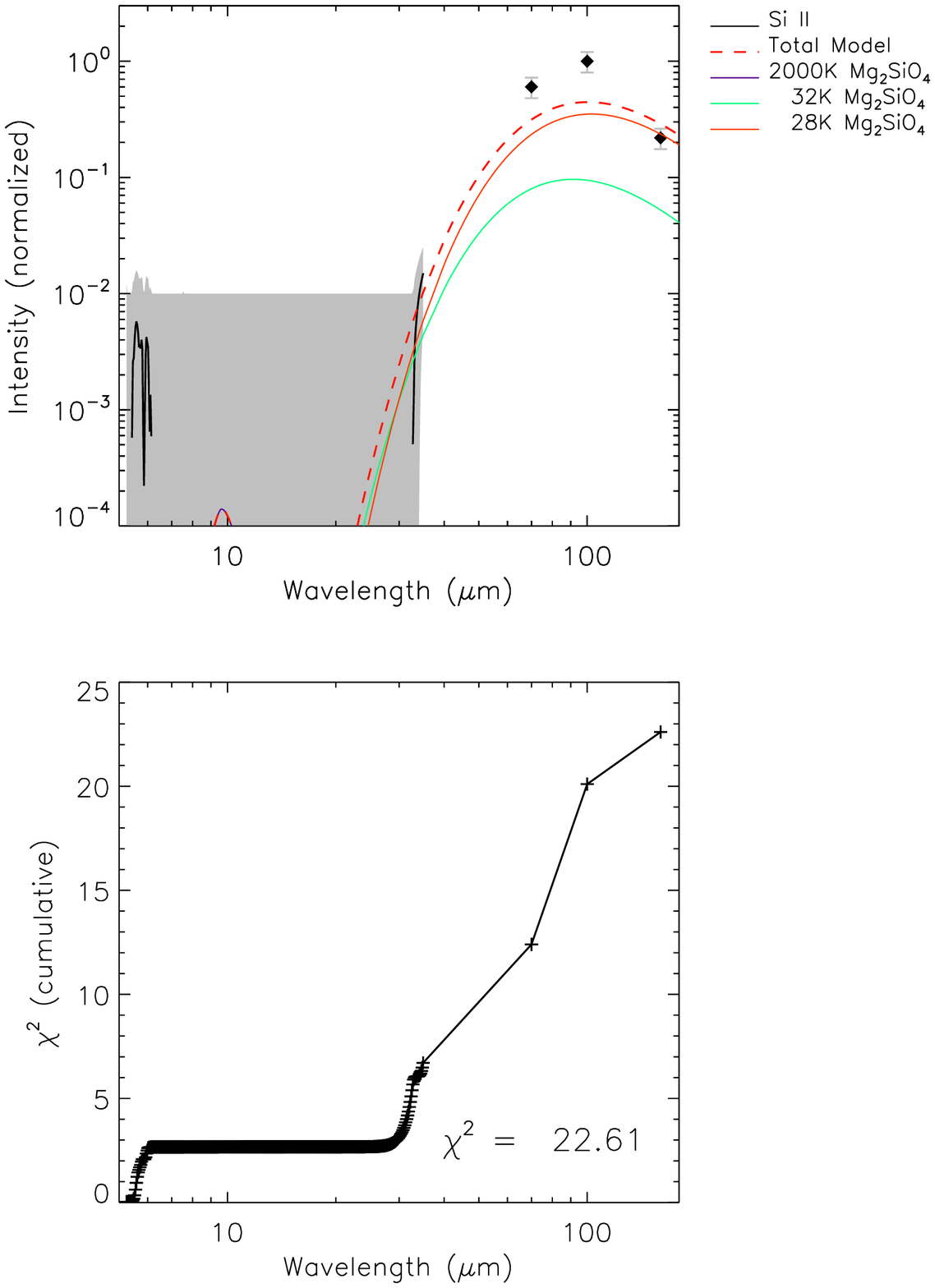}\\
   \caption{(left) The best 1-component fit to the \ion{Si}{2} spectrum uses 
   CaAl$_{12}$O$_{19}$, which happens to have a broad peak in its
   absorption efficiency at $\sim80$ $\micron$
   (right) This alternate fit uses 
   Mg$_2$SiO$_4$ which has a more typical power law absorption efficiency 
   with an index of $-2$ at long wavelengths.
   \label{fig:siiidust}}
\end{figure}

The MPFIT code \citep{Markwardt:2009} was used to determine coefficients $M_{n,i,j}$
needed to fit each normalized spectrum with the constraint that $M_{n,i,j} \geq 0$. 
In the fitting, the dust temperature is constrained to lie on a grid of values 
ranging logarithmically from $T_i=20$ to $T_i=2000$ with $\Delta \log_{10}(T_i) = 0.05$ 
($0 \leq i \leq40$).
Only components with intensities exceeding $10^{-4}$ times the peak of the fitted 
spectrum are kept in the final result.
For the purpose of fitting, the uncertainties of the characteristic spectra (normalized to a peak of 1.0) 
were taken to be $\sigma_S = 0.2 S$ with a floor of $\sigma_S \ge 0.01$.
The 20\% fractional uncertainty is chosen to represent both the noise in the measurement and 
systematic errors associated with the data reduction (e.g. joining spectral orders, subtracting
spectral lines). The 1\% (of peak flux) floor on the uncertainties represents the random 
noise terms, which generally dominate the spectra at $\lambda \lesssim 10$ $\micron$. These 
uncertainties are indicated as the gray bands in Figures \ref{fig:ariidust}-\ref{fig:siiidust}.
Given these uncertainties, spectral features at $\sim8-10$, 12, and $\sim 18-20$ $\micron$ 
reflect real features of the dust spectra, but smaller spectral details and details at 
shorter wavelengths carry little weight in the fitting.

For each spectrum, an initial set of fits was performed using only one 
dust composition, $n$, at a time. No single composition provided a good fit to the spectrum.
The fits were repeated using all possible combinations of pairs of compositions
(i.e. the left summation in Equation 3 includes only two of the possible values of $n$). 
Some combinations explored in this way result in substantial improvements in the fits.
Usually the improved fits involved the pairing of a dust composition with strong
spectral features with another composition that is relatively ``featureless.''
To check if a second ``strong--featured'' dust component was warranted, 
we also performed a set of fits using 3 compositions, but 
with the constraint that one was required to have a ``featureless''
emission spectrum. Since all the featureless compositions produce similar 
results, we fixed either graphite or the ``ac'' amorphous carbon composition 
as the required featureless component in these tests. 

\begin{figure}[t]
  \includegraphics[height=3in]{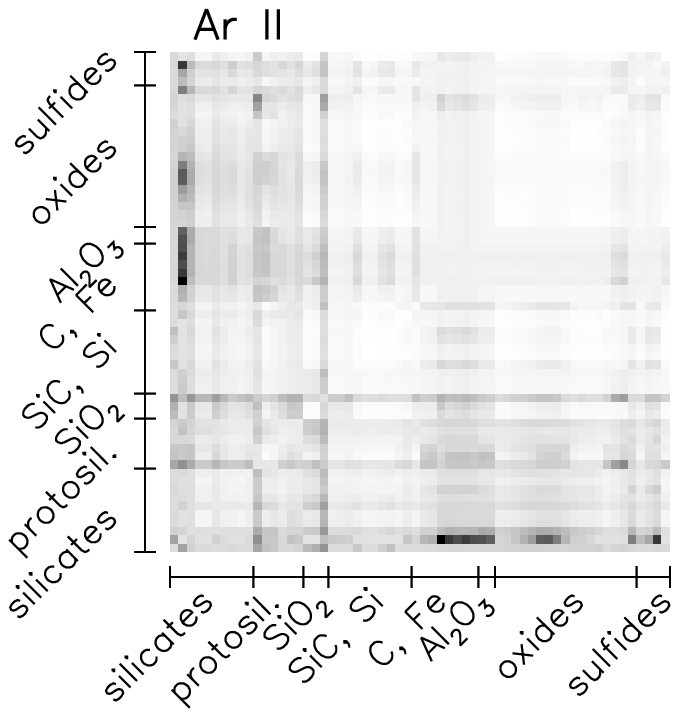}
  \includegraphics[height=3in]{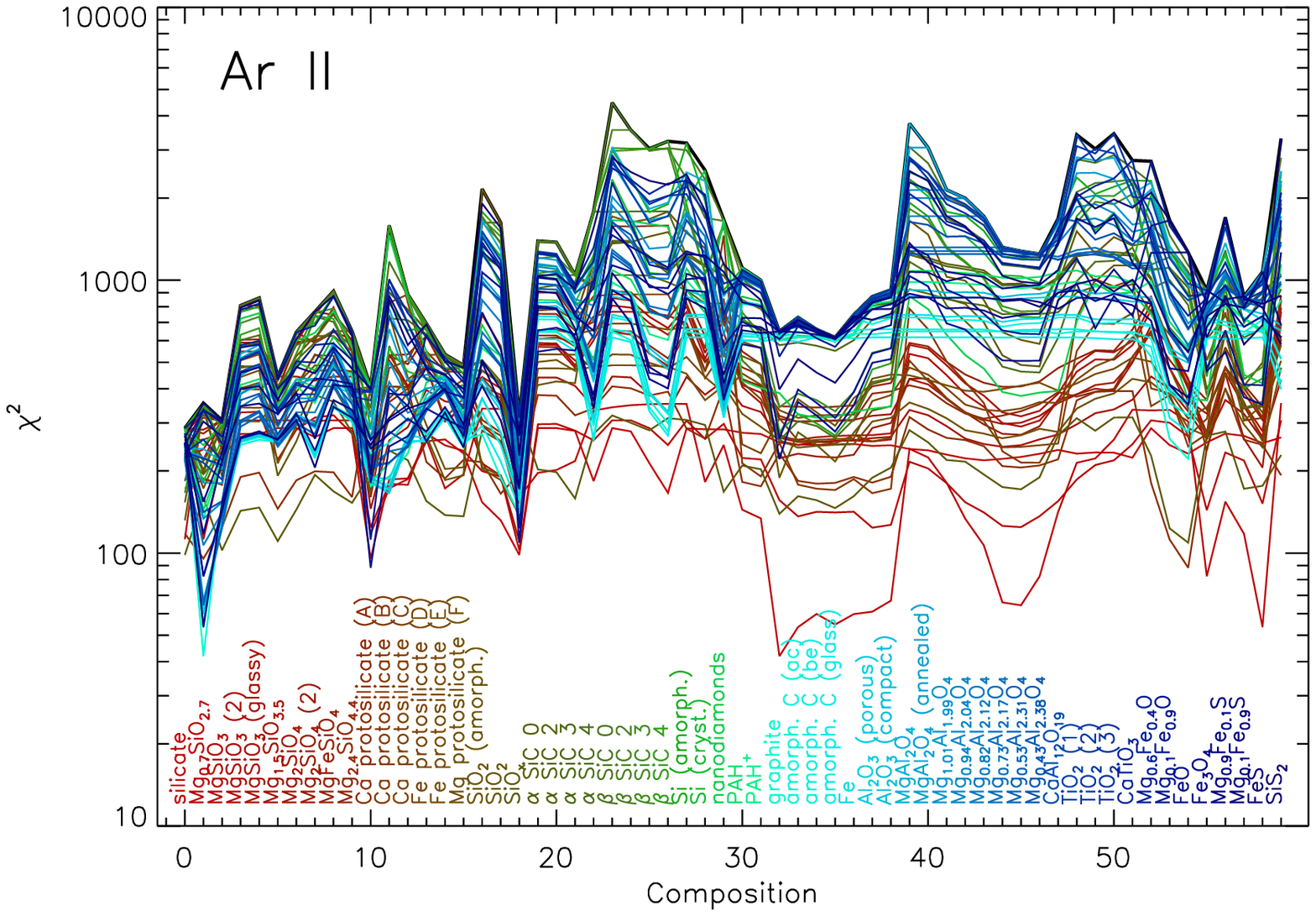}\\
   \caption{
   (left) A map of $\chi^2$ for 2-composition models of 
   the \ion{Ar}{2} dust spectrum. Darker shading indicates lower values.
   The map shows that the lowest values of $\chi^2$ are achieved when 
   one of the silicates (second row from bottom) is paired with the relatively 
   featureless dust such as C, Fe, or Al$_2$O$_3$. Good fits are also obtained by
   pairing this silicate with nonstoichiometric spinels (MgAl$_2$O$_4$, 
   i.e. oxides) with low Mg/Al ratios.
   (right) In this alternate depiction of $\chi^2$ for 2-composition models,
   each line plotted shows $\chi^2$ for a given ``primary'' dust component (as coded
   by line/label colors) when paired with each possible ``secondary'' component (as listed
   in order along the abscissa). Thus each plotted line corresponds to a row in the matrix
   shown to the left. Here one can see more specifically that $40 < \chi^2 < 80$ for 
   Mg$_{0.7}$SiO$_{2.7}$ when paired with the relatively featureless 
   graphite, amorphous C, Fe, Al$_2$O$_3$, Fe$_3$O$_4$, and FeS, or with the 
   nonstoichiometric spinels with low Mg/Al ratios.
   \label{fig:arii_chimap}}
\end{figure}

\begin{figure}[h]
  \includegraphics[height=3in]{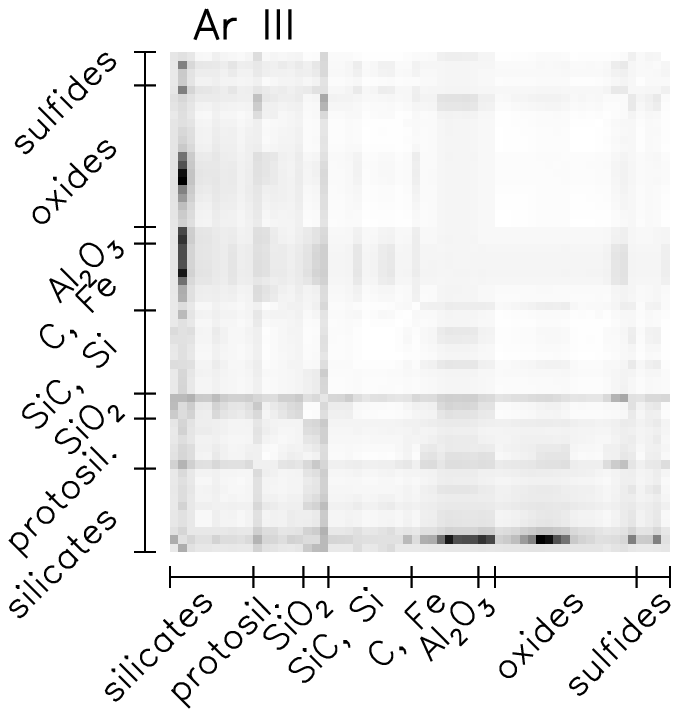}
  \includegraphics[height=3in]{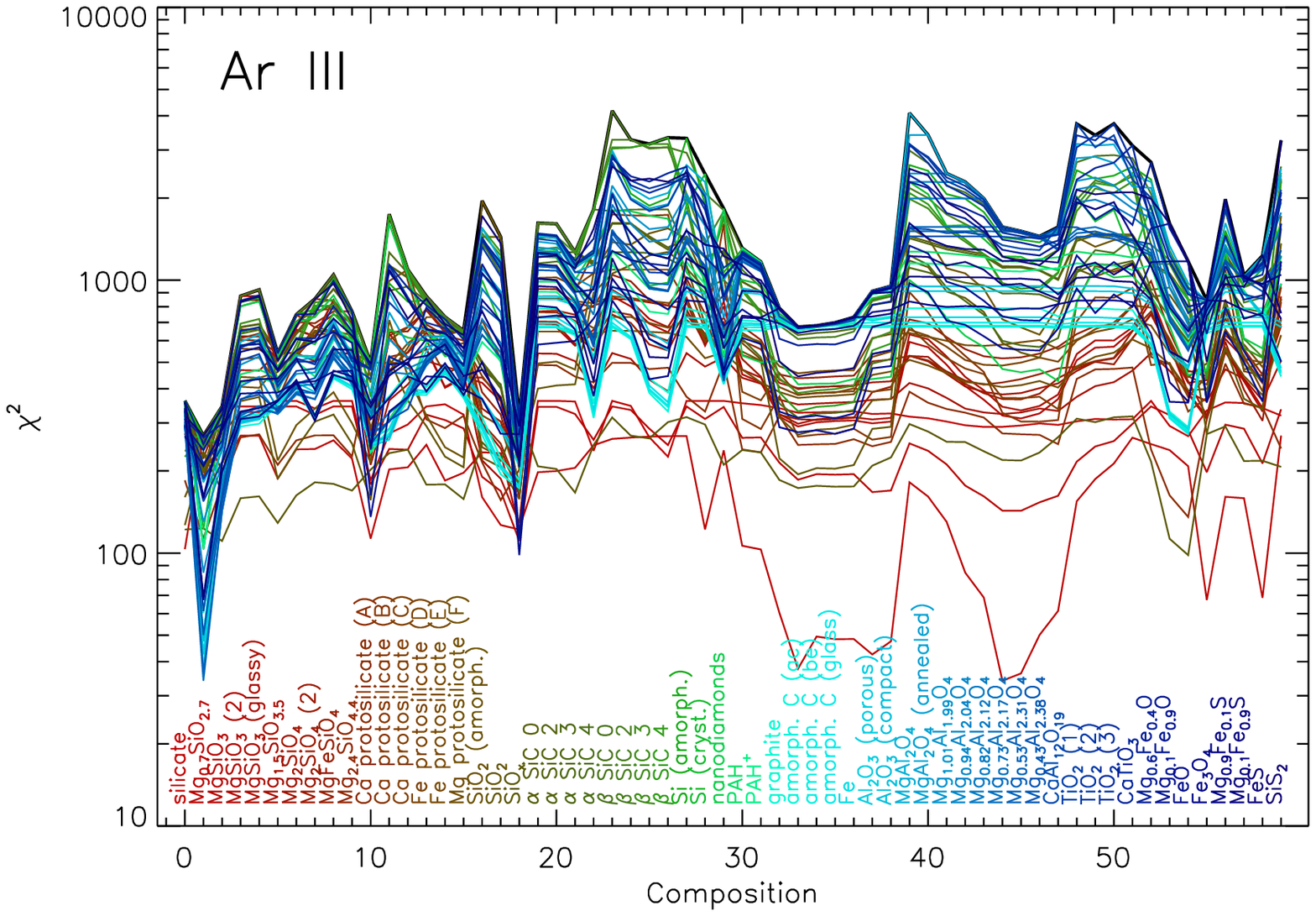}\\
   \caption{$\chi^2$ as in Fig. \ref{fig:arii_chimap}, but for the \ion{Ar}{3} dust spectrum.
   Because the \ion{Ar}{3} dust spectrum is very similar to the \ion{Ar}{2} dust spectrum, 
   $\chi^2$ for each pair of compositions is very similar in both cases.
   \label{fig:ariii_chimap}}
\end{figure}

\begin{figure}[h]
  \includegraphics[height=3in]{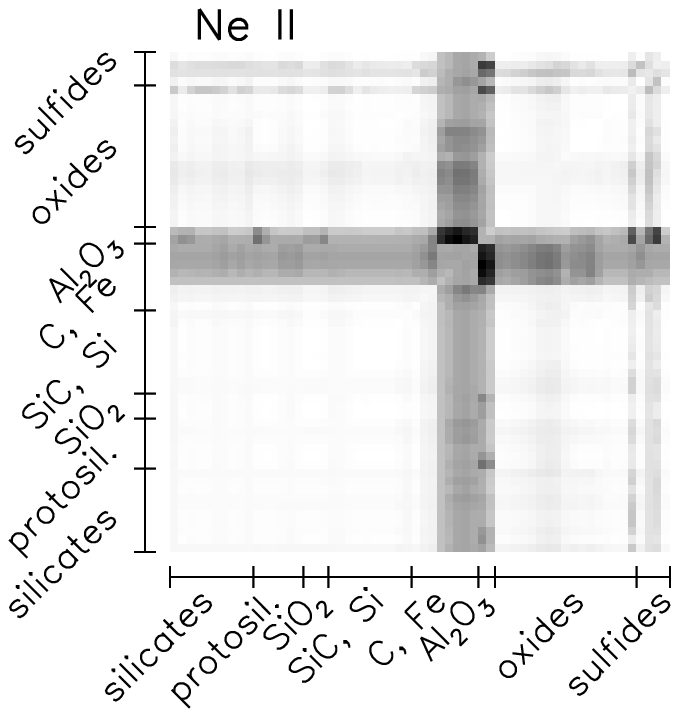}
  \includegraphics[height=3in]{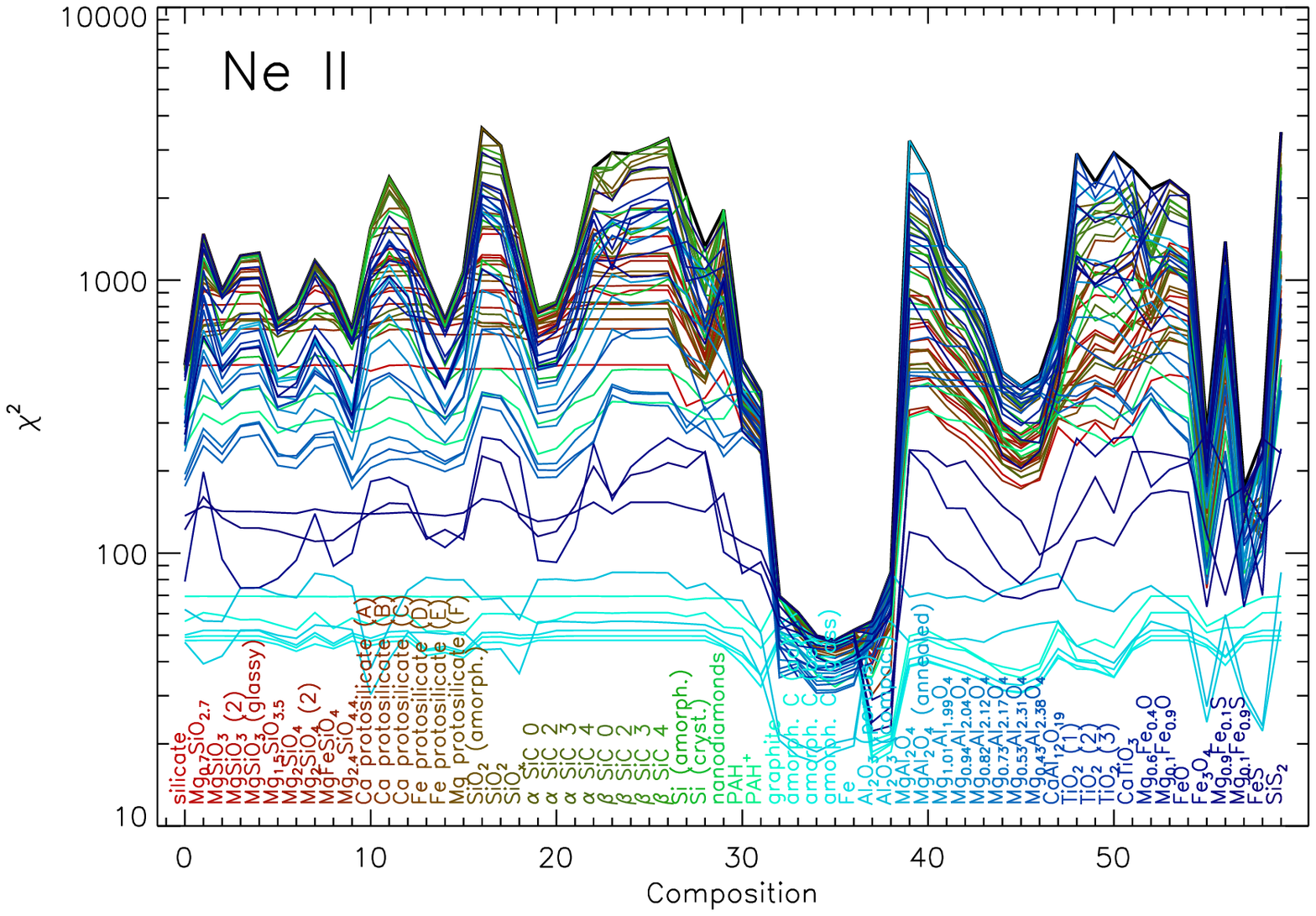}\\
   \caption{$\chi^2$ as in Fig. \ref{fig:arii_chimap}, but for the \ion{Ne}{2} dust spectrum.
   The best models contain either porous or compact Al$_2$O$_3$ paired with any of the more 
   featureless dust components.
   \label{fig:neii_chimap}}
\end{figure}

\begin{figure}[h]
  \includegraphics[height=3in]{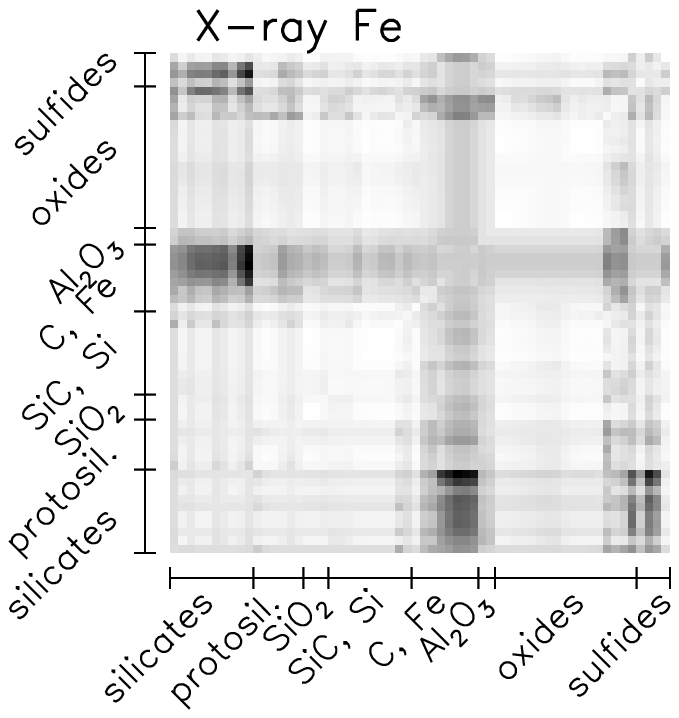}
  \includegraphics[height=3in]{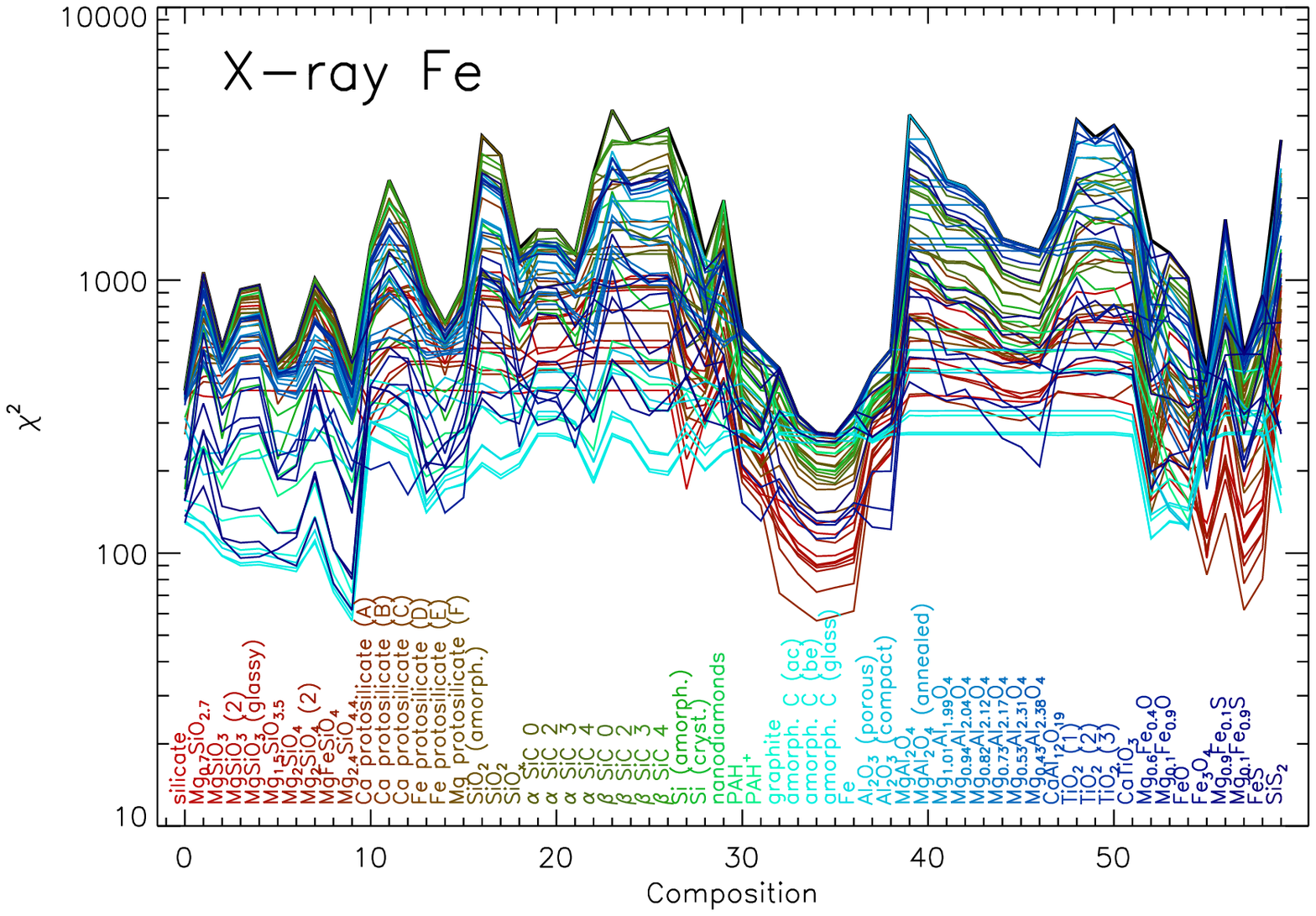}\\
   \caption{$\chi^2$ as in Fig. \ref{fig:arii_chimap}, but for the X-ray Fe dust spectrum.
   This spectrum is best fit by silicates with high Mg/Si ratios and paired with the 
   relatively featureless compositions. 
   \label{fig:xrayfe_chimap}}
\end{figure}

\begin{figure}[h]
  \includegraphics[height=3in]{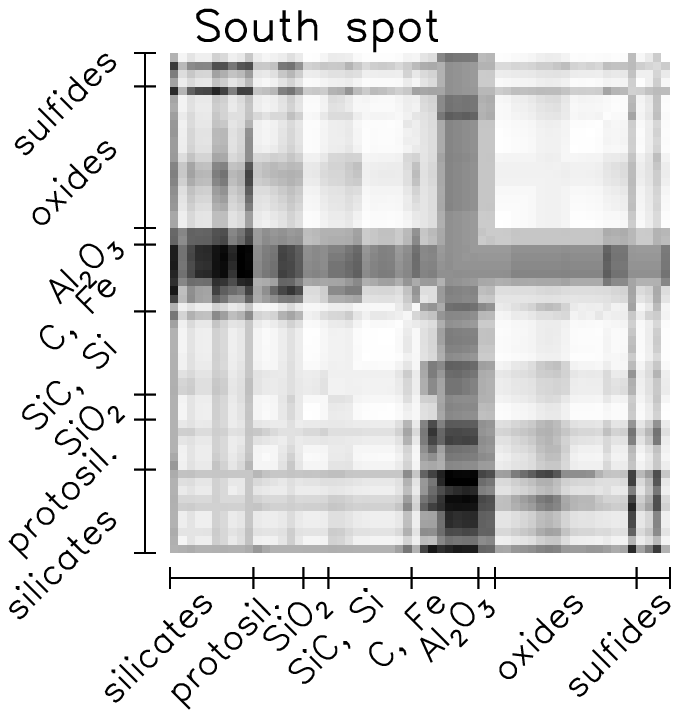}
  \includegraphics[height=3in]{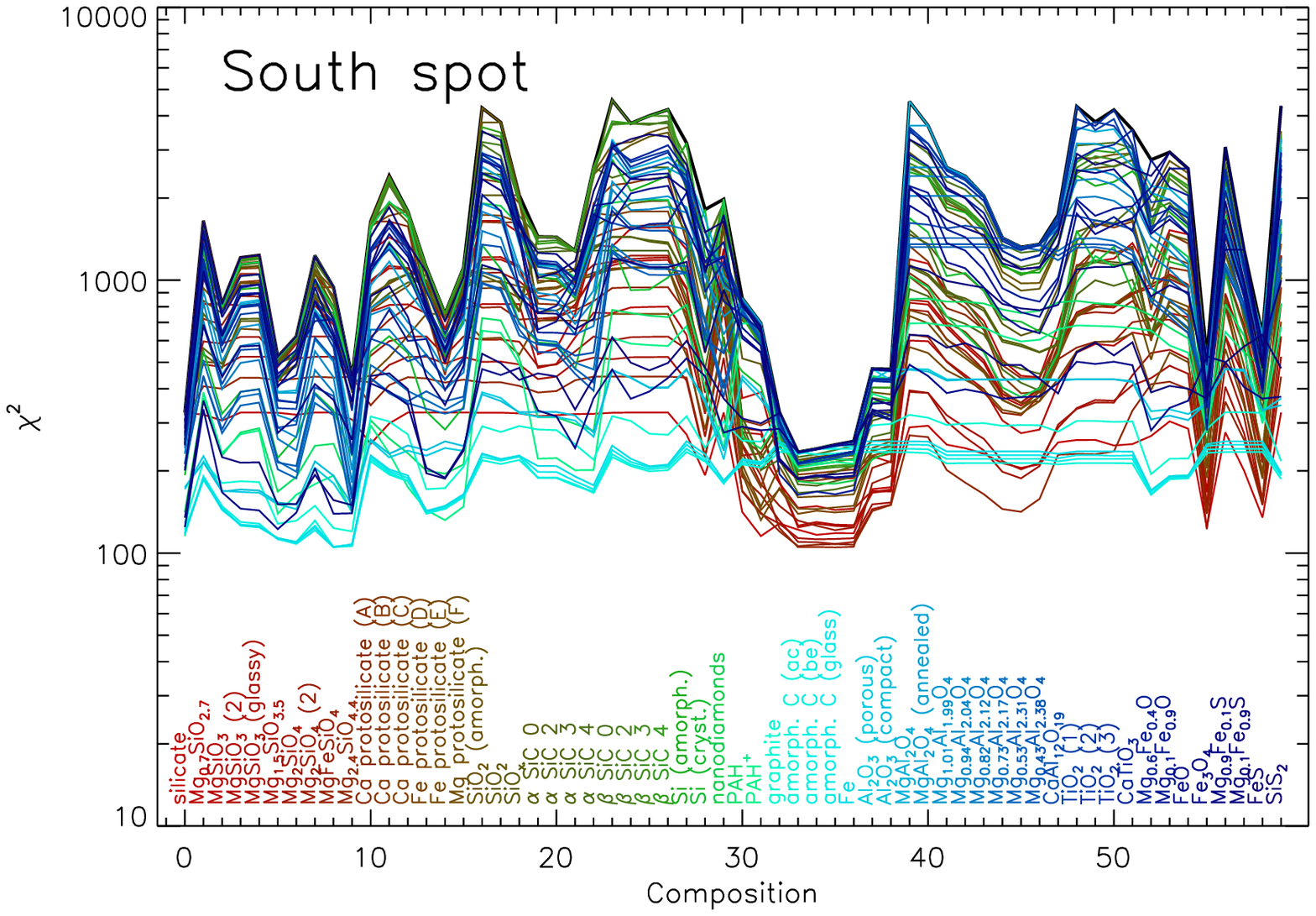}\\
   \caption{$\chi^2$ as in Fig. \ref{fig:arii_chimap}, but for the South Spot dust spectrum. 
   The results here are very similar to those for the X-ray Fe spectrum, but with somewhat less 
   contrast between the best fits and the more marginal ones.
   \label{fig:ss_chimap}}
\end{figure}

\begin{figure}[h]
  \includegraphics[height=3in]{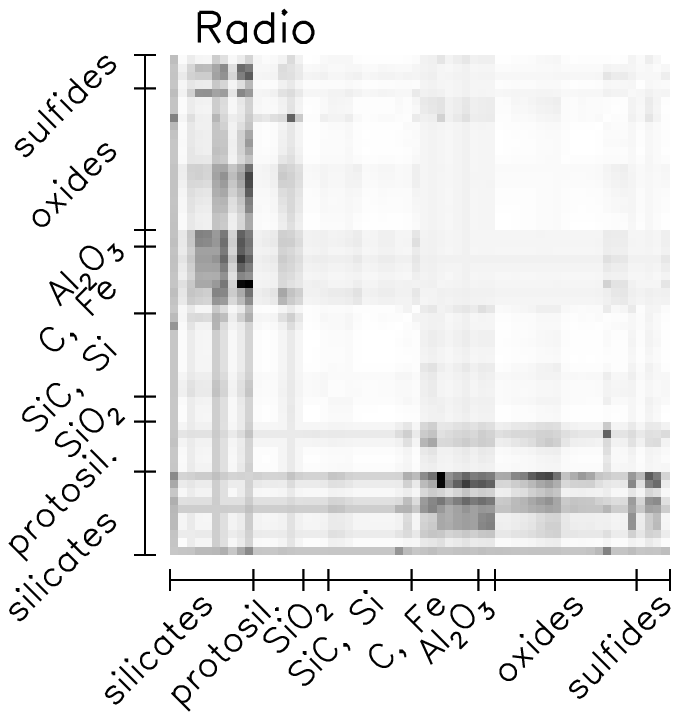}
  \includegraphics[height=3in]{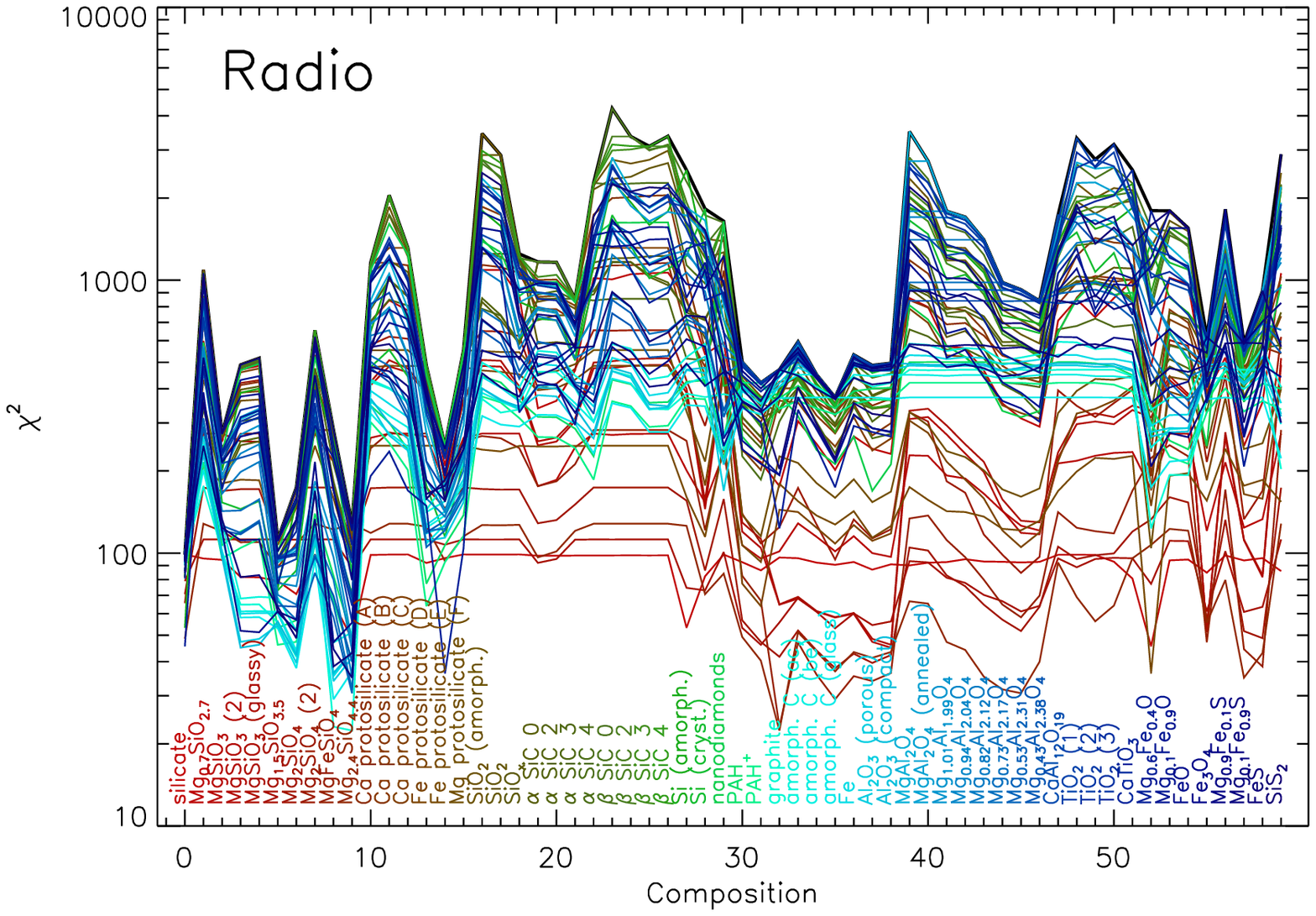}\\
   \caption{$\chi^2$ as in Fig. \ref{fig:arii_chimap}, but for the radio dust spectrum.
   The best fits here are for silicates with high Mg/Si ratios paired with the more featureless 
   compositions. 
   \label{fig:radio_chimap}}
\end{figure}

The quality of the 1- and 2-composition fits as measured by $\chi^2$ is shown in Figures
\ref{fig:arii_chimap}-\ref{fig:radio_chimap}. The left panel of each figure shows a 
matrix of the $\chi^2$ values, with generalized axes corresponding to different compositions. 
This is useful to get an overview of which pairs of compositions provide 
the best fits. The general groupings are based on chemical and/or spectral
similarity. The right panel of each of the figures displays the same $\chi^2$ values 
but now plotted as a function of the second composition and color coded according to the 
first composition. This plot quantifies $\chi^2$ and allows specific compositions to be identified. 
The results for single composition fits correspond to the upper envelope of the
curves in these plots (or the diagonal of the matrix in the left panels).
Examples of the best fitting 2- and 3-composition models, as well as contrasting models 
that also provide reasonably good fits, are shown in Figures \ref{fig:ariidust}-\ref{fig:siiidust}.
The results for the 7 different characteristic spectra are discussed in more detail below.

\subsection{\ion{Ar}{2} Dust}
The characteristic dust spectrum 
associated with \ion{Ar}{2} exhibits the sharpest peaks at 9 and 21 $\micron$
and has a third, weaker peak at $\sim$12 $\micron$. The astronomical silicate of the 
typical ISM does not have silicate peaks sharp enough to produce the observed features. 
Magnesium silicates (particularly Mg$_{0.7}$SiO$_{2.7}$, characterized by Mg/Si = 0.7) 
have peaks that are in roughly the 
right locations, but their shapes are not a good match to the data.
However, when two-composition fits are performed, the addition of a second component
lacking strong spectral features (e.g. graphite, amorphous C, or Al$_2$O$_3$), can 
soften and smooth the appearance of the silicate peaks to provide a much better fit to the spectrum. 

\begin{figure}[hp]
  \includegraphics[height=3.5in]{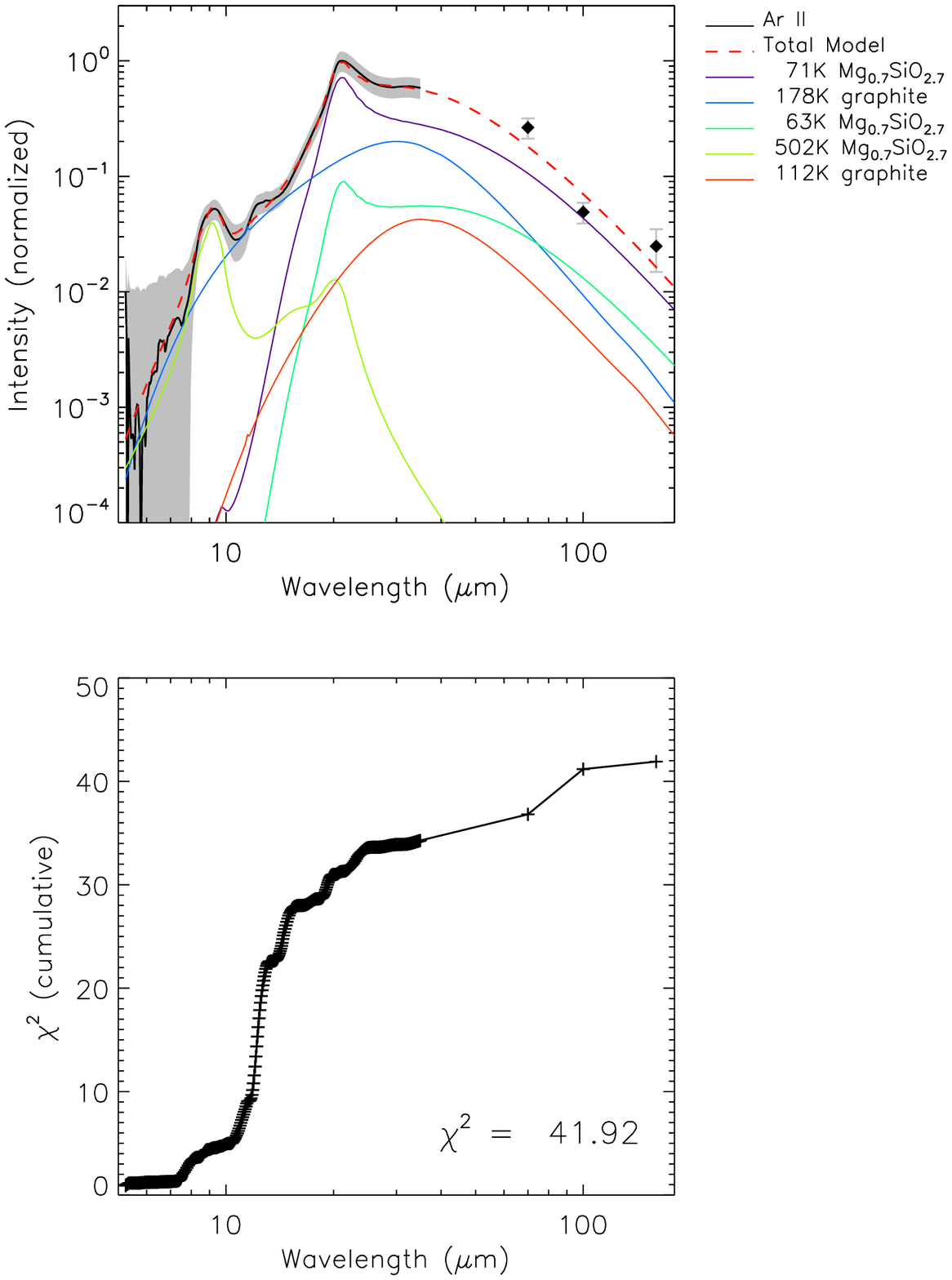}~~~~~~
  \includegraphics[height=3.5in]{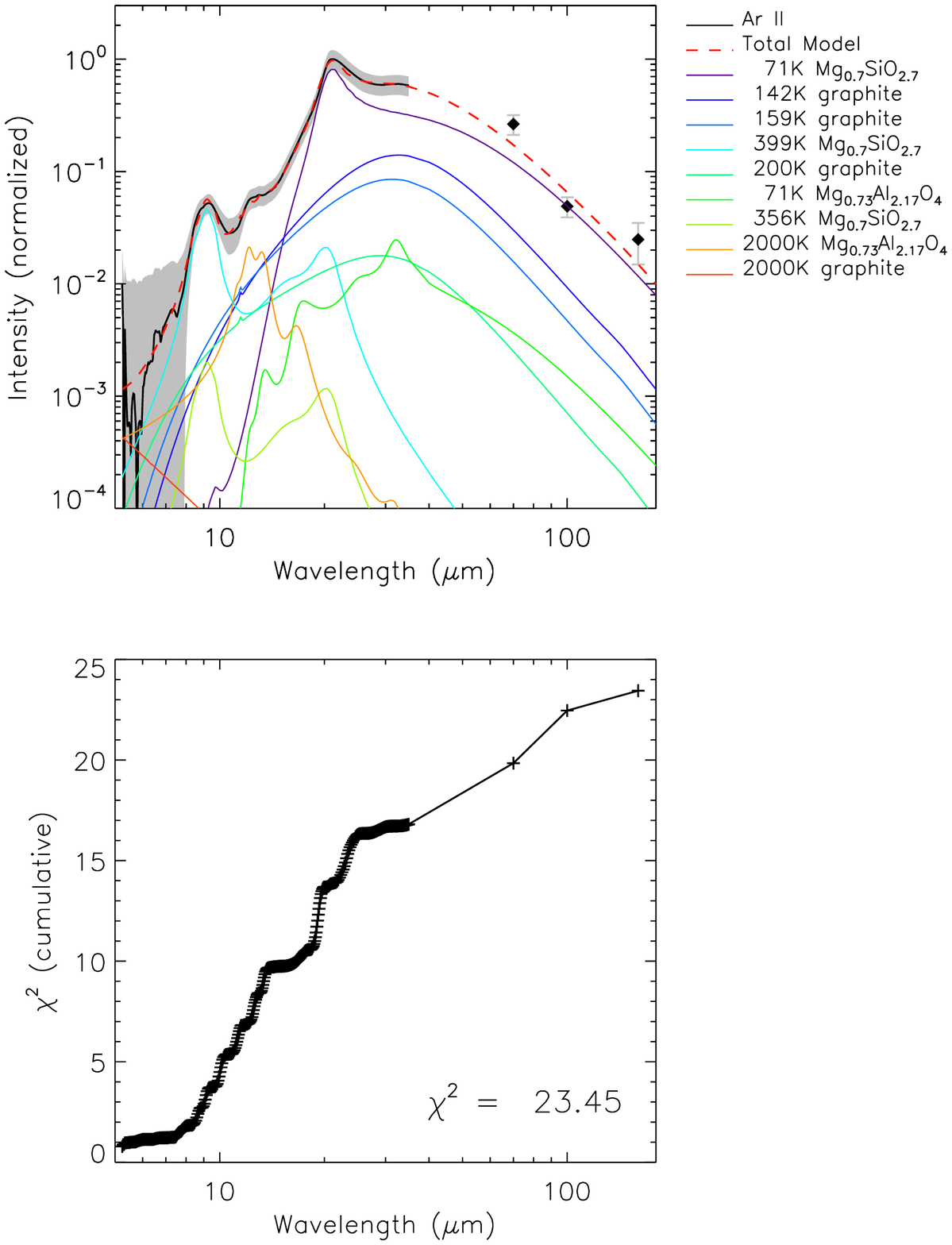}\\
  ~~\\
  \includegraphics[height=3.5in]{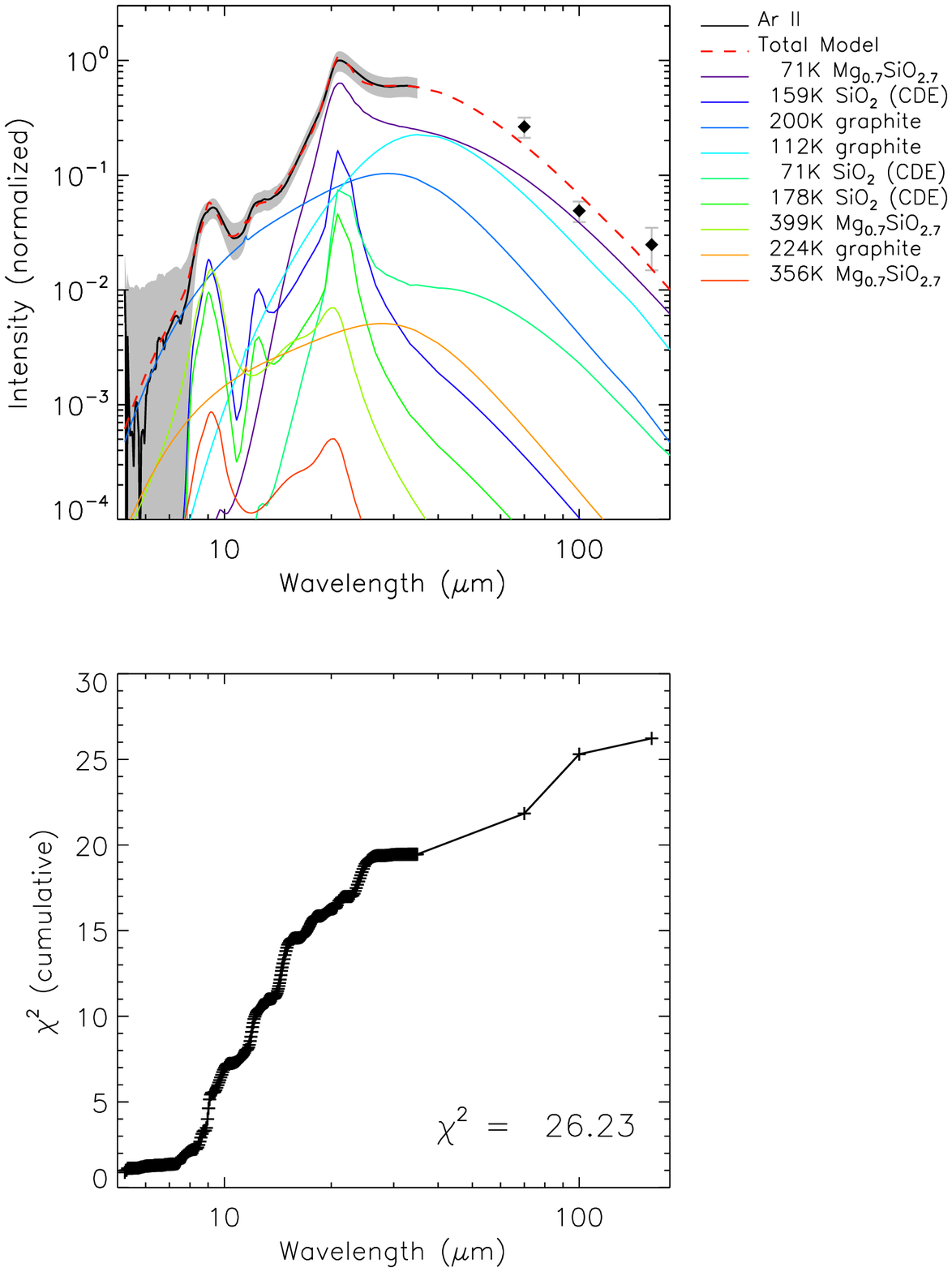}~~~~~~
  \includegraphics[height=3.5in]{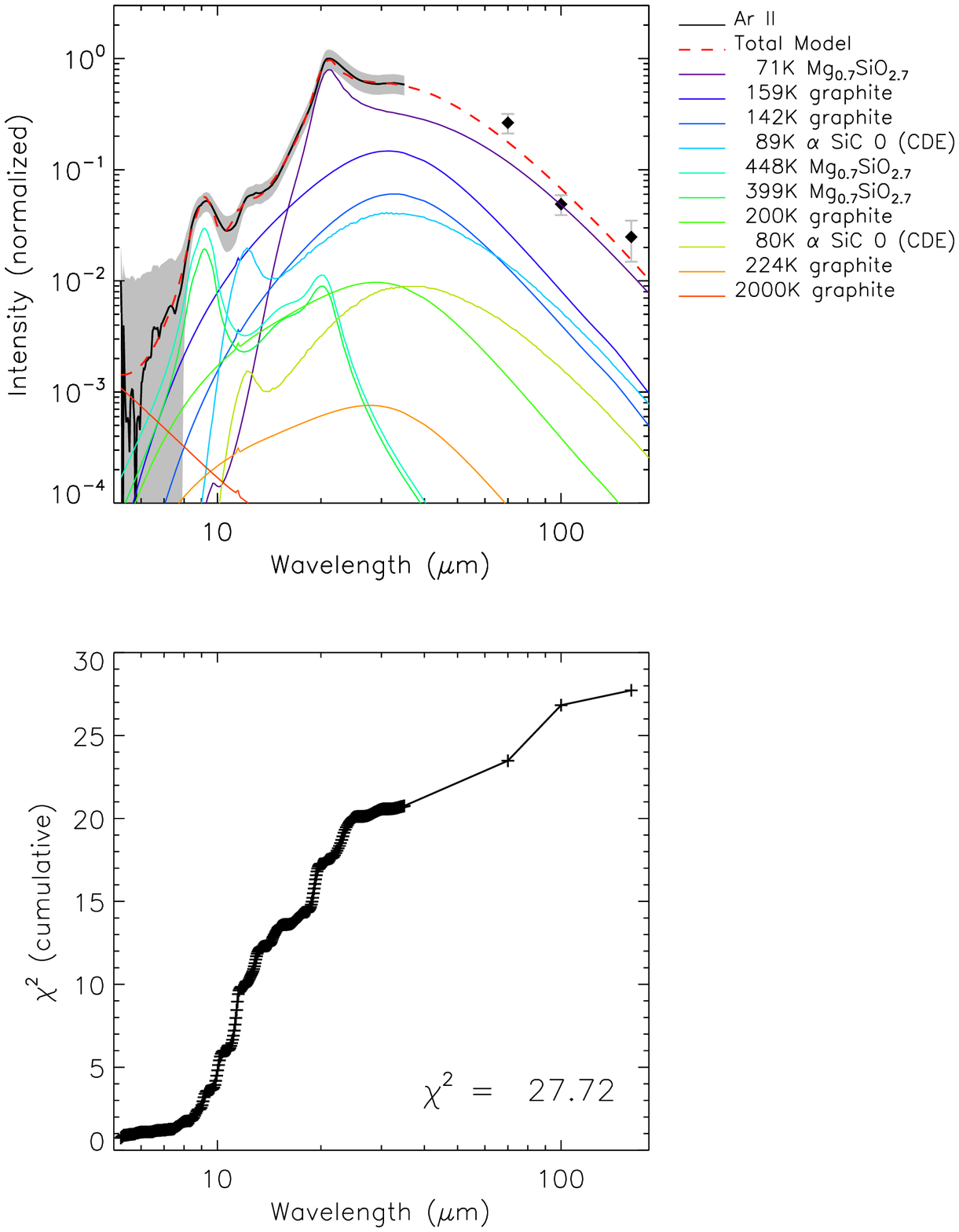}\\
\vspace{-0.25in}  
   \caption{(upper left) The best 2-composition fit to the \ion{Ar}{2} spectrum includes Mg$_{0.7}$SiO$_{2.7}$
   and a featureless dust composition. Uncertainties of the IRS data are indicated by the gray band. 
   The lower panel shows how $\chi^2$ accumulates as a function of wavelength.
   (upper right) The best 3-component fit to the spectrum is a good match to all
   of the features at 9, 12 and 21 $\micron$. However, the 12 $\micron$ feature is 
   provided by implausibly hot nonstoichiometric spinel.
   (lower left) A very good 3-component fit can also 
   be obtained with SiO$_2$ as the third component, if its mass absorption 
   coefficient is calculated via CDE rather than Mie theory. SiO$_2$ has features
   at each of the observed peaks.
   (lower right) Another very good fit can be obtained with SiC as the third component, again 
   if its mass absorption coefficient is calculated via CDE rather than Mie theory. 
   SiC contributes only at 12 $\micron$, but may also provide a partial match of the 21 $\micron$
   feature \citep{Jiang:2005}.
   \label{fig:ariidust}}
\end{figure}

The best 2-component fit (see Figure \ref{fig:ariidust}) employs Mg$_{0.7}$SiO$_{2.7}$
to fit the 9 and 21 $\micron$ silicate peaks, and graphite as the featureless component
that weakens the relative strength of the peaks. The fit to these peaks is very good,
although the weaker 12 $\micron$ peak is not fit at all. Despite the overall good
fit, the temperature components needed for the Mg silicate are somewhat suspicious. Components at 
71--63 K are needed to fit the 21 $\micron$ feature, but are too cold to contribute 
to the 9 $\micron$ feature. Conversely a 500 K component produces the 9 $\micron$ 
peak while contributing little to the 21 $\micron$ peak. Small, stochastically heated grains 
might be expected to be heated to this range of temperatures, but it is 
problematic that there is no indication of grains at intermediate
temperatures (70 K$ < T < 400$ K). This suggests that the absorption efficiencies used here are not 
a true physical representation of the \ion{Ar}{2} dust. The actual dust must either have 
a relatively stronger 9 $\micron$ peak so that it can produce the observed spectrum from 
cooler dust, or it must have a somewhat lower absorption efficiency at roughly 10--20 $\micron$ so that 
the presence of grains at intermediate temperatures would be required.

As suggested by the best 2-component fits, the best 3-component fit combines 
the Mg$_{0.7}$SiO$_{2.7}$ with graphite (to weaken the silicate peaks)
and adds a nonstoichiometric spinel to provide the 12 $\micron$ peak 
(see Fig. \ref{fig:ariidust}). However there is still a tendency in this fit to use 
distinctly different temperatures and/or compositions to fit each different 
spectral feature. The spinel material that produces the 12 $\micron$ peak is 
required to be at extremely high temperatures so that it only contributes to this peak. 
Alternate sources of the 
12 $\micron$ feature are SiO$_2$ and SiC. For both of these materials, the spectral 
features are too sharp and slightly too blue to provide a good match when the
mass absorption coefficients, $\kappa(\lambda)$, 
are calculated from Mie theory for spherical grains. 
However, calculations assuming a continuous distribution of ellipsoids (CDE) 
for particle shapes \citep{Bohren:1983}, 
result in spectral features that are broader and redder.
The dust temperatures required to produce the features are also much more reasonable.
Figure \ref{fig:ariidust} also shows the fits when 
CDE calculations are used for a third composition of SiO$_2$ or $\alpha$-SiC.
SiO$_2$ provides some contribution to all three of the observed peaks. 
SiC only provides the 12 $\micron$ peak. 

\subsection{\ion{Ar}{3} Dust}
The \ion{Ar}{3} dust spectrum is very similar to that of the \ion{Ar}{2} dust. 
It was derived separately
to check if there might be an identifiable shift in dust temperature or composition 
in these closely related regions. Not surprisingly, the combinations that produced good fits to
the \ion{Ar}{2} spectrum also produce good fits to the \ion{Ar}{3} spectrum
(see Figure \ref{fig:ariiidust}).
The increased strength of the 9 $\micron$ peak relative to the 21 $\micron$ peak 
would usually suggest warmer dust temperatures in association with the \ion{Ar}{3}.
However, because the dust features arise largely from separate components of the models,
the temperature of the components are largely unchanged, but the relative mass in the 
hotter components is higher for the \ion{Ar}{3} than for the \ion{Ar}{2} spectrum.

For the \ion{Ar}{3} spectrum we consistently find that the amorphous C (ac) provides
a slightly better fit than the graphite that was somewhat preferred for \ion{Ar}{2}
spectrum. The \ion{Ar}{3} spectrum is also slightly better fit when hibonite 
(CaAl$_{12}$O$_{19}$) rather than a spinel composition 
is used as a third component, although if CDE calculations 
are applied, SiO$_2$ and SiC can again provide equally good fits as the 
third component.

\subsection{\ion{Ne}{2} Dust}
\begin{figure}[hp]
  \includegraphics[height=3.5in]{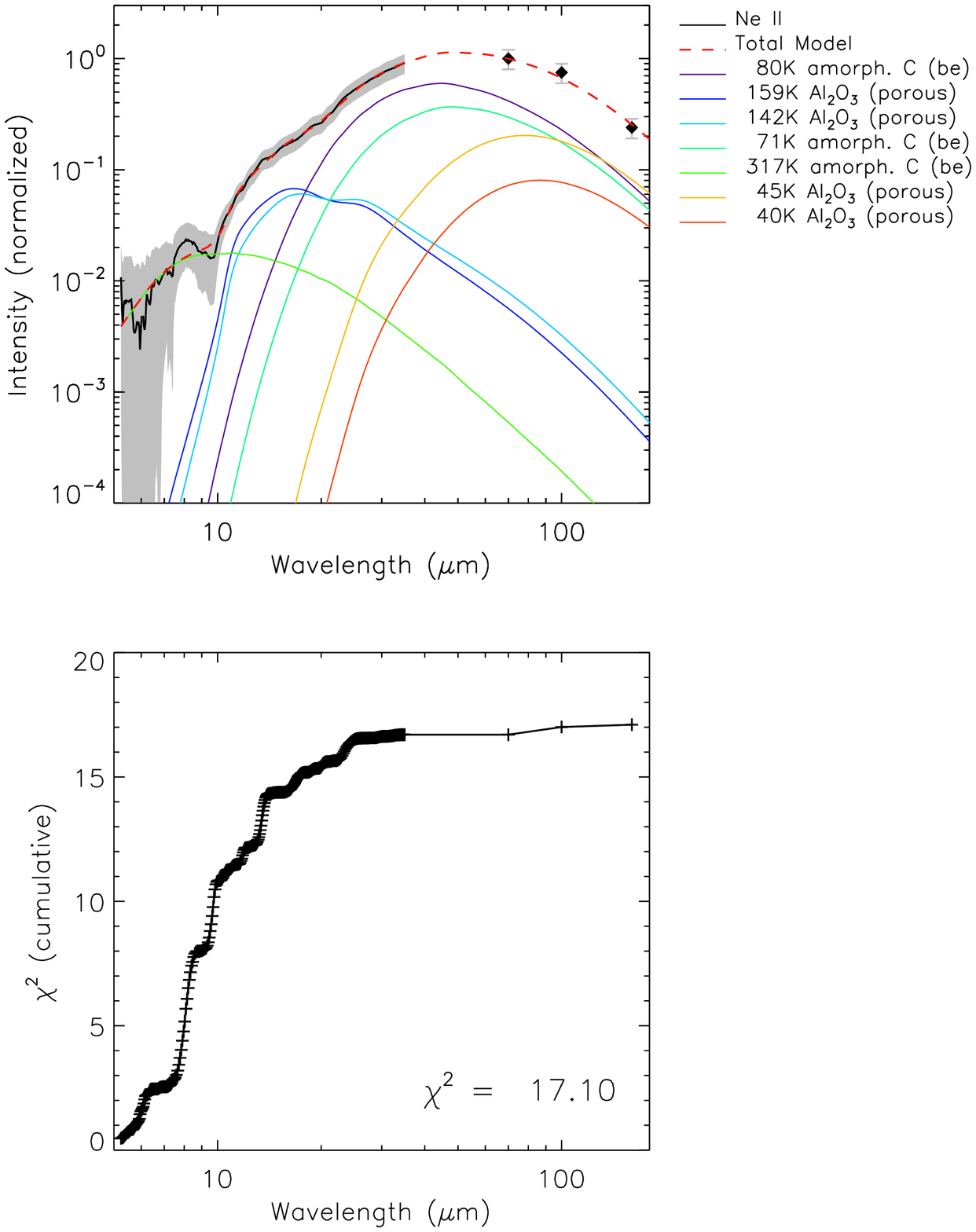}~~~~~~
  \includegraphics[height=3.5in]{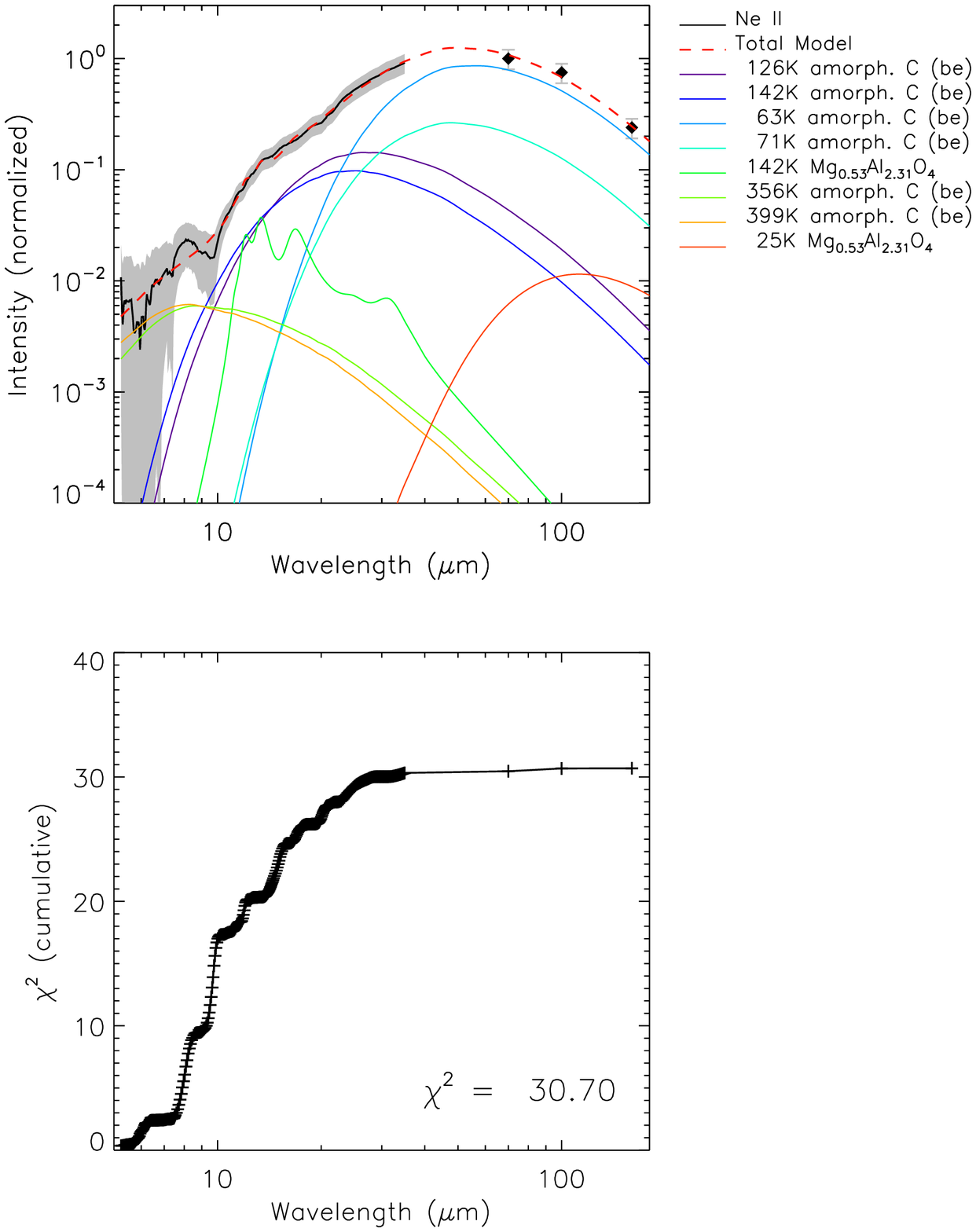}\\
  ~~\\
  \includegraphics[height=3.5in]{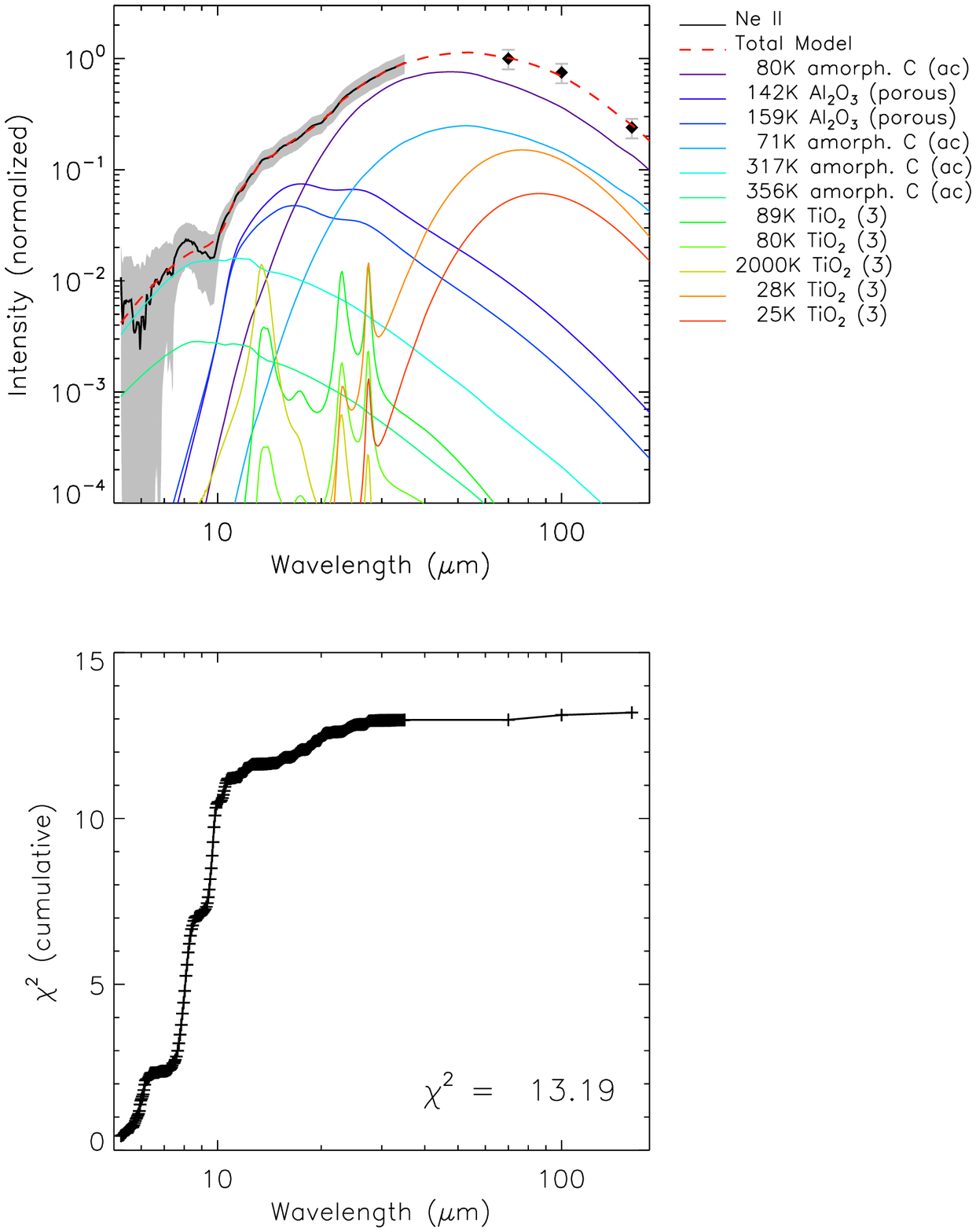}\\
   \caption{(upper left) The best 2-component fits to the \ion{Ne}{2} spectrum use Al$_2$O$_3$ in combination 
   with a more featureless component. None of the silicates are a good match for 
   this very smooth spectrum.
   (upper right) An alternate 2-component fit to the spectrum can be found using
   nonstoichiometric spinel instead of Al$_2$O$_3$.
   (lower left) The best 3-component fit adds TiO$_2$, but the 
   improvement in the fit is very small and does not justify the addition of the third composition.
   \label{fig:neiidust}}
\end{figure}

The dust associated with \ion{Ne}{2} exhibits an unusually smooth spectrum.
This spectrum excludes significant amounts of warm silicates, but can be fit 
 by compositions with relatively featureless spectra, e.g. graphite, 
and amorphous carbon. The best fits are found by combining featureless dust 
with Al$_2$O$_3$. The latter component has a very broad emission peak between 10 
and 20 $\micron$ that improves the fit. Figure \ref{fig:neiidust} shows the fit for 
an amorphous carbon (be) and Al$_2$O$_3$ mixture.
The fit suggests that some fraction of the Al$_2$O$_3$ component is fairly warm ($T \approx 150$ K),
while the brightest carbon dust has a temperature
of $\sim75$ K. A significant cool ($\sim40$ K) dust component 
helps to fit the {\it Herschel} data at 70 -- 160 $\micron$. The model attributes 
this to Al$_2$O$_3$, but this identification is degenerate because neither of
these compositions have any characteristic spectral features at these wavelengths.
The composition of very cool components cannot be identified very conclusively,
because the shape of the spectra are hardly affected by composition. 

Figure \ref{fig:neiidust} also shows the best fit model that does not contain Al$_2$O$_3$.
In this case a nonstoichiometric spinel provides a relatively small perturbation 
to an otherwise smooth amorphous C spectrum. Amorphous C picks up the 
intermediate temperature ($\sim135$K) components that account for the bulk 
of the 10 -- 20 $\micron$ emission. 

Finally, Figure \ref{fig:neiidust} also shows the best 3-component model for the \ion{Ne}{2} dust.
The addition of TiO$_2$ nominally improves the fit over the best 2-component model,
but the changes to the spectrum are too minor to suggest that any third component is warranted. 

\subsection{X-Ray Fe Dust}
The spectrum of dust associated with the X-ray Fe emission does exhibit apparent 
silicate peaks at $\sim10$ and 20 $\micron$. The 10 $\micron$ peak is relatively 
weak, suggesting a lack of hot silicate grains. As with the \ion{Ar}{2} dust,
mixing silicate and more featureless emission (e.g. amorphous C) improves the 
quality of the fit. The preferred silicates here tend to have higher 
Mg/Si ratios than those that fit the \ion{Ar}{2} and \ion{Ar}{3} dust. 
Figure \ref{fig:xrayfedust} shows the best 2--composition fit. 
The temperatures for both components are warm, $T \approx 100-112$ K, 
with additional hot and cold components to account for the
short and long wavelength emission. 
The figure also shows the best 2--component fit that can be obtained without
the use of silicate compositions. This marginal fit uses Mg$_{0.6}$Fe$_{0.4}$O 
to adjust the shape of the model spectrum near 20 $\micron$, but contains no 
replacement for a 10 $\micron$ silicate feature.
The best 3--component fit (see Fig. \ref{fig:xrayfedust}) merely adds PAH$^+$ 
emission to reproduce a bump at 8 $\micron$ and sharpen the spectral feature 
at $\sim10$ $\micron$. The overall change in $\chi^2$ is small.

\subsection{South Spot Dust}
The South Spot spectrum looks rather different than the X-ray Fe spectrum, yet
it tends to be best fit by similar dust compositions.
Here the temperatures of the components that 
produce the $<20$ $\micron$ emission are higher
than those for the X-ray Fe dust. Figure \ref{fig:ssdust} shows the 
best 2--component fit which uses similar compositions as the X-ray Fe spectrum 
(cf. Fig. \ref{fig:xrayfedust}). It also shows a marginal 
fit obtained without the use of silicates, and thus without any component 
that provides a 10 $\micron$ feature in the spectrum. The best 3--component fit
(see Fig. \ref{fig:ssdust}) uses nonstoichiometric spinel to make minor adjustments 
to the shape of a good 2--component model.

\subsection{Radio (ISM) Dust}
\begin{figure}[hp]
  \includegraphics[height=3.5in]{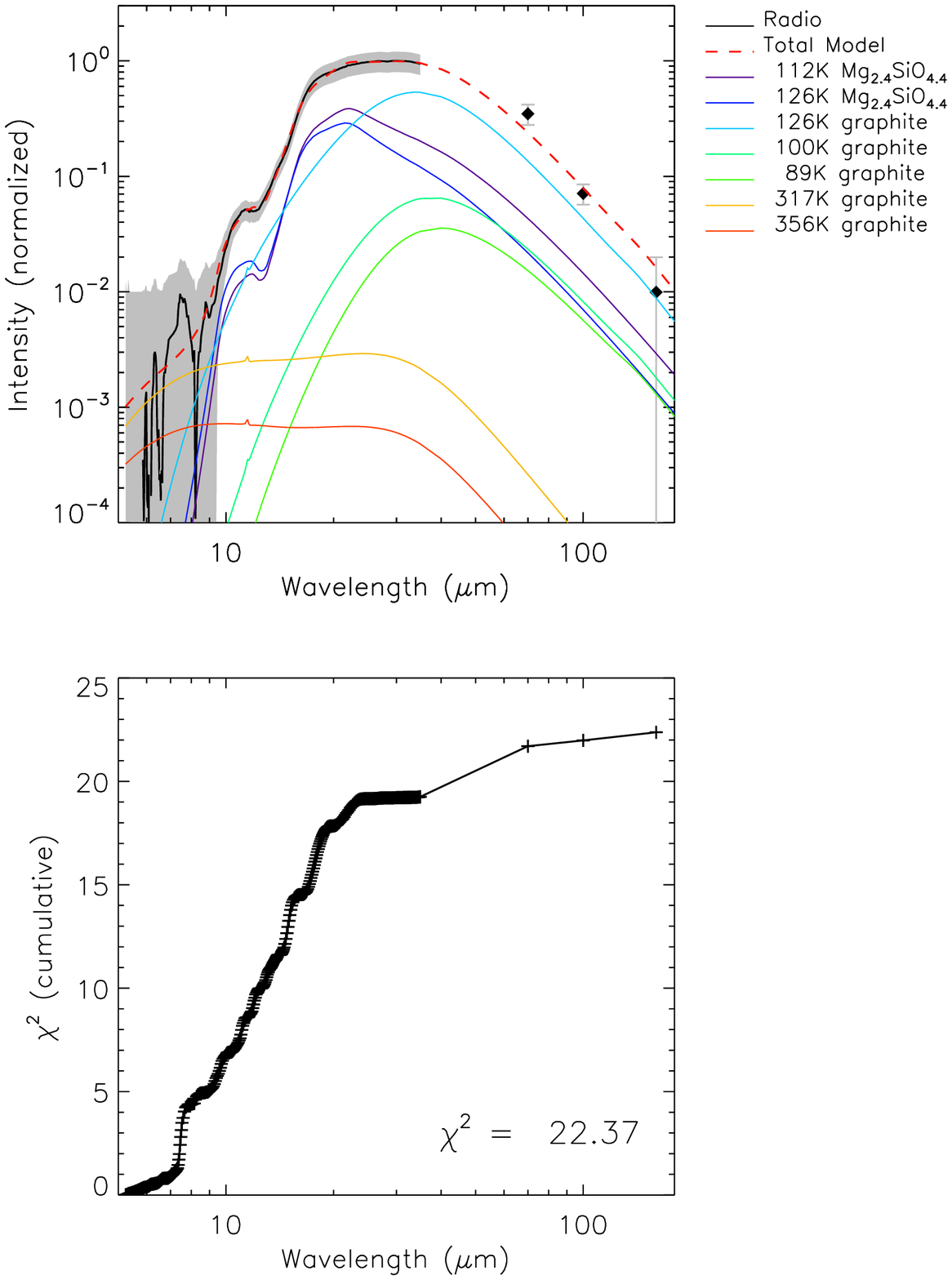}~~~~~~
  \includegraphics[height=3.5in]{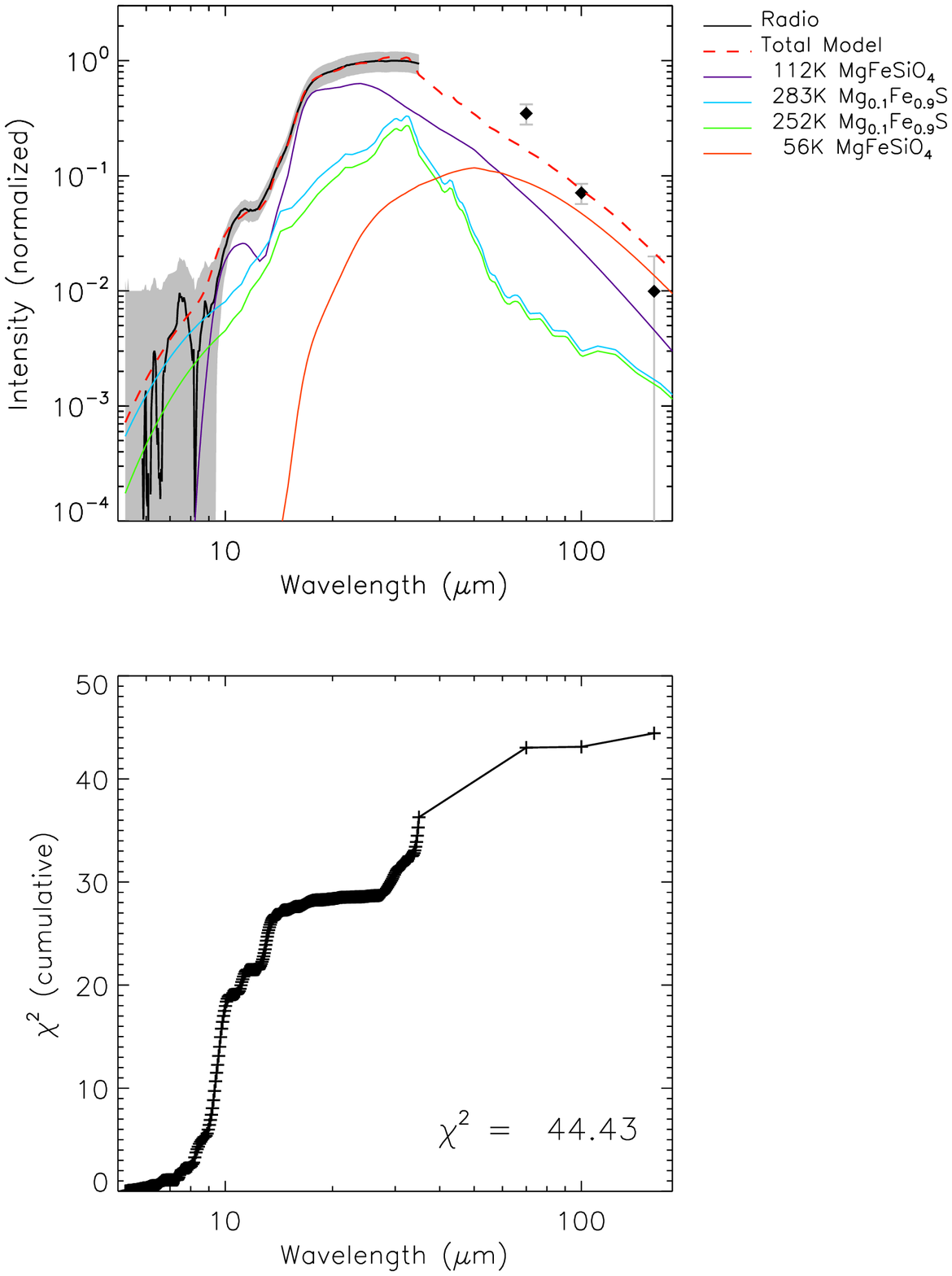}\\
  ~~\\
  \includegraphics[height=3.5in]{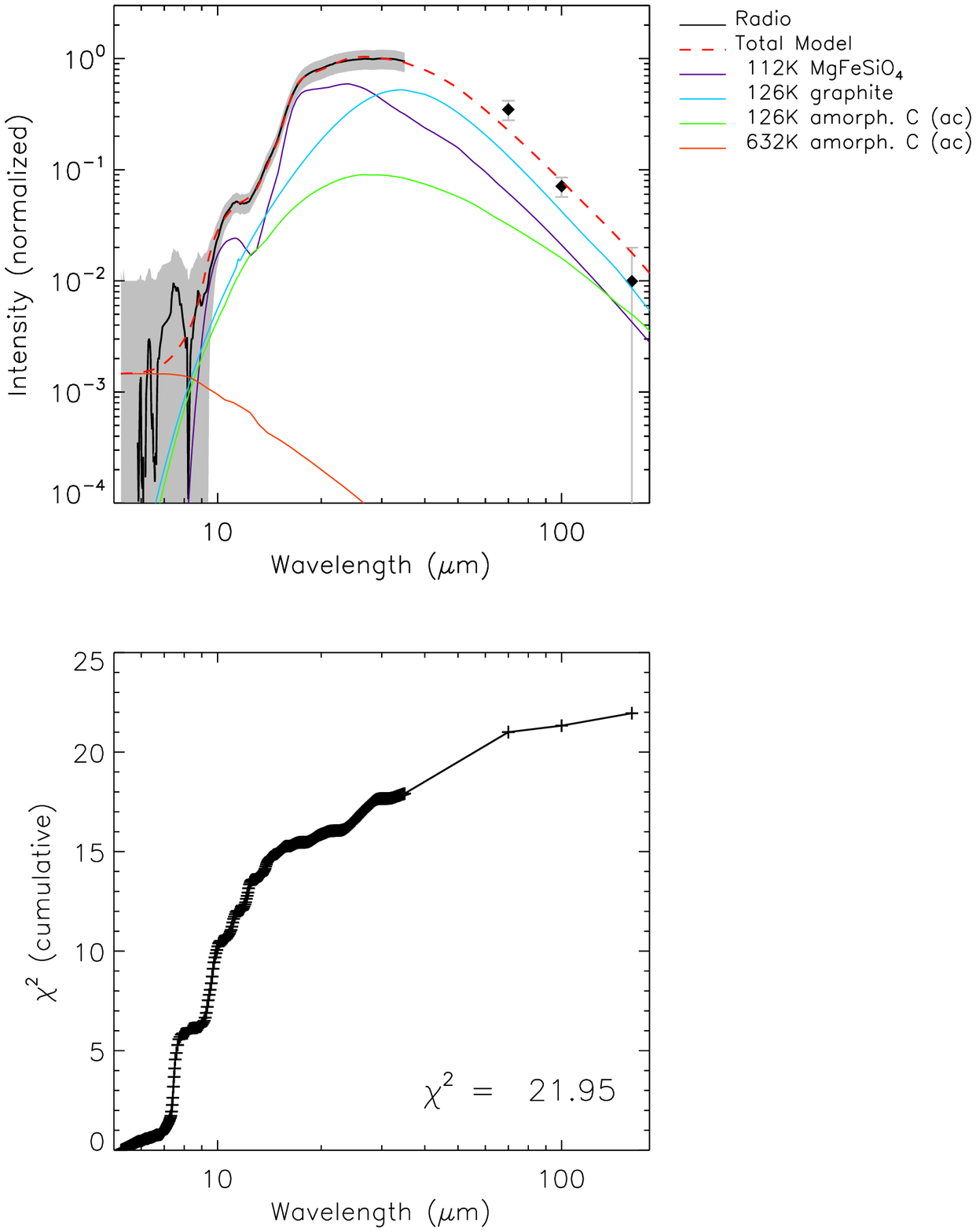}~~~~~~
  \includegraphics[height=3.5in]{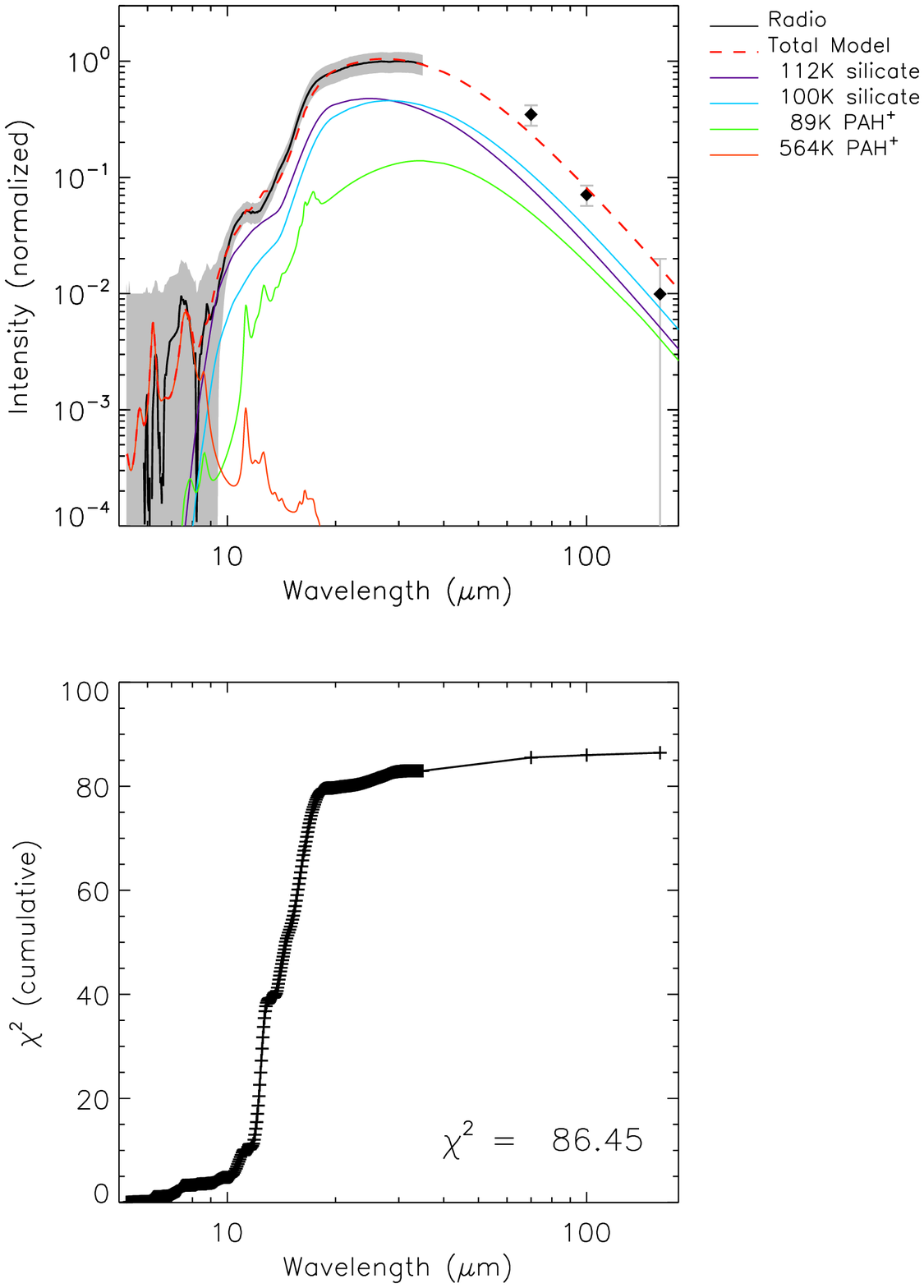}\\
   \caption{(upper left) The best 2-component fit to the Radio spectrum uses 
   graphite and a Mg silicate. Combinations of other silicates with other 
   featureless components are nearly as good.
   (upper right) This alternate fit to the Radio spectrum shows a marginal model
   using a sulfide composition instead of a one of the more featureless 
   dust compositions.
   (lower left) The best 3-component fit is only very slightly better
   than the 2-component fit.
   (lower right) The fit using the standard astronomical silicate, 
   graphite, PAH, and PAH$^+$ combination. Only two of the 
   four components are actually needed for the fit since the PAH$^+$ component is sufficient 
   to provide a relatively featureless component at $\gtrsim20$ $\micron$ and the PAH emission
   band at $\sim8$ $\micron$.
   \label{fig:radiodust}}
\end{figure}

As for the X-ray Fe and South Spot spectra, the spectrum associated with the 
radio emission tends to be best fit with a mix of silicate and featureless dust 
(Figure \ref{fig:radiodust}). A marginal 2--composition fit, in which 
Mg$_{0.1}$Fe$_{0.9}$S serves as the ``featureless'' component across the 
10--20 $\micron$ part of the spectrum is also shown in Figure \ref{fig:radiodust}.
The best 3--component fit 
(see Fig. \ref{fig:radiodust}) is not significantly different from the best
2-component fit, using two featureless components instead of one.
The last example in Figure \ref{fig:radiodust} shows the best results when 
dust compositions are restricted to the 
astronomical silicate, graphite, PAH and PAH$^+$,
which are commonly used to fit general interstellar material 
\citep[e.g.][]{Draine:1984,Zubko:2004,Draine:2007}.
The shape of the astronomical silicate's 10 $\micron$ 
emission peak is not as sharp as the observed spectrum.
Cool PAH$^+$ emission provides sufficient adjustment to the 
shape of the silicate spectrum such that the model does not require
graphite or neutral PAH components.

\subsection{\ion{Si}{2} Dust}
\begin{figure}[h]
  \includegraphics[height=3in]{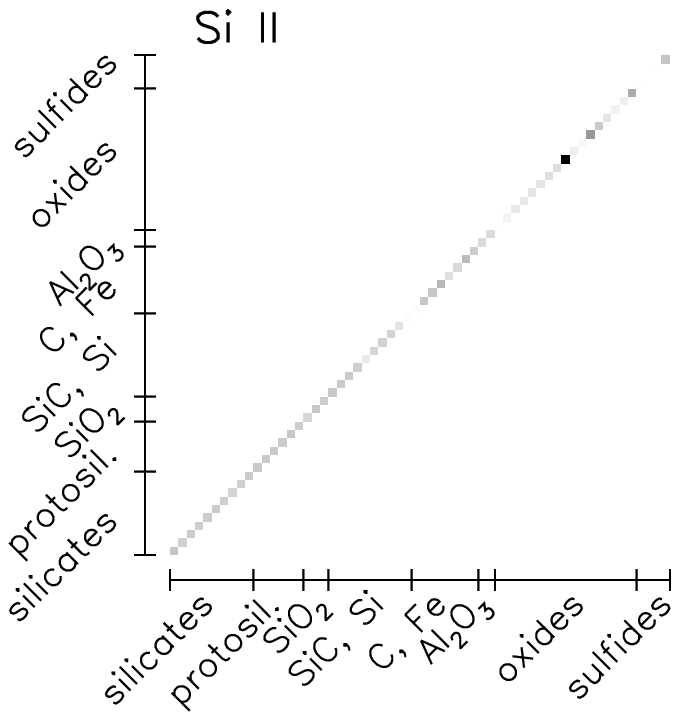}
  \includegraphics[height=3in]{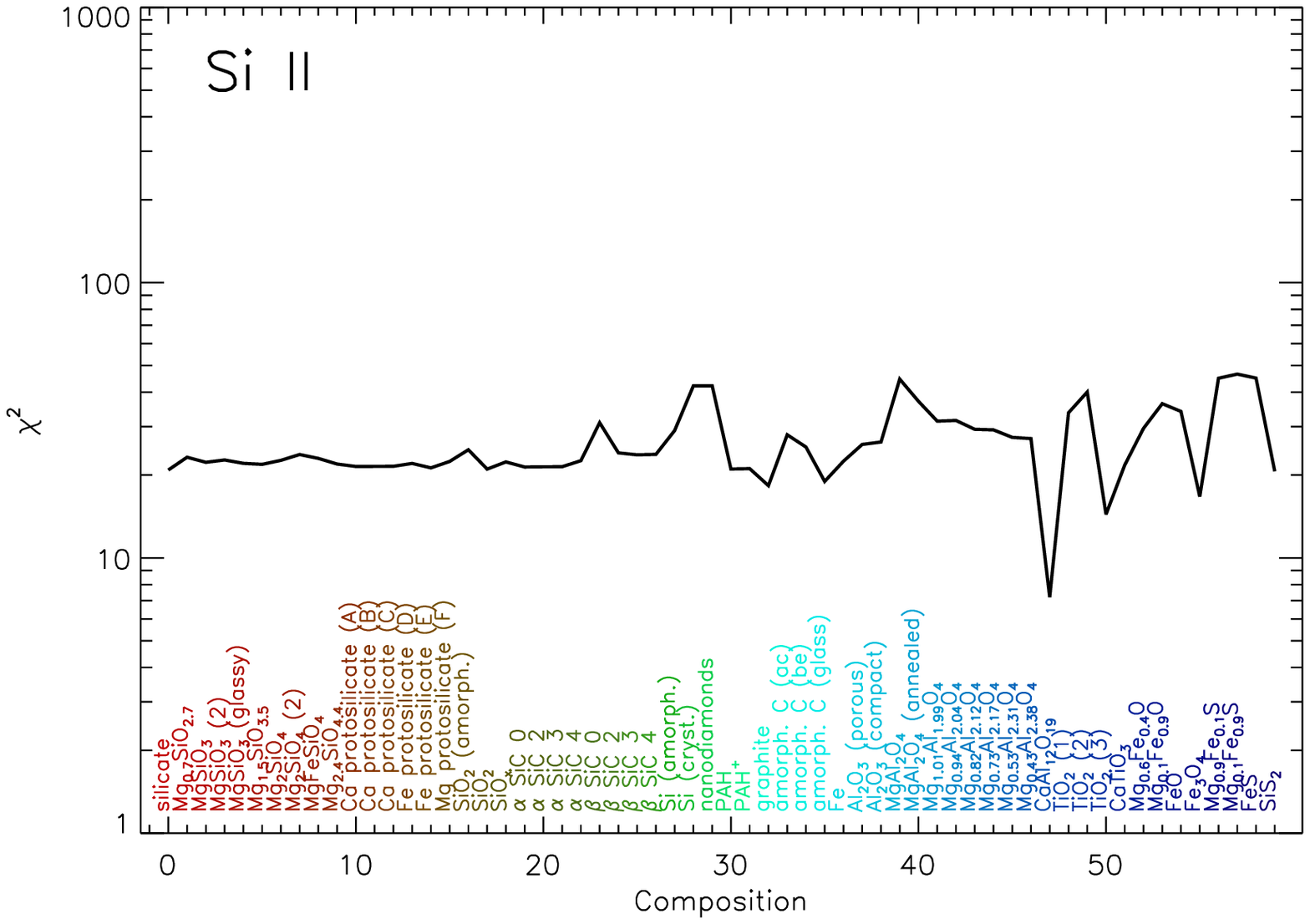}\\
   \caption{$\chi^2$ as in Fig. \ref{fig:arii_chimap}, but for the \ion{Si}{2} 
   dust spectrum. Only single composition fits were performed because of the lack of
   detailed spectral constraints. The best fits are for compositions that have 
   steeper long-wavelength absorption efficiencies, and for hibonite (CaAl$_{12}$O$_{19}$)
   which has a broad feature at $\sim 80$ $\micron$. 
   \label{fig:siii_chimap}}
\end{figure}

The \ion{Si}{2} dust spectrum essentially contains only the {\it Herschel} 
PACS measurements at 70, 100, and 160 $\micron$, with only upper limits 
on the emission at shorter wavelengths. Because of this lack of detailed 
spectral information, the spectrum was only fit with single composition 
dust models. The measured \ion{Si}{2} spectrum apparently has a relatively sharp 
peak compared to a black body spectrum. 
The best fit is provided by hibonite (CaAl$_{12}$O$_{19}$) which
has a broad emission feature at $\sim 80$ $\micron$ making it an 
especially good fit to the data. 
Figure \ref{fig:siiidust} shows the best fit spectrum (CaAl$_{12}$O$_{19}$), 
and a fit using a more typical Mg silicate (Mg$_2$SiO$_4$).
Compositions that have a steeper spectral index at long wavelengths
provide better fits, but the relatively good fit 
for TiO$_2\ (3)$ (rutile; second lowest $\chi^2$ in Fig. \ref{fig:siii_chimap}) 
is an artifact of an unphysically steep 
extrapolation ($\lambda^{-3.5}$) of the 
measured optical constants which are not published for $\lambda \ge 70$ $\micron$. 
For the \ion{Si}{2} spectrum, any models that are within a factor of 4 of the minimum 
$\chi^2$ are deemed acceptable. This includes most dust compositions,
and only excludes compositions that have strong features near $\sim 40$ 
$\micron$ that would have exceeded the upper limits provided by the 
{\it Spitzer} IRS data. 

%%%%%%%%%%%%%%%%%%%%%%%%%%%%%%%%%%
\section{DISCUSSION}
\subsection{Dust Compositions}

The identifications of possible dust compositions in the different environments
in Cas A are summarized in Table \ref{tab:families}. 
A single dust composition can never provide a good fit to the observed spectra,
except for the \ion{Si}{2} spectrum which has only upper limits at $<70$ $\micron$.
In general, 2 compositions are 
sufficient to get acceptable fits to the spectra. The \ion{Ar}{2} and \ion{Ar}{3}
spectra are the only ones that show significant (though small) benefit 
for the addition of a third composition. 
Among the 7 characteristic dust spectra examined, the results can be 
grouped into 3 different families of dust. 

The first family is found in association with the \ion{Ar}{2} and \ion{Ar}{3} 
emitting ejecta, which are distributed widely across the SNR. 
This family consists of the Mg silicate with the lowest Mg/Si ratio
in combination with one or more other compositions that have featureless
spectra. The Mg$_{0.7}$SiO$_{2.7}$ is a good fit to the 9 and 21 $\micron$ peaks 
in the observed spectrum, but only if a featureless dust composition is also
present to reduce the apparent strength of these features. However, the Ar dust
spectra also contain a weaker 12 $\micron$ feature which is not accounted for 
by any Mg silicate. Nonstoichiometric spinel with low Mg/Al ratios can provide a 
feature at approximately the correct wavelength, but only when the dust grains are 
extremely hot, such that longer wavelength spinel features are relatively faint.
A better explanation for the 12 $\micron$ feature may be provided by SiO$_2$.
The spectrum of SiO$_2$ exhibits all three peaks seen in \ion{Ar}{2}
and \ion{Ar}{3} spectra, although they are  significantly sharper 
and slightly bluer than the observed ones, and even in combination with other 
materials the fits are not very good. However, this is when Mie theory is
used to calculate the absorption cross sections assuming small spherical grains. 
Using a continuous distribution of ellipsoids (CDE) approximation instead
broadens and shifts the SiO$_2$ features to be a better match as a third component. 
\cite{Jager:2003} point out that the 12 $\micron$ SiO$_2$
feature disappears in Mg silicates when Mg/Si $> 0.5$. Therefore,
it seems likely that the dust associated with the Ar emission is a mixture of silica and 
Mg silicate (with Mg/Si $\lesssim 0.5$) in combination with a separate 
featureless dust component. SiC can also provide the 12 $\micron$ feature,
but again only if CDE calculations are applied to this component. If SiC is
present, this would be the only direct evidence of carbon-bearing dust.
Table \ref{tab:12um} lists the possible origins for the 12 $\micron$ feature
of the Ar dust.

The strong 21 $\micron$ peak in this family has been the hallmark of Cas A IR spectra
since it was first observed using the Kuiper Airborne Observatory (KAO) and the 
{\it Infrared Space Observatory (ISO)}. 
On the basis of those data, 
the peak was suggested to arise from Mg protosilicate \citep{Arendt:1999}. 
Subsequent analysis by \cite{Douvion:2001} modeled an {\it ISO} spectrum 
as MgSiO$_3$, SiO$_2$ and Al$_2$O$_3$ with the weak 12 
$\micron$ feature largely produced by the Al$_2$O$_3$
as proposed for the feature in ``Spectrum 2'' of \cite{Douvion:1999}.
\cite{Ennis:2006} using {\it Spitzer} IRS data 
noted the distinction of several different dust spectra
in different parts of Cas A and referred to this as the ``Strong 21 $\micron$''
spectrum, and calling it ``21 $\micron$ peak dust'' \cite{Rho:2008} 
modeled it as Mg protosilicate and MgSiO$_3$ with secondary components of 
SiO$_2$, FeO, FeS, Si, and Al$_2$O$_3$ and/or Fe. Later work indicated that the 21 $\micron$ 
feature may be fit primarily by SiO$_2$ grains if CDE calculations rather than 
spherical grain approximations are used \citep{Rho:2009}.

The second dust family is associated with the \ion{Ne}{2} emission, which is 
especially prominent in two opposing ``Ne crescents'' 
\citep{Ennis:2006,Smith:2009} 
in the N and S parts of the SNR. These morphological features are evident in the derived
spatial distribution of the \ion{Ne}{2} dust shown in Figure \ref{fig:spatial_distributions}.
This dust has a very smooth spectrum that 
does not suggest any silicate material. The best fits to the spectrum are 
found with Al$_2$O$_3$ in combination with other featureless dust. Featureless
dust alone can provide moderately good fits, but a broad asymmetric feature in the 
10 -- 20 $\micron$ portion of the Al$_2$O$_3$ absorption cross section seems
to match the observed spectrum especially well. Alternately, nonstoichiometric 
spinel can provide the needed emission at 10 -- 20 $\micron$, although the
spinel absorption efficiency has more detailed substructure that is not evident 
in the observed spectrum. 

This second family corresponds to the
``weak 21 $\micron$'' components noted by \citep{Ennis:2006} and 
\cite{Rho:2008}, which they also associate with relatively 
strong Ne emission. However, the fact that 
they see even a weak 21 $\micron$ peak in their spectrum suggests that it is 
a mixture of what we identify as very distinct Ar and Ne dust families.
As in our fitting, \citep{Rho:2008} fit this spectrum with hot and cool
featureless dust (C glass) and with an intermediate temperature Al$_2$O$_3$
components. They included other components to add the 
weak 21 $\micron$ peak that appears in their spectrum.
\cite{Douvion:1999} had also noted an anti-correlation 
between the 9 $\micron$ silicate emission and \ion{Ne}{2} and 
\ion{Ne}{3} emission. 

The third dust family is associated with the X-ray Fe emission and the South Spot.
This family also seems to match the dust associated with the radio emission, which
is expected to be dust that has been swept up from the interstellar or circumstellar 
medium by the forward shock. The primary component in this family is one of the 
Mg silicates with high Mg/Si ratios or MgFe silicate:
Mg$_2$SiO$_4$, Mg$_{2.4}$SiO$_{4.4}$, MgFeSiO$_4$. Other Mg silicates with 
Mg/Si $\geq 1$ can provide acceptable fits, but the silicates with 
lower ratios that were needed for the Ar dust are not suitable here
because of the changing placement and shape of the 9 and 21 $\micron$
features.

This dust family is matched by the ``Broad'' component identified by \cite{Ennis:2006} 
and the ``Featureless'' component modeled by \cite{Rho:2008}.
In both cases this component is {\it not} associated with Ar or Ne emission lines,
just as we also find. \cite{Rho:2008} modeled this component as 
MgSiO$_3$, FeS, and Si combined with Al$_2$O$_3$, Mg$_2$SiO$_4$ and/or Fe.
Using 5--17 $\micron$ ISO data, \cite{Douvion:1999} extracted a
spectrum (``Spectrum 3'') from a region that should match our Radio spectrum.
Despite the limited wavelength coverage, they also found that the spectrum
could be fit with astronomical silicate \citep{Draine:1984}
at $T \sim 105$ K, as we confirm (in Fig. \ref{fig:radiodust}).

The \ion{Si}{2} dust has no significant emission across the 5-40 $\micron$
wavelength range of the IRS. The three {\it Herschel} PACS measurements at 
70, 100, and 160 are insufficient to constrain the composition of the 
associated dust. The \ion{Si}{2} dust may belong to one of the above families,
or it may be an entirely different composition.

\subsection{Dust Masses}
The modeled dust temperatures and compositions allow the determination of 
the dust mass in each of the components. 
The dust mass, $M_d$, is calculated as:
\begin{equation}
M_d = {D^2 S_\nu(\lambda)\over{\kappa_\nu(\lambda) B_\nu(\lambda,T_d)}}
\end{equation}
where $D = 3.4$ kpc is the distance to Cas A \citep{Reed:1995}, 
$S_\nu(\lambda)$ is the total flux density of the component in question,
$\kappa_\nu(\lambda)$ is the mass absorption coefficient appropriate for the derived composition of the dust, and 
$B_\nu(\lambda,T_d)$ is the Planck function evaluated at the derived temperature.
For each of the models fit to the characteristic spectra, the total mass was calculated by 
summing Eq. 4 over all temperature components of each of the compositions used in the fit.
Generally the total dust mass is 
dominated by a warm component (60 - 130K) that also produces the bulk of the 
luminosity. 
However the X-ray Fe spectrum is an exception where an additional cold 
component dominates the mass, because 
the {\it Herschel} PACS data at 70--160 $\micron$ are elevated
relative to the shorter wavelength emission. The composition of this cold component is uncertain 
because absorption efficiencies are smooth and the spectral resolution is poor
at the long wavelengths. 
This situation is worse for the \ion{Si}{2} dust which is only detected 
at $\geq$ 70 $\micron$. Its temperature is relatively well constrained, but 
its composition, and therefore its mass, is not. If the dust is composed of Mg 
silicates, then the total mass of the \ion{Si}{2} dust is $\lesssim 0.1$ M$_{\sun}$, but this value can 
be much lower if other compositions are appropriate. Much higher masses are 
ruled out by the expected nucleosynthetic yields of various elements, 
and would imply very high condensation efficiencies given that the 
mass of the unshocked gas is only $\sim0.4$ M$_{\sun}$ \citep{DeLaney:2014}.

\begin{figure}[h]
  \includegraphics[width=5.25in]{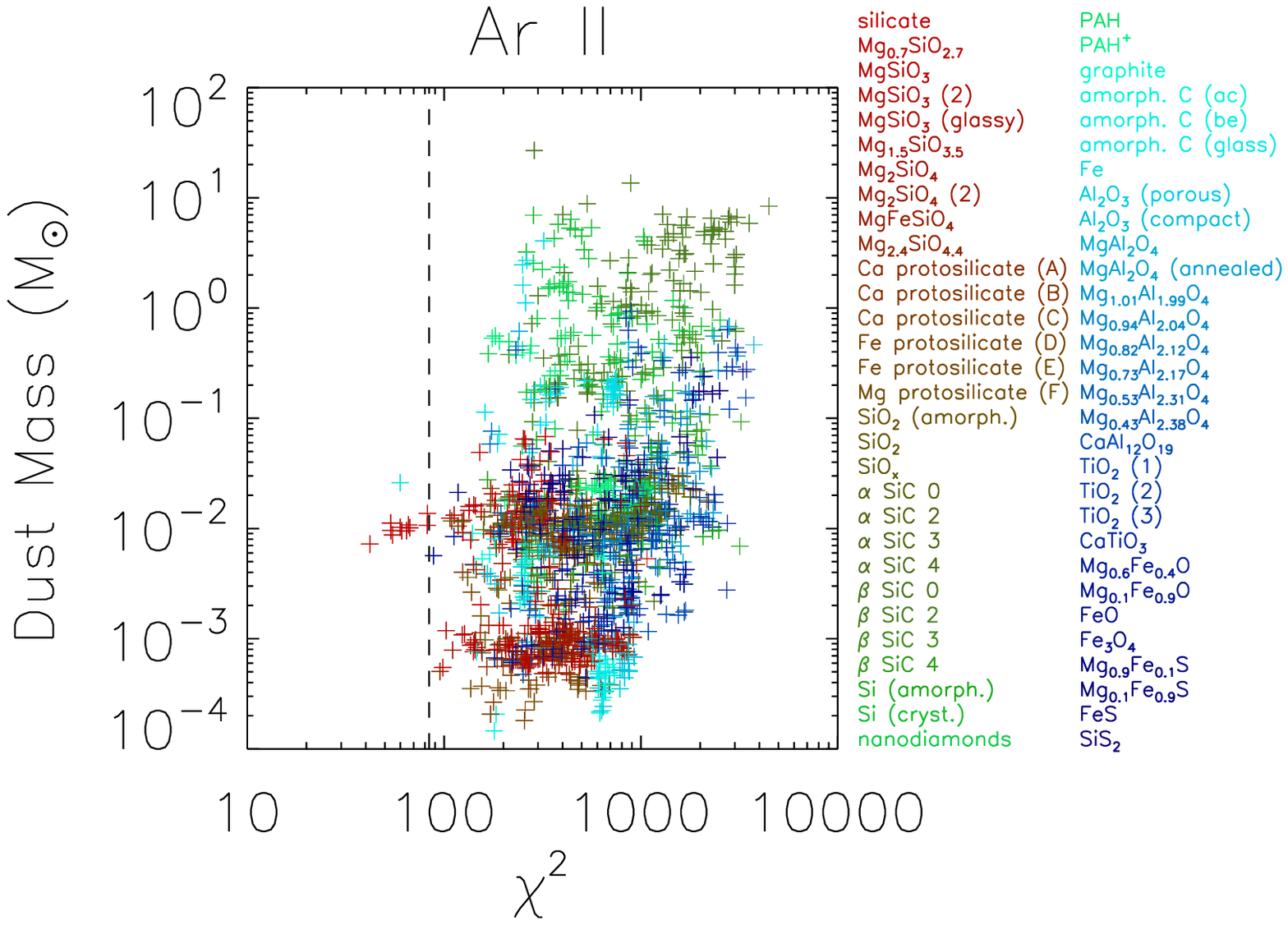}
  \includegraphics[width=5.25in]{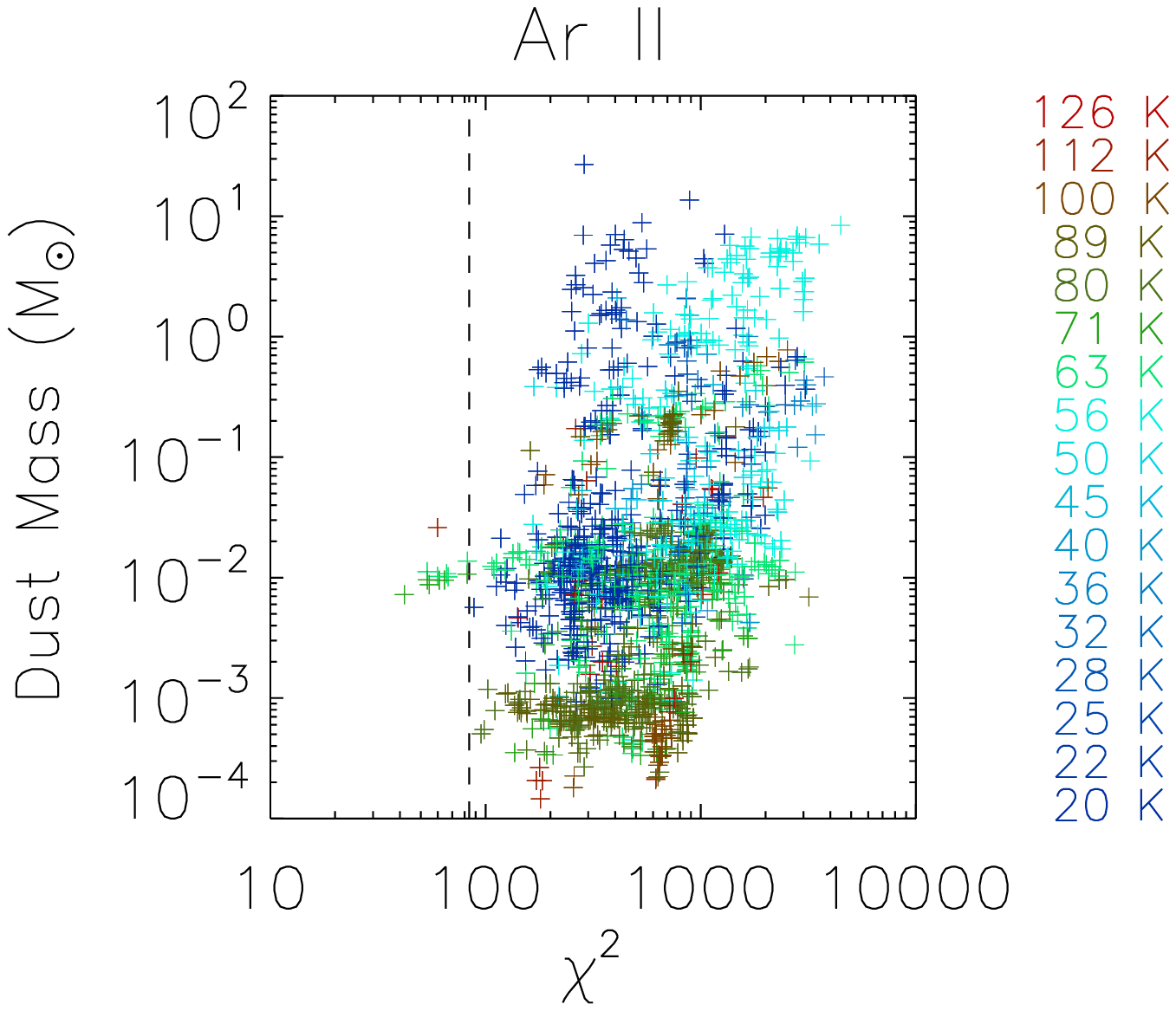}\\
   \caption{Total dust masses for \ion{Ar}{2} models. In the top panel the points are color coded
   by the composition that provides the largest mass component. In the bottom panel 
   the masses are color coded by the temperature of the dominant mass component.
   Only models to the left of the dashed line (within a factor of 2 in $\chi^2$ of 
   the best model) are deemed acceptable. 
   These good models indicate that $\sim10^{-2}$ M$_{\sun}$ of dust, mostly silicates
   at $\sim 70$K, is associated with the \ion{Ar}{2} emission in Cas A.
   \label{fig:arii_mass}}
\end{figure}

The derived dust masses 
averaged over all models that fit within a factor of 2 of the minimum $\chi^2$
and are dominated by Mg silicates (or Al$_2$O$_3$ for the \ion{Ne}{2} spectrum)
are listed in Table \ref{tab:families}. 
Figure \ref{fig:arii_mass} 
plots the total dust mass 
for {\it all} 2--component models of the \ion{Ar}{2} dust spectrum as a 
function of $\chi^2$ with 
color coding to indicate compositions and temperatures. The ``good'' models
are within a factor of 2 of the minimum $\chi^2$ (to the left of the dashed line).
The figure shows that although there are correlations between composition
and temperatures, the derived mass is more strongly dependent on the composition
than the temperature. %{\bf **]}

The total mass of warm and hot dust for all components that contribute to the $<35$ $\micron$
spectrum is found to be $0.04\pm0.01\ M_\sun$ (see Table 3). This is consistent with other measurements
from {\it IRAS} \citep{Braun:1987, Arendt:1989,Saken:1992} and other analysis of the {\it Spitzer} IRS data
\citep{Rho:2008} because the mid-IR (12--100 $\micron$) 
flux density $S_\nu(\lambda)$ remains basically the unchanged since the {\it IRAS} measurements. This is the primary 
observable factor in determining the mass. When corrected for the same distance, previously published
mass estimates contain modest differences (factors of $\sim2$) due to different assumptions of the
mass absorption coefficients, $\kappa_\nu(\lambda)$, and dust temperature, $T_d$. 
{\it ISO}, {\it Akari}, BLAST, and {\it Herschel} have revealed an additional cool dust component in 
Cas A which contains $\sim 0.08\ M_\sun$ of dust \citep[e.g.][]{Tuffs:1999, Tuffs:2005, Sibthorpe:2010,
Barlow:2010}. This cool component is what we associated with the \ion{Si}{2}. 
Our mass estimate of this component is consistent,  
but is very uncertain due to the unknown composition of the dust, and the difficulty in 
distinguishing the SNR dust from the heavy confusion of the line of sight ISM at these
wavelengths.

%%%%%%%%%%%%%%%%%%%%%%%%%%%%%%%%%%
\section{SUMMARY}
We have decomposed the spatially resolved IR continuum spectra of 
dust in the Cas~A SNR into contributions from distinct regions of the 
remnant, comprising: 
shocked circumstellar and interstellar medium swept up by the 
forward shock; hot X-ray emitting gas and cooler, denser 
fine-structure IR line emitting regions that result when the SN ejecta 
passes through the reverse shock; 
and unshocked regions of the ejecta that have not yet reached the reverse shock.
We then calculated the composition and mass of the dust associated with each region of the 
ejecta. The methodology and results of our paper can be summarized as follows:
\begin{itemize}
%-----
\item We started by identifying a set of {\it spatial templates} to represent various IR emitting 
regions of Cas~A. These were defined by the fine 
structure IR lines from \ion{Ar}{2}, \ion{Ar}{3}, \ion{Ne}{2}, and \ion{Si}{2}, by the 
X-ray (Fe line), and by the radio synchrotron emission. The spatial 
templates used for the decomposition are shown in Figure~\ref{fig:spatial_templates}. 
%-----
\item We then identified spatially distinct {\it zones}, shown in 
Figure~\ref{fig:zones}, in which a given template emission was dominant. We 
extracted the {\it characteristic spectra} of the dust in each zone using the 
procedure described in Section 3. The resulting IR spectra were 
assumed to represent the different spatial templates, and are 
presented in Figure~\ref{fig:spectral_distributions}. Figure~\ref{fig:spatial_distributions} shows the derived {\it spatial distribution} 
of the dust associated with each of the spectral templates.
 %-----
 \item We compiled an extensive list of dust compositions with 
 measured optical constants to calculate their possible contribution to the 
 IR emission from each spatial templates. The dust species fall 
 into 8 broad categories defined by similarity of chemical
 and/or optical properties: silicates, protosilicates, silica, 
 silicon carbide, carbon and metallic iron, aluminum oxides, 
 oxides, and sulfides. The dust compositions used in the 
 analysis are listed in Table~2.
%-----
 \item We fit each characteristic spectrum with all possible 
 combinations of 2 distinct dust species (1711 combinations),
 allowing each dust species to emit at a range of temperatures.
 Figures \ref{fig:arii_chimap}--\ref{fig:siii_chimap} depict 
 the $\chi^2$ values to give an overview of which compositions 
 can provide good fits to the characteristic spectra. 
 Figures \ref{fig:ariidust}--\ref{fig:siiidust} 
 provide examples of how well the spectra can be fit using 2 (and
 sometimes 3) dust species. 
 %-----
 \item The primary dust composition could be readily identified in 
 spectra that have strong IR dust features. The \ion{Ar}{2} and \ion{Ar}{3} 
 characteristic spectra exhibited strong features at $\sim9$ and 21~$\micron$. 
 Magnesium silicates, characterized by a Mg/Si 
 ratio of 0.7 (i.e. Mg$_{0.7}$SiO$_{2.7}$) 
 provided the best fit to the spectra of these regions. The 
 \ion{Ne}{2} spectrum was best fit with Al$_2$O$_3$ dust, and the 
 X-ray Fe spectrum, which exhibit peaks at $\sim 10$ and 20~$\micron$ 
 was best fit with silicates having a higher Mg/Si ratio of 
 2.4 (i.e. Mg$_{2.4}$SiO$_{4.4}$). A region called the South Spot 
 had similar composition. The dust composition associated with 
 the spectrum of the radio synchrotron emitting region behind the SNR's 
 forward shock was consistent with that expected for typical interstellar dust. 
 The IR spectrum from the \ion{Si}{2} emitting 
 region arises from a cold dust component that is only seen at long
 wavelengths. Since very few dust compositions have distinguishing 
 features at these wavelengths, and because these broad band data at long wavelengths 
 provide little detail on the spectrum, the composition of this cold dust
 could not be determined.
 %-----
 \item Secondary dust components are needed 
 to improve the fit for most spectra, but these components are not 
 uniquely identified. Table 3 lists the primary dust composition 
 for each spectrum, and all possible secondary dust species 
 that they need to be paired with. The secondary dust species 
 generally have a featureless dust spectrum.
 %-----
 \item A minimal number of only 4 dust species: 
 Mg$_{0.7}$SiO$_{2.7}$, Mg$_{2.4}$SiO$_{4.4}$, 
 Al$_2$O$_3$, and amorphous carbon, would be sufficient to fit the entire 
 spectrum of Cas~A. These compositions suggest that the seed dust 
 particles that formed the more complex species 
 are MgO, SiO$_2$, Al$_2$O$_3$, and carbon.
 %-----
 \item The total mass of dust is about 0.04~$M_{\sun}$ 
 in the shocked CSM/ISM and ejecta regions, and $\lesssim 0.1$~$M_{\sun}$ 
 in the unshocked ejecta characterized by the \ion{Si}{2} emission.
 This dust mass is similar to that derived by \cite{Rho:2008}
 from the {\it Spitzer} data, and the $\sim 0.08$~$M_{\sun}$ derived 
 for dust in the cold unshocked ejecta derived from the {\it Herschel}, BLAST, and 
 {\it Akari} data \citep{Barlow:2010,Sibthorpe:2010}.
\end{itemize}

\acknowledgements 
This work is based on observations made with the {\it Spitzer Space Telescope}, 
which is operated by the Jet Propulsion Laboratory, California Institute of Technology 
under a contract with NASA. Support for this work was provided by NASA Program NNH09ZDA001N-ADP-0032. 
This research made use of Tiny Tim/Spitzer, developed by John Krist for the Spitzer Science 
Center. The Center is managed by the California Institute of Technology under a contract with NASA
This research has made use of NASA's Astrophysics Data System Bibliographic Services. 
We thank T. Kozasa for providing digitized (and extrapolated and interpolated) versions of the 
optical constants for several dust species as noted in Table \ref{tab:emiss}.
We also thank the referee for constructive comments on the manuscript.

{\it Facilities:} \facility{Spitzer}, \facility{Herschel}, \facility{CXO}, \facility{VLA} 

\appendix
%\section{Spatial Templates}
%\subsection{Selection of Spatial Templates}
\section{Selection of Spatial Templates}
A set of spatial templates are needed to distinguish different physical 
environments within the SNR. We want to be able to distinguish swept-up ISM or CSM
dust from ejecta dust, to probe the nature of ejecta dust created in 
different nucleosynthetic layers of the SN, and to investigate ejecta dust in 
preshock and postshock regions having different gas temperatures and densities.
The primary source for templates are the 
line emission maps that are generated by subtracting the continuum cube from the 
full data cube, and then integrating over the [-6000,+8000] km s$^{-1}$ velocity 
range for each of the emission lines in the IRS spectrum (see Table 1). 
Most of these lines are associated with the fast-moving knots
of ejecta. These knots tend to be O-rich, but there are some variations in composition
that tend to be correlated with spatial and kinematic differences.
Because of the large range in velocity in the Cas A ejecta, 
closely spaced [\ion{O}{4}] 25.89 $\micron$ and [\ion{Fe}{2}] 
25.99 $\micron$ lines would be confused in the low resolution spectra. 
However, \cite{Isensee:2010,Isensee:2012} report that high resolution spectra
show that Cas A's 26 $\micron$ line is entirely produced by [\ion{O}{4}] 
without any contribution from [\ion{Fe}{2}].

Several other spatial templates were also tested. These are intended to trace the 
presence of dust in a wider range of environments than the relatively cool and 
dense line-emitting knots:\\
(1) A radio template was derived from 6 cm VLA image
of Cas A \citep{DeLaney:2004}.
The region of the forward shock is traced more clearly in the radio than at most 
other wavelengths, and the bright radio knots in the reverse shock region 
are a different population than line-emitting ejecta knots.\\
(2) and (3) X-ray Fe and Si templates were derived from {\it Chandra} observations 
\citep{Hwang:2004}, integrated over the widths of the 6.4 keV Fe and 1.7 keV Si K $\alpha$ lines. 
The X-ray continuum is included in these integrations. These templates trace ejecta
in regions of lower density and much higher temperature. These regions include the bulk 
of the mass of the ejecta of the SN. \\
(4) The IRAC 4.5 $\micron$ image lies outside 
the range of the IRS spectral coverage and contains both a synchrotron component
and a probable line emission component. The synchrotron 
emission was subtracted using a scaled version of the IRAC 3.6 $\micron$ image
which is dominated by synchrotron emission and has only weak line emission. 
However, this over-subtracts stellar sources in the image. Thus regions where negative 
values resulted were replaced by nearby background values. \\
(5) The IRAC 8 $\micron$
image is dominated by the [\ion{Ar}{2}] 6.99 $\micron$ emission, but was considered as a 
potentially cleaner map of this line than the IRS line maps. The background
ISM emission at 8 $\micron$ was removed by subtracting a scaled version of the 
5.8 $\micron$ IRAC image, where the SNR emission is relatively weaker. As with the 
4.5 $\micron$ template, the over-subtracted stellar sources were replaced by the 
local background levels. \\
Each of these templates is convolved and reprojected to match the IRS 
continuum data cube.

\begin{figure}[h]
  \includegraphics[height=6in,angle=90]{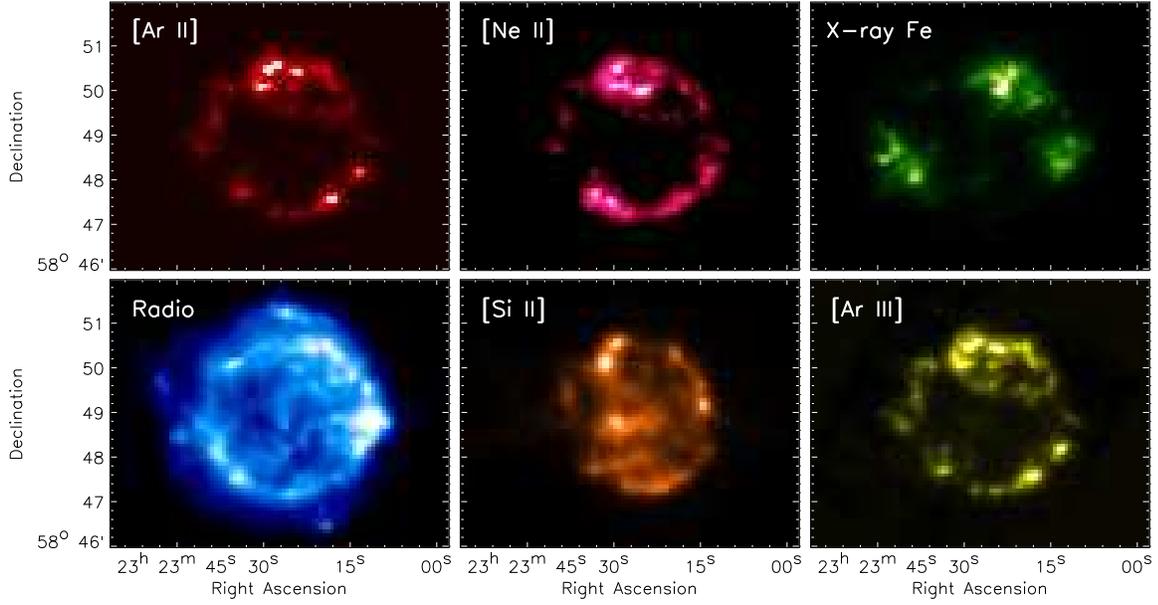}
   \caption{Spatial templates used to signify regions that may contain dust 
   distinguished by different composition, origin (ejecta vs. ISM), and/or heating. 
   The [\ion{Ar}{2}] 6.99 $\micron$,
   [\ion{Ne}{2}] 12.81 $\micron$, [\ion{Si}{2}] 34.8 $\micron$, [\ion{Ar}{3}] 8.99 $\micron$
   lines are derived from the IRS data cube. The X-ray Fe emission is from \cite{Hwang:2004}
   and the VLA 6 cm radio map is from \cite{DeLaney:2004}.
   \label{fig:spatial_templates}}
\end{figure}

\begin{figure}[h]
\begin{center}
  \includegraphics[height=6in,angle=90]{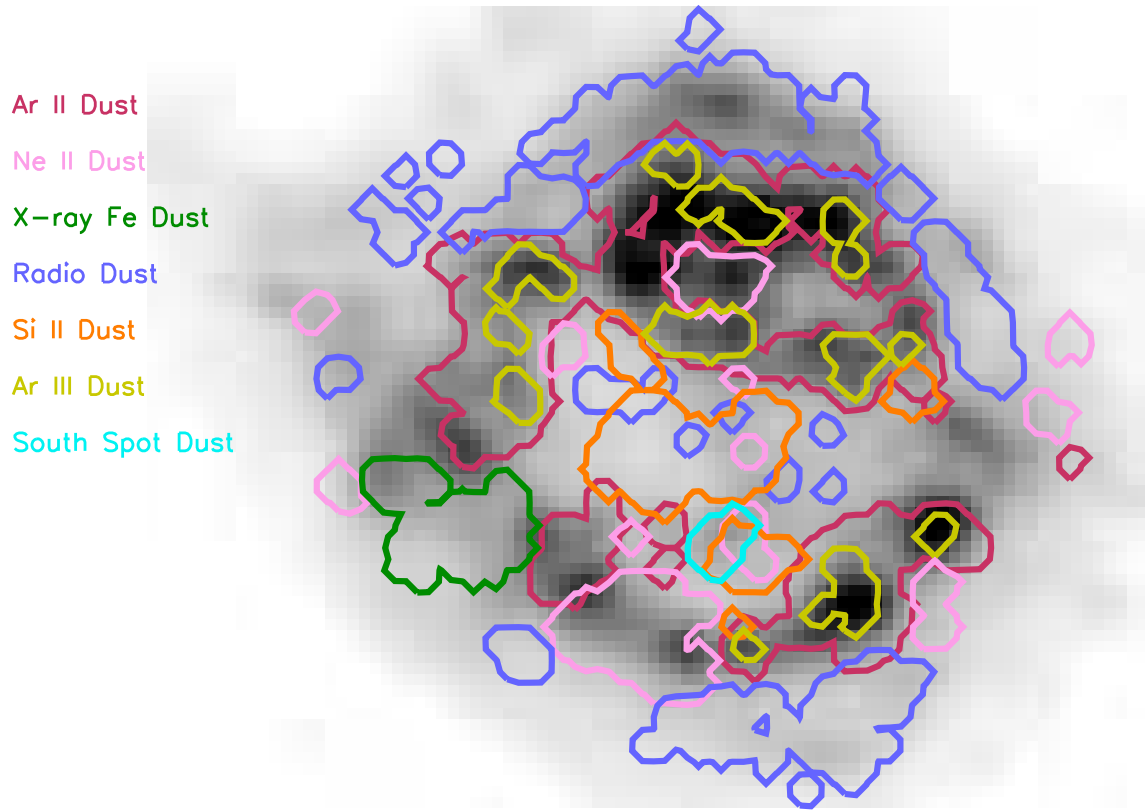}
\end{center}  
   \caption{Depiction of the zones within which the characteristic spectra of each spatial template 
   were extracted. These zones
   correspond to the regions where the templates shown in Figure \ref{fig:spatial_templates} are
   dominant (apart from the nearly ubiquitous \ion{Ar}{2}). The additional ``South Spot'' zone 
   was identified as the primary region where there is relatively strong continuum emission that 
   is not traced by any of the spatial templates.
   \label{fig:zones}}
\end{figure}

\begin{figure}[h]
  \includegraphics[width=6in]{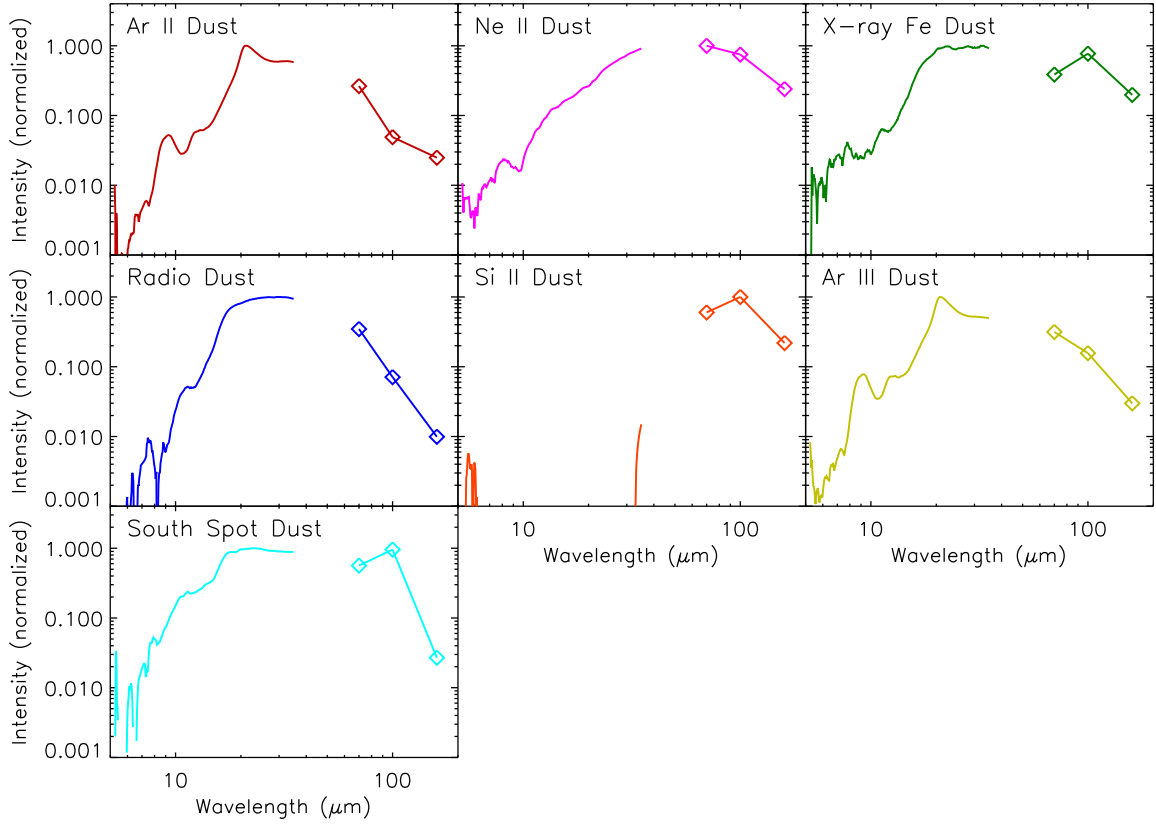}
   \caption{The characteristic spectra extracted for each of the spectral zones defined in Figure
   \ref{fig:zones}. In each case, the solid line from 5 -- 35 $\micron$ is the IRS data. The three points at 70, 100, and 
   160 $\micron$ are broadband {\it Herschel} PACS data. (The \ion{Si}{2} dust has
   no significant emission at $\leq 35$ $\micron$).
   \label{fig:spectral_distributions}}
\end{figure}

\begin{figure}[h]
\includegraphics[height=6in,angle=90]{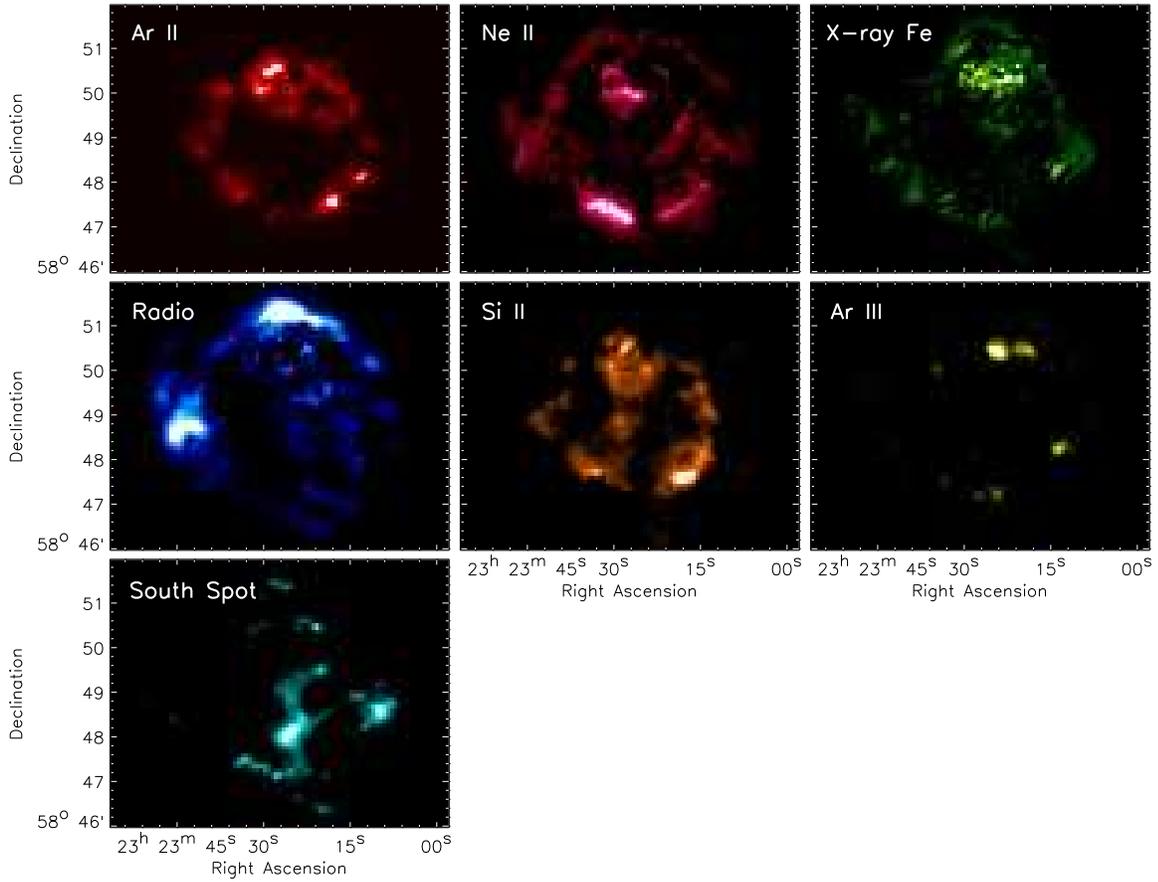}
    \caption{Spatial distribution of the dust associated with each of the characteristic spectra
    shown in Figure \ref{fig:spectral_distributions}.
   \label{fig:spatial_distributions}}
\end{figure}

\begin{figure}[h]
  \includegraphics[]{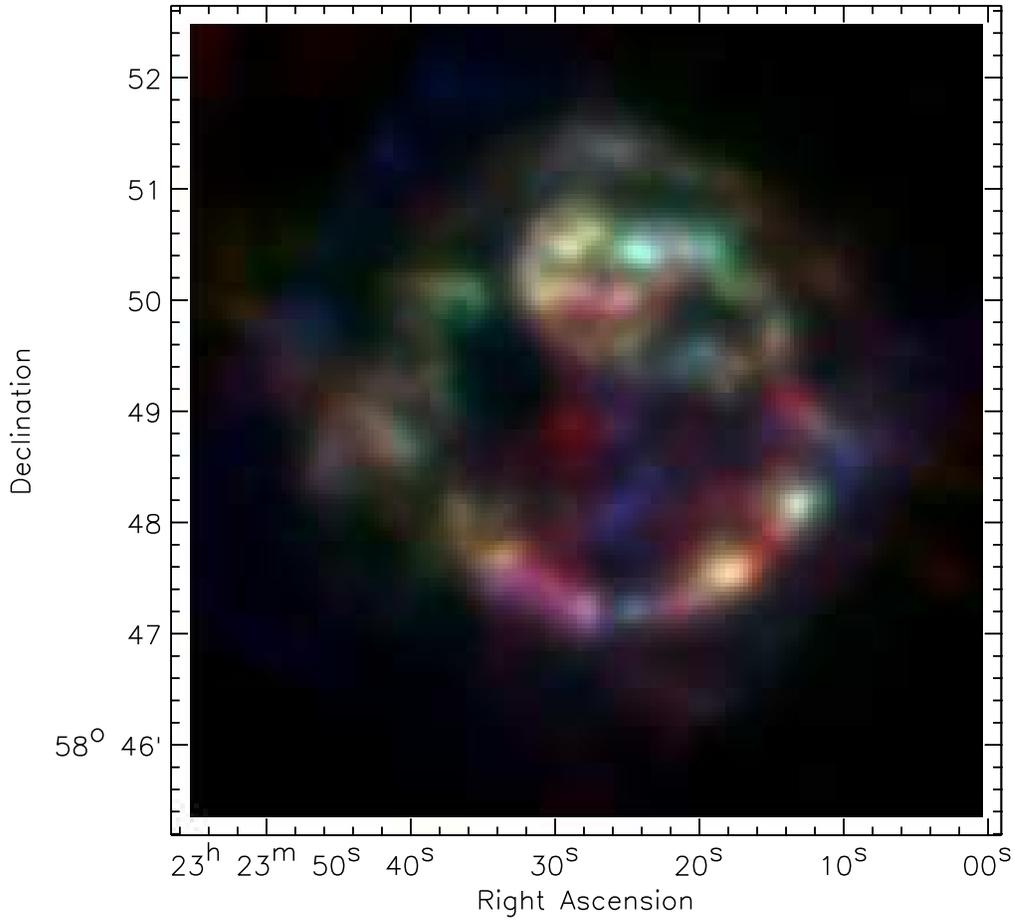}
   \caption{This color image of Cas A dust displays continuum emission at 
   11.8, 20.8, 70 $\micron$ in the blue, green, and red channels, respectively. 
   Comparison with 
   Figure \ref{fig:spatial_distributions} shows that these wavelengths provide 
   good illustration of the different dust types: \ion{Ar}{2} = green, 
   \ion{Ne}{2} = pink, Radio = grey, \ion{Si}{2} = red, 
   \ion{Ar}{3} = pale cyan, South Spot = purple.
   (X-ray Fe dust is not clearly distinguished.)
   \label{fig:3color}}
\end{figure}

This full set of templates would be expected to contain some degeneracies, where
different lines (especially from the same species or element) would be tracing
the same physical component of the SNR. To select a useful subset of the possible templates,
we calculated the 
linear correlation coefficient for each pair of templates. These are listed in 
Table \ref{tab:cc}. The table shows which templates are well-correlated with others.
Based on these correlations we sorted the templates in to similar groups and selected 
one template to represent each group. Images of the selected templates
are shown in Figure \ref{fig:spatial_templates}.

There are several points of interest in the correlations listed in Table \ref{tab:cc}. 
Even though the [\ion{Ar}{2}] and [\ion{Ar}{3}] emission are
very well correlated (as expected), 
we chose to keep both these templates in order to monitor for changes in the dust that might be 
related to the ionization state of the gas. It was not expected that highly ionized species 
such as [\ion{S}{4}] would correlate so well with [\ion{Ar}{2}]. This correlation may become
significantly weaker if higher spatial resolution data were available. Correlations 
involving the [\ion{Fe}{2}] line are weaker than most due to the low signal to noise ratio 
of this line. The [\ion{Ne}{5}] line is also relatively weak, yet surprisingly it correlates
better with the [\ion{Ar}{2}] group than the [\ion{Ne}{2}] and [\ion{Ne}{3}] lines. 
Despite an absence of Ne lines in the band, the 4.5 $\micron$ IRAC $\micron$ emission is somewhat 
better correlated with the Ne group templates than the Ar group. In particular, the 
``Ne crescents'' noted by \cite{Ennis:2006} are present in the IRAC 4.5 $\micron$ image.
Although the [\ion{S}{3}] 33.48 $\micron$ correlates moderately well with the [\ion{S}{3}] 
18.71 $\micron$ line, it does not correlate as well with the other templates of the Ar group.
This is an indication of variation in the \ion{S}{3} line ratios, which would indicate 
large scale variations in the density of the ejecta \citep{Smith:2009}. 
The [\ion{O}{4}] 25.89 $\micron$ and [\ion{S}{3}] 33.48 $\micron$ templates are moderately correlated with the 
both the \ion{Si}{2} template and the Ar-like templates. Because the 
[\ion{O}{4}] and [\ion{S}{3}] templates did not have distinct features lacking in these other 
templates, we did not use either in the analysis. 
The X-ray Si template is in fact distinct from the X-ray Fe template, but it was not used because
initial analysis using only the X-ray Fe template did not indicate the presence of any 
residual emission that would have been reduced by the addition of an X-ray Si template.

%%%%%%%%%%%%%%%%%%%%%%%%%%%%%%%%%%
\bibliographystyle{apj}
\bibliography{casa}

\begin{thebibliography}{}
\expandafter\ifx\csname natexlab\endcsname\relax\def\natexlab#1{#1}\fi

\bibitem[{{Arendt}(1989)}]{Arendt:1989}
{Arendt}, R.~G. 1989, \apjs, 70, 181

\bibitem[{{Arendt} {et~al.}(1999){Arendt}, {Dwek}, \& {Moseley}}]{Arendt:1999}
{Arendt}, R.~G., {Dwek}, E., \& {Moseley}, S.~H. 1999, \apj, 521, 234

\bibitem[{{Barlow} {et~al.}(2010){Barlow}, {Krause}, {Swinyard}, {Sibthorpe},
  {Besel}, {Wesson}, {Ivison}, {Dunne}, {Gear}, {Gomez}, {Hargrave}, {Henning},
  {Leeks}, {Lim}, {Olofsson}, \& {Polehampton}}]{Barlow:2010}
{Barlow}, M.~J., {Krause}, O., {Swinyard}, B.~M., {et~al.} 2010, \aap, 518,
  L138

\bibitem[{{Begemann} {et~al.}(1996){Begemann}, {Dorschner}, {Henning}, \&
  {Mutschke}}]{Begemann:1996}
{Begemann}, B., {Dorschner}, J., {Henning}, T., \& {Mutschke}, H. 1996, \apjl,
  464, L195

\bibitem[{{Begemann} {et~al.}(1997){Begemann}, {Dorschner}, {Henning},
  {Mutschke}, {Guertler}, {Koempe}, \& {Nass}}]{Begemann:1997}
{Begemann}, B., {Dorschner}, J., {Henning}, T., {et~al.} 1997, \apj, 476, 199

\bibitem[{{Begemann} {et~al.}(1994){Begemann}, {Dorschner}, {Henning},
  {Mutschke}, \& {Thamm}}]{Begemann:1994}
{Begemann}, B., {Dorschner}, J., {Henning}, T., {Mutschke}, H., \& {Thamm}, E.
  1994, \apjl, 423, L71

\bibitem[{{Bohren} \& {Huffman}(1983)}]{Bohren:1983}
{Bohren}, C.~F., \& {Huffman}, D.~R. 1983, {Absorption and Scattering of Light
  by Small Particles} (New York: Wiley)

\bibitem[{{Braatz} {et~al.}(2000){Braatz}, {Ott}, {Henning}, {J{\"a}ger}, \&
  {Jeschke}}]{Braatz:2000}
{Braatz}, A., {Ott}, U., {Henning}, T., {J{\"a}ger}, C., \& {Jeschke}, G. 2000,
  Meteoritics and Planetary Science, 35, 75

\bibitem[{{Brandt} \& {Draine}(2012)}]{Brandt:2012}
{Brandt}, T.~D., \& {Draine}, B.~T. 2012, \apj, 744, 129

\bibitem[{{Braun}(1987)}]{Braun:1987}
{Braun}, R. 1987, \aap, 171, 233

\bibitem[{{Cherchneff}(2012)}]{Cherchneff:2012}
{Cherchneff}, I. 2012, \aap, 545, A12

\bibitem[{{DeLaney} {et~al.}(2014){DeLaney}, {Kassim}, {Rudnick}, \&
  {Perley}}]{DeLaney:2014}
{DeLaney}, T., {Kassim}, N.~E., {Rudnick}, L., \& {Perley}, R.~A. 2014, \apj,
  arXiv:1403.0032

\bibitem[{{DeLaney}(2004)}]{DeLaney:2004}
{DeLaney}, T.~A. 2004, PhD thesis, University of Minnesota

\bibitem[{{Dorschner} {et~al.}(1995){Dorschner}, {Begemann}, {Henning},
  {Jaeger}, \& {Mutschke}}]{Dorschner:1995}
{Dorschner}, J., {Begemann}, B., {Henning}, T., {Jaeger}, C., \& {Mutschke}, H.
  1995, \aap, 300, 503

\bibitem[{{Dorschner} {et~al.}(1980){Dorschner}, {Friedemann}, {Guertler}, \&
  {Duley}}]{Dorschner:1980}
{Dorschner}, J., {Friedemann}, C., {Guertler}, J., \& {Duley}, W.~W. 1980,
  \apss, 68, 159

\bibitem[{{Douvion} {et~al.}(1999){Douvion}, {Lagage}, \&
  {Cesarsky}}]{Douvion:1999}
{Douvion}, T., {Lagage}, P.~O., \& {Cesarsky}, C.~J. 1999, \aap, 352, L111

\bibitem[{{Douvion} {et~al.}(2001){Douvion}, {Lagage}, \&
  {Pantin}}]{Douvion:2001}
{Douvion}, T., {Lagage}, P.~O., \& {Pantin}, E. 2001, \aap, 369, 589

\bibitem[{{Draine} \& {Lee}(1984)}]{Draine:1984}
{Draine}, B.~T., \& {Lee}, H.~M. 1984, \apj, 285, 89

\bibitem[{{Draine} \& {Li}(2007)}]{Draine:2007}
{Draine}, B.~T., \& {Li}, A. 2007, \apj, 657, 810

\bibitem[{{Dwek} \& {Cherchneff}(2011)}]{Dwek:2011}
{Dwek}, E., \& {Cherchneff}, I. 2011, \apj, 727, 63

\bibitem[{{Edoh}(1983)}]{Edoh:1983}
{Edoh}, O. 1983, PhD thesis, The University of Arizona

\bibitem[{Edwards(1985)}]{Edwards:1985}
Edwards, D.~F. 1985, Handbook of Optical Constants of Solids, ed. E.~D. Palik
  (San Diego, CA: Acedemic Press), 547

\bibitem[{{Ennis} {et~al.}(2006){Ennis}, {Rudnick}, {Reach}, {Smith}, {Rho},
  {DeLaney}, {Gomez}, \& {Kozasa}}]{Ennis:2006}
{Ennis}, J.~A., {Rudnick}, L., {Reach}, W.~T., {et~al.} 2006, \apj, 652, 376

\bibitem[{{Fabian} {et~al.}(2001){Fabian}, {Henning}, {J{\"a}ger}, {Mutschke},
  {Dorschner}, \& {Wehrhan}}]{Fabian:2001}
{Fabian}, D., {Henning}, T., {J{\"a}ger}, C., {et~al.} 2001, \aap, 378, 228

\bibitem[{{Fesen} {et~al.}(2006){Fesen}, {Hammell}, {Morse}, {Chevalier},
  {Borkowski}, {Dopita}, {Gerardy}, {Lawrence}, {Raymond}, \& {van den
  Bergh}}]{Fesen:2006}
{Fesen}, R.~A., {Hammell}, M.~C., {Morse}, J., {et~al.} 2006, \apj, 645, 283

\bibitem[{{Forrest} {et~al.}(1981){Forrest}, {Houck}, \&
  {McCarthy}}]{Forrest:1981}
{Forrest}, W.~J., {Houck}, J.~R., \& {McCarthy}, J.~F. 1981, \apj, 248, 195

\bibitem[{{Gordon} {et~al.}(2008){Gordon}, {Engelbracht}, {Rieke}, {Misselt},
  {Smith}, \& {Kennicutt}}]{Gordon:2008}
{Gordon}, K.~D., {Engelbracht}, C.~W., {Rieke}, G.~H., {et~al.} 2008, \apj,
  682, 336

\bibitem[{{Griffin} {et~al.}(2010){Griffin}, {Abergel}, {Abreu}, {Ade},
  {Andr{\'e}}, {Augueres}, {Babbedge}, {Bae}, {Baillie}, {Baluteau}, {Barlow},
  {Bendo}, {Benielli}, {Bock}, {Bonhomme}, {Brisbin}, {Brockley-Blatt},
  {Caldwell}, {Cara}, {Castro-Rodriguez}, {Cerulli}, {Chanial}, {Chen},
  {Clark}, {Clements}, {Clerc}, {Coker}, {Communal}, {Conversi}, {Cox},
  {Crumb}, {Cunningham}, {Daly}, {Davis}, {de Antoni}, {Delderfield}, {Devin},
  {di Giorgio}, {Didschuns}, {Dohlen}, {Donati}, {Dowell}, {Dowell}, {Duband},
  {Dumaye}, {Emery}, {Ferlet}, {Ferrand}, {Fontignie}, {Fox}, {Franceschini},
  {Frerking}, {Fulton}, {Garcia}, {Gastaud}, {Gear}, {Glenn}, {Goizel},
  {Griffin}, {Grundy}, {Guest}, {Guillemet}, {Hargrave}, {Harwit}, {Hastings},
  {Hatziminaoglou}, {Herman}, {Hinde}, {Hristov}, {Huang}, {Imhof}, {Isaak},
  {Israelsson}, {Ivison}, {Jennings}, {Kiernan}, {King}, {Lange}, {Latter},
  {Laurent}, {Laurent}, {Leeks}, {Lellouch}, {Levenson}, {Li}, {Li},
  {Lilienthal}, {Lim}, {Liu}, {Lu}, {Madden}, {Mainetti}, {Marliani}, {McKay},
  {Mercier}, {Molinari}, {Morris}, {Moseley}, {Mulder}, {Mur}, {Naylor},
  {Nguyen}, {O'Halloran}, {Oliver}, {Olofsson}, {Olofsson}, {Orfei}, {Page},
  {Pain}, {Panuzzo}, {Papageorgiou}, {Parks}, {Parr-Burman}, {Pearce},
  {Pearson}, {P{\'e}rez-Fournon}, {Pinsard}, {Pisano}, {Podosek}, {Pohlen},
  {Polehampton}, {Pouliquen}, {Rigopoulou}, {Rizzo}, {Roseboom}, {Roussel},
  {Rowan-Robinson}, {Rownd}, {Saraceno}, {Sauvage}, {Savage}, {Savini},
  {Sawyer}, {Scharmberg}, {Schmitt}, {Schneider}, {Schulz}, {Schwartz},
  {Shafer}, {Shupe}, {Sibthorpe}, {Sidher}, {Smith}, {Smith}, {Smith},
  {Spencer}, {Stobie}, {Sudiwala}, {Sukhatme}, {Surace}, {Stevens}, {Swinyard},
  {Trichas}, {Tourette}, {Triou}, {Tseng}, {Tucker}, {Turner}, {Vaccari},
  {Valtchanov}, {Vigroux}, {Virique}, {Voellmer}, {Walker}, {Ward}, {Waskett},
  {Weilert}, {Wesson}, {White}, {Whitehouse}, {Wilson}, {Winter}, {Woodcraft},
  {Wright}, {Xu}, {Zavagno}, {Zemcov}, {Zhang}, \& {Zonca}}]{Griffin:2010}
{Griffin}, M.~J., {Abergel}, A., {Abreu}, A., {et~al.} 2010, \aap, 518, L3

\bibitem[{{Henning} {et~al.}(1995){Henning}, {Begemann}, {Mutschke}, \&
  {Dorschner}}]{Henning:1995}
{Henning}, T., {Begemann}, B., {Mutschke}, H., \& {Dorschner}, J. 1995, \aaps,
  112, 143

\bibitem[{{Hony} {et~al.}(2002){Hony}, {Waters}, \& {Tielens}}]{Hony:2002}
{Hony}, S., {Waters}, L.~B.~F.~M., \& {Tielens}, A.~G.~G.~M. 2002, \aap, 390,
  533

\bibitem[{{Hwang} {et~al.}(2004){Hwang}, {Laming}, {Badenes}, {Berendse},
  {Blondin}, {Cioffi}, {DeLaney}, {Dewey}, {Fesen}, {Flanagan}, {Fryer},
  {Ghavamian}, {Hughes}, {Morse}, {Plucinsky}, {Petre}, {Pohl}, {Rudnick},
  {Sankrit}, {Slane}, {Smith}, {Vink}, \& {Warren}}]{Hwang:2004}
{Hwang}, U., {Laming}, J.~M., {Badenes}, C., {et~al.} 2004, \apjl, 615, L117

\bibitem[{{Isensee} {et~al.}(2010){Isensee}, {Rudnick}, {DeLaney}, {Smith},
  {Rho}, {Reach}, {Kozasa}, \& {Gomez}}]{Isensee:2010}
{Isensee}, K., {Rudnick}, L., {DeLaney}, T., {et~al.} 2010, \apj, 725, 2059

\bibitem[{{Isensee} {et~al.}(2012){Isensee}, {Olmschenk}, {Rudnick}, {DeLaney},
  {Rho}, {Smith}, {Reach}, {Kozasa}, \& {Gomez}}]{Isensee:2012}
{Isensee}, K., {Olmschenk}, G., {Rudnick}, L., {et~al.} 2012, \apj, 757, 126

\bibitem[{{Jaeger} {et~al.}(1994){Jaeger}, {Mutschke}, {Begemann}, {Dorschner},
  \& {Henning}}]{Jaeger:1994}
{Jaeger}, C., {Mutschke}, H., {Begemann}, B., {Dorschner}, J., \& {Henning}, T.
  1994, \aap, 292, 641

\bibitem[{{J{\"a}ger} {et~al.}(2003){J{\"a}ger}, {Dorschner}, {Mutschke},
  {Posch}, \& {Henning}}]{Jager:2003}
{J{\"a}ger}, C., {Dorschner}, J., {Mutschke}, H., {Posch}, T., \& {Henning}, T.
  2003, \aap, 408, 193

\bibitem[{{Jiang} {et~al.}(2005){Jiang}, {Zhang}, \& {Li}}]{Jiang:2005}
{Jiang}, B.~W., {Zhang}, K., \& {Li}, A. 2005, \apjl, 630, L77

\bibitem[{{Krause} {et~al.}(2008){Krause}, {Birkmann}, {Usuda}, {Hattori},
  {Goto}, {Rieke}, \& {Misselt}}]{Krause:2008}
{Krause}, O., {Birkmann}, S.~M., {Usuda}, T., {et~al.} 2008, Science, 320, 1195

\bibitem[{{Li} \& {Draine}(2001)}]{Li:2001}
{Li}, A., \& {Draine}, B.~T. 2001, \apj, 554, 778

\bibitem[{{Lynch} \& {Hunter}(1991)}]{Lynch:1991}
{Lynch}, D.~W., \& {Hunter}, W.~R. 1991, Handbook of Optical Constants of
  Solids II, ed. E.~D. Palik (San Diego, CA: Acedemic Press), 341

\bibitem[{{Markwardt}(2009)}]{Markwardt:2009}
{Markwardt}, C.~B. 2009, in Astronomical Society of the Pacific Conference
  Series, Vol. 411, Astronomical Data Analysis Software and Systems XVIII, ed.
  D.~A. {Bohlender}, D.~{Durand}, \& P.~{Dowler}, 251

\bibitem[{{Mukai}(1989)}]{Mukai:1989}
{Mukai}, T. 1989, in Evolution of Interstellar Dust and Related Topics, ed.
  A.~{Bonetti}, J.~M. {Greenberg}, \& S.~{Aiello}, 397

\bibitem[{{Mutschke} {et~al.}(2004){Mutschke}, {Andersen}, {J{\"a}ger},
  {Henning}, \& {Braatz}}]{Mutschke:2004}
{Mutschke}, H., {Andersen}, A.~C., {J{\"a}ger}, C., {Henning}, T., \& {Braatz},
  A. 2004, \aap, 423, 983

\bibitem[{{Mutschke} {et~al.}(2002){Mutschke}, {Posch}, {Fabian}, \&
  {Dorschner}}]{Mutschke:2002}
{Mutschke}, H., {Posch}, T., {Fabian}, D., \& {Dorschner}, J. 2002, \aap, 392,
  1047

\bibitem[{{Nozawa} {et~al.}(2010){Nozawa}, {Kozasa}, {Tominaga}, {Maeda},
  {Umeda}, {Nomoto}, \& {Krause}}]{nozawa:2010}
{Nozawa}, T., {Kozasa}, T., {Tominaga}, N., {et~al.} 2010, \apj, 713, 356

\bibitem[{{Nozawa} {et~al.}(2008){Nozawa}, {Kozasa}, {Tominaga}, {Sakon},
  {Tanaka}, {Suzuki}, {Nomoto}, {Maeda}, {Umeda}, {Limongi}, \&
  {Onaka}}]{nozawa:2008}
---. 2008, \apj, 684, 1343

\bibitem[{{Omont} {et~al.}(1995){Omont}, {Moseley}, {Cox}, {Glaccum}, {Casey},
  {Forveille}, {Chan}, {Szczerba}, {Loewenstein}, {Harvey}, \&
  {Kwok}}]{Omont:1995}
{Omont}, A., {Moseley}, S.~H., {Cox}, P., {et~al.} 1995, \apj, 454, 819

\bibitem[{Philipp(1985)}]{Philipp:1985}
Philipp, H.~R. 1985, Handbook of Optical Constants of Solids, ed. E.~D. Palik
  (San Diego, CA: Acedemic Press), 749

\bibitem[{{Pilbratt} {et~al.}(2010){Pilbratt}, {Riedinger}, {Passvogel},
  {Crone}, {Doyle}, {Gageur}, {Heras}, {Jewell}, {Metcalfe}, {Ott}, \&
  {Schmidt}}]{Pilbratt:2010}
{Pilbratt}, G.~L., {Riedinger}, J.~R., {Passvogel}, T., {et~al.} 2010, \aap,
  518, L1

\bibitem[{Piller(1985)}]{Piller:1985}
Piller, H. 1985, Handbook of Optical Constants of Solids, ed. E.~D. Palik (San
  Diego, CA: Acedemic Press), 571

\bibitem[{{Poglitsch} {et~al.}(2010){Poglitsch}, {Waelkens}, {Geis},
  {Feuchtgruber}, {Vandenbussche}, {Rodriguez}, {Krause}, {Renotte}, {van
  Hoof}, {Saraceno}, {Cepa}, {Kerschbaum}, {Agn{\`e}se}, {Ali}, {Altieri},
  {Andreani}, {Augueres}, {Balog}, {Barl}, {Bauer}, {Belbachir}, {Benedettini},
  {Billot}, {Boulade}, {Bischof}, {Blommaert}, {Callut}, {Cara}, {Cerulli},
  {Cesarsky}, {Contursi}, {Creten}, {De Meester}, {Doublier}, {Doumayrou},
  {Duband}, {Exter}, {Genzel}, {Gillis}, {Gr{\"o}zinger}, {Henning},
  {Herreros}, {Huygen}, {Inguscio}, {Jakob}, {Jamar}, {Jean}, {de Jong},
  {Katterloher}, {Kiss}, {Klaas}, {Lemke}, {Lutz}, {Madden}, {Marquet},
  {Martignac}, {Mazy}, {Merken}, {Montfort}, {Morbidelli}, {M{\"u}ller},
  {Nielbock}, {Okumura}, {Orfei}, {Ottensamer}, {Pezzuto}, {Popesso},
  {Putzeys}, {Regibo}, {Reveret}, {Royer}, {Sauvage}, {Schreiber}, {Stegmaier},
  {Schmitt}, {Schubert}, {Sturm}, {Thiel}, {Tofani}, {Vavrek}, {Wetzstein},
  {Wieprecht}, \& {Wiezorrek}}]{Poglitsch:2010}
{Poglitsch}, A., {Waelkens}, C., {Geis}, N., {et~al.} 2010, \aap, 518, L2

\bibitem[{{Posch} {et~al.}(2003){Posch}, {Kerschbaum}, {Fabian}, {Mutschke},
  {Dorschner}, {Tamanai}, \& {Henning}}]{Posch:2003}
{Posch}, T., {Kerschbaum}, F., {Fabian}, D., {et~al.} 2003, \apjs, 149, 437

\bibitem[{{Reed} {et~al.}(1995){Reed}, {Hester}, {Fabian}, \&
  {Winkler}}]{Reed:1995}
{Reed}, J.~E., {Hester}, J.~J., {Fabian}, A.~C., \& {Winkler}, P.~F. 1995,
  \apj, 440, 706

\bibitem[{{Rest} {et~al.}(2011){Rest}, {Foley}, {Sinnott}, {Welch}, {Badenes},
  {Filippenko}, {Bergmann}, {Bhatti}, {Blondin}, {Challis}, {Damke}, {Finley},
  {Huber}, {Kasen}, {Kirshner}, {Matheson}, {Mazzali}, {Minniti}, {Nakajima},
  {Narayan}, {Olsen}, {Sauer}, {Smith}, \& {Suntzeff}}]{Rest:2011}
{Rest}, A., {Foley}, R.~J., {Sinnott}, B., {et~al.} 2011, \apj, 732, 3

\bibitem[{{Rho} {et~al.}(2008){Rho}, {Kozasa}, {Reach}, {Smith}, {Rudnick},
  {DeLaney}, {Ennis}, {Gomez}, \& {Tappe}}]{Rho:2008}
{Rho}, J., {Kozasa}, T., {Reach}, W.~T., {et~al.} 2008, \apj, 673, 271

\bibitem[{{Rho} {et~al.}(2009){Rho}, {Reach}, {Tappe}, {Rudnick}, {Kozasa},
  {Hwang}, {Andersen}, {Gomez}, {Delaney}, {Dunne}, \& {Slavin}}]{Rho:2009}
{Rho}, J., {Reach}, W.~T., {Tappe}, A., {et~al.} 2009, in Astronomical Society
  of the Pacific Conference Series, Vol. 414, Cosmic Dust - Near and Far, ed.
  T.~{Henning}, E.~{Gr{\"u}n}, \& J.~{Steinacker}, 22

\bibitem[{{Rinehart} {et~al.}(2011){Rinehart}, {Benford}, {Cataldo}, {Dwek},
  {Henry}, {Kinzer}, {Nuth}, {Silverberg}, {Wheeler}, \&
  {Wollack}}]{Rinehart:2011}
{Rinehart}, S.~A., {Benford}, D.~J., {Cataldo}, G., {et~al.} 2011, \ao, 50,
  4115

\bibitem[{{Rouleau} \& {Martin}(1991)}]{Rouleau:1991}
{Rouleau}, F., \& {Martin}, P.~G. 1991, \apj, 377, 526

\bibitem[{{Saken} {et~al.}(1992){Saken}, {Fesen}, \& {Shull}}]{Saken:1992}
{Saken}, J.~M., {Fesen}, R.~A., \& {Shull}, J.~M. 1992, \apjs, 81, 715

\bibitem[{{Semenov} {et~al.}(2003){Semenov}, {Henning}, {Helling}, {Ilgner}, \&
  {Sedlmayr}}]{Semenov:2003}
{Semenov}, D., {Henning}, T., {Helling}, C., {Ilgner}, M., \& {Sedlmayr}, E.
  2003, \aap, 410, 611

\bibitem[{{Sibthorpe} {et~al.}(2010){Sibthorpe}, {Ade}, {Bock}, {Chapin},
  {Devlin}, {Dicker}, {Griffin}, {Gundersen}, {Halpern}, {Hargrave}, {Hughes},
  {Jeong}, {Kaneda}, {Klein}, {Koo}, {Lee}, {Marsden}, {Martin}, {Mauskopf},
  {Moon}, {Netterfield}, {Olmi}, {Pascale}, {Patanchon}, {Rex}, {Roy}, {Scott},
  {Semisch}, {Truch}, {Tucker}, {Tucker}, {Viero}, \& {Wiebe}}]{Sibthorpe:2010}
{Sibthorpe}, B., {Ade}, P.~A.~R., {Bock}, J.~J., {et~al.} 2010, \apj, 719, 1553

\bibitem[{{Siebenmorgen} {et~al.}(2014){Siebenmorgen}, {Voshchinnikov}, \&
  {Bagnulo}}]{Siebenmorgen:2014}
{Siebenmorgen}, R., {Voshchinnikov}, N.~V., \& {Bagnulo}, S. 2014, \aap, 561,
  A82

\bibitem[{{Smith} {et~al.}(2009){Smith}, {Rudnick}, {Delaney}, {Rho}, {Gomez},
  {Kozasa}, {Reach}, \& {Isensee}}]{Smith:2009}
{Smith}, J.~D.~T., {Rudnick}, L., {Delaney}, T., {et~al.} 2009, \apj, 693, 713

\bibitem[{{Tuffs} {et~al.}(1999){Tuffs}, {Fischera}, {Drury}, {Gabriel},
  {Heinrichsen}, {Rasmussen}, \& {Volk}}]{Tuffs:1999}
{Tuffs}, R.~J., {Fischera}, J., {Drury}, L.~O., {et~al.} 1999, in ESA Special
  Publication, Vol. 427, The Universe as Seen by ISO, ed. P.~{Cox} \&
  M.~{Kessler}, 241

\bibitem[{{Tuffs} {et~al.}(2005){Tuffs}, {Popescu}, \& {V{\"o}lk}}]{Tuffs:2005}
{Tuffs}, R.~J., {Popescu}, C.~C., \& {V{\"o}lk}, H.~J. 2005, in ESA Special
  Publication, Vol. 577, ESA Special Publication, ed. A.~{Wilson}, 427--428

\bibitem[{{Valiante} {et~al.}(2011){Valiante}, {Schneider}, {Salvadori}, \&
  {Bianchi}}]{Valiante:2011}
{Valiante}, R., {Schneider}, R., {Salvadori}, S., \& {Bianchi}, S. 2011,
  \mnras, 416, 1916

\bibitem[{{Zeidler} {et~al.}(2011){Zeidler}, {Posch}, {Mutschke}, {Richter}, \&
  {Wehrhan}}]{Zeidler:2011}
{Zeidler}, S., {Posch}, T., {Mutschke}, H., {Richter}, H., \& {Wehrhan}, O.
  2011, \aap, 526, A68

\bibitem[{{Zinner}(2008)}]{Zinner:2008}
{Zinner}, E. 2008, \pasa, 25, 7

\bibitem[{{Zubko} {et~al.}(2004){Zubko}, {Dwek}, \& {Arendt}}]{Zubko:2004}
{Zubko}, V., {Dwek}, E., \& {Arendt}, R.~G. 2004, \apjs, 152, 211

\end{thebibliography}

\newpage
%--------------------------------------------------------------

\clearpage

%CHI2 MAPS
\clearpage

%MODELS

\clearpage

%%MASSES

%%%%%%%%%%%%%%%%%%%%%%%%%%%%%%%%%%%%%%%%%%%%%%%%%%%%%%%%%%%%%%%%%%%%% 
\begin{deluxetable}{c|ccccccc|ccc|cc|c|cc|c}
\rotate
\tabletypesize{\tiny}
\tablewidth{0pt}
\tablecaption{Linear Correlation Coefficients between the Spatial Templates}
\tablehead{
\colhead{}&
\colhead{}&
\colhead{}&
\colhead{}&
\colhead{}&
\colhead{}&
\colhead{}&
\colhead{IRAC}&
\colhead{}&
\colhead{}&
\colhead{IRAC}&
\colhead{}& 
\colhead{}&
\colhead{}&
\colhead{}&
\colhead{}&
\colhead{}\\
\colhead{}&
\colhead{\bf [Ar II]}&
\colhead{\bf [Ar III]}&
\colhead{[S IV]}&
\colhead{[Ne V]}&
\colhead{[Fe II]}&
\colhead{[S III]}&
\colhead{8 $\micron$}&
\colhead{\bf [Ne II]}&
\colhead{[Ne III]}&
\colhead{4.5 $\micron$}&
\colhead{[O IV]}&
\colhead{[S III]}&
\colhead{\bf [Si II]}&
\colhead{\bf X-ray Fe}&
\colhead{X-ray Si}&
\colhead{\bf Radio}

}
\startdata
%                          \ion{Ar}{2}    ArIII    S IV      NeV       FeII      SIII     IRAC4     NeII      \ion{Ne}{3}   IRAC2      OIV       SIII      SiII      XFE       XSi       Radio
$[$Ar II$]$ 6.99 $\micron$            & 1.00    & 0.99    & 0.97    & 0.89    & 0.77    & 0.97    & 0.97    & 0.76    & 0.76    & 0.88    & 0.83    & 0.84    & 0.54    & 0.44    & 0.67    & 0.54 \\
$[$Ar III$]$ 8.99 $\micron$          & \nodata & 1.00    & 0.97    & 0.88    & 0.78    & 0.95    & 0.97    & 0.73    & 0.73    & 0.87    & 0.82    & 0.83    & 0.56    & 0.46    & 0.68    & 0.54 \\
$[$S IV$]$ 10.51 $\micron$            & \nodata & \nodata & 1.00    & 0.85    & 0.74    & 0.96    & 0.93    & 0.74    & 0.75    & 0.84    & 0.88    & 0.89    & 0.66    & 0.41    & 0.65    & 0.55 \\
$[$Ne V$]$ 14.32 $\micron$            & \nodata & \nodata & \nodata & 1.00    & 0.73    & 0.90    & 0.85    & 0.78    & 0.81    & 0.84    & 0.82    & 0.82    & 0.56    & 0.45    & 0.65    & 0.53 \\
$[$Fe II$]$ 17.94 $\micron$           & \nodata & \nodata & \nodata & \nodata & 1.00    & 0.74    & 0.76    & 0.58    & 0.61    & 0.73    & 0.67    & 0.65    & 0.45    & 0.34    & 0.48    & 0.45 \\
$[$S III$]$ 18.71 $\micron$    & \nodata & \nodata & \nodata & \nodata & \nodata & 1.00    & 0.92    & 0.77    & 0.79    & 0.85    & 0.88    & 0.90    & 0.64    & 0.46    & 0.70    & 0.58 \\
IRAC 8 $\micron$       & \nodata & \nodata & \nodata & \nodata & \nodata & \nodata & 1.00    & 0.70    & 0.70    & 0.87    & 0.75    & 0.76    & 0.45    & 0.43    & 0.63    & 0.47 \\ \hline
$[$Ne II$]$ 12.81 $\micron$           & \nodata & \nodata & \nodata & \nodata & \nodata & \nodata & \nodata & 1.00    & 0.96    & 0.90    & 0.81    & 0.73    & 0.50    & 0.37    & 0.50    & 0.48 \\
$[$Ne III$]$ 15.56 $\micron$          & \nodata & \nodata & \nodata & \nodata & \nodata & \nodata & \nodata & \nodata & 1.00    & 0.88    & 0.83    & 0.75    & 0.54    & 0.42    & 0.52    & 0.50 \\
IRAC 4.5 $\micron$     & \nodata & \nodata & \nodata & \nodata & \nodata & \nodata & \nodata & \nodata & \nodata & 1.00    & 0.81    & 0.76    & 0.49    & 0.45    & 0.58    & 0.54 \\ \hline
$[$O IV$]$ 25.89 $\micron$ & \nodata & \nodata & \nodata & \nodata & \nodata & \nodata & \nodata & \nodata & \nodata & \nodata & 1.00    & 0.93    & 0.83    & 0.39    & 0.64    & 0.62 \\
$[$S III$]$ 33.48 $\micron$  & \nodata & \nodata & \nodata & \nodata & \nodata & \nodata & \nodata & \nodata & \nodata & \nodata & \nodata & 1.00    & 0.84    & 0.41    & 0.62    & 0.62 \\ \hline
$[$Si II$]$ 34.82 $\micron$           & \nodata & \nodata & \nodata & \nodata & \nodata & \nodata & \nodata & \nodata & \nodata & \nodata & \nodata & \nodata & 1.00    & 0.23    & 0.42    & 0.55 \\ \hline
X-ray Fe               & \nodata & \nodata & \nodata & \nodata & \nodata & \nodata & \nodata & \nodata & \nodata & \nodata & \nodata & \nodata & \nodata & 1.00    & 0.71    & 0.65 \\
X-ray Si               & \nodata & \nodata & \nodata & \nodata & \nodata & \nodata & \nodata & \nodata & \nodata & \nodata & \nodata & \nodata & \nodata & \nodata & 1.00    & 0.75 \\ \hline
Radio                  & \nodata & \nodata & \nodata & \nodata & \nodata & \nodata & \nodata & \nodata & \nodata & \nodata & \nodata & \nodata & \nodata & \nodata & \nodata & 1.00 \\
\enddata
\label{tab:cc}
\tablecomments{Spatial templates in boldface were selected as the basis for further analysis.}
\end{deluxetable}

\begin{deluxetable}{ll}
\tabletypesize{\small}
\tablewidth{0pt}
\tablecaption{Chosen Spatial Templates}
\tablehead{
\colhead{Template} &
\colhead{Purpose}
}
\startdata
$[$\ion{Ar}{2}$]$ & trace dust in the most dominant line emitting ejecta\\
$[$\ion{Ne}{2}$]$ & trace dust in a distinctly different ejecta component from higher layers of the SN\\
X-ray Fe      & trace dust in hotter, low-density ejecta from the inner portion of the SN\\
Radio         & trace ISM or CSM dust swept up behind the forward shock\\
$[$\ion{Si}{2}$]$ & trace dust in slow-moving ejecta, yet to cross the reverse shock\\
$[$\ion{Ar}{3}$]$ & test if dust properties differ in regions where ionization state of the gas is higher\\
\enddata
\label{tab:purpose}
\end{deluxetable}

\begin{deluxetable}{llcll}
\tabletypesize{\small}
\tablewidth{0pt}
\tablecaption{Dust Compositions Considered for Fitting Spectra}
\tablehead{
\colhead{$n$} &
\colhead{Composition} &
\colhead{$\lambda$ Range (\micron)} & 
\colhead{Note} &
\colhead{Reference}
}
\startdata
%\cutinhead{Silicates}
\sidehead{Silicates}
0 & Silicate              & 0.01--9400 & & \cite{Draine:1984}\\
1 & Mg$_{0.7}$SiO$_{2.7}$ & 0.2--470 & & \cite{Jager:2003}\\
2 & MgSiO$_{3}$           & 0.2--10000 & & \cite{Jager:2003}\\
3 & MgSiO$_{3}$           & 0.1--100000 &  & \cite{Dorschner:1995}; Kozasa (priv. comm.)\\
4 & MgSiO$_{3}$           & 0.2--500 & glassy & \cite{Jaeger:1994,Dorschner:1995}\\
5 & Mg$_{1.5}$SiO$_{3.5}$ & 0.2--6000 & & \cite{Jager:2003}\\
6 & Mg$_2$SiO$_4$         & 0.2--950 & & \cite{Jager:2003}\\
7 & Mg$_2$SiO$_{4}$       & 0.1--100000 &  & \cite{Jager:2003}; Kozasa (priv. comm.)\\
8 & MgFeSiO$_4$           & 0.2--500 & glassy & \cite{Jaeger:1994,Dorschner:1995}\\
9 & Mg$_{2.4}$SiO$_{4.4}$ & 0.2--8200 & & \cite{Jager:2003}\\
%\cutinhead{Protosilicates}
\sidehead{Protosilicates}
10 & Ca Protosilicate      & 7.6--25 & A -- unheated & \cite{Dorschner:1980}\\
11 & Ca Protosilicate      & 7.7--25 & B -- 450C & \cite{Dorschner:1980}\\
12 & Ca Protosilicate      & 7.8--25 & C -- 695C & \cite{Dorschner:1980}\\
13 & Fe Protosilicate      & 8.2--40 & D -- unheated & \cite{Dorschner:1980}\\
14 & Fe Protosilicate      & 8--40 & E -- 490C & \cite{Dorschner:1980}\\
15 & Mg Protosilicate      & 8--40 & F -- 485C & \cite{Dorschner:1980}\\
%\cutinhead{Silica}
\sidehead{Silica}
16 & SiO$_2$               & 0.1--500 & am & \cite{Philipp:1985}; Kozasa (priv. comm.)\\
17 & SiO$_2$               & 5--500 & & \cite{Philipp:1985}\\
18 & SiO$_x$               & 5.6--520 & extrap. at $\lambda > 65$ $\micron$ & \cite{Rinehart:2011}\\
%\cutinhead{SiC, Si}
\sidehead{SiC, Si}
19 & $\alpha$ SiC          & 0.1--1000 & $21\ \micron: Q_{abs}/a = 0$ & \cite{Jiang:2005}\\
20 & $\alpha$ SiC          & 0.1--1000 & $21\ \micron: Q_{abs}/a = 100$ & \cite{Jiang:2005}\\
21 & $\alpha$ SiC          & 0.1--1000 & $21\ \micron: Q_{abs}/a = 10^3$ & \cite{Jiang:2005}\\
22 & $\alpha$ SiC          & 0.1--1000 & $21\ \micron: Q_{abs}/a = 10^4$ & \cite{Jiang:2005}\\
23 & $\beta$ SiC           & 0.1--1000 & $21\ \micron: Q_{abs}/a = 0$ & \cite{Jiang:2005}\\
24 & $\beta$ SiC           & 0.1--1000 & $21\ \micron: Q_{abs}/a = 100$ & \cite{Jiang:2005}\\
25 & $\beta$ SiC           & 0.1--1000 & $21\ \micron: Q_{abs}/a = 10^3$ & \cite{Jiang:2005}\\
26 & $\beta$ SiC           & 0.1--1000 & $21\ \micron: Q_{abs}/a = 10^4$ & \cite{Jiang:2005}\\
27 & Si                    & 0.02--247.7 & am & \cite{Piller:1985}; Kozasa (priv. comm.)\\
28 & Si                    & 0.07--433.3 & cr & \cite{Edwards:1985}; Kozasa (priv. comm.)\\
%\cutinhead{C, Fe}
\sidehead{C, Fe}
29 & Meteoritic Diamond    & 0.02--110 & & \cite{Braatz:2000}; \cite{Mutschke:2004}\\
30 & PAH                   & 0.01--9400 & & \cite{Draine:2007}\\
31 & PAH$^+$               & 0.01--9400 & & \cite{Draine:2007}\\
32 & Graphite              & 0.01--9400 & & \cite{Draine:1984}\\
33 & Carbon      & 0.01--9400 & amorphous ``ac'' & \cite{Rouleau:1991}\\
34 & Carbon      & 0.01--9400 & amorphous ``be'' & \cite{Rouleau:1991}\\
35 & Carbon      & 0.1--3000 & glass & \cite{Edoh:1983}; Kozasa (priv. comm.)\\
36 & Fe                    & 0.1--100000 &  & \cite{Lynch:1991}; Kozasa (priv. comm.)\\
%\cutinhead{Al Oxides}
\sidehead{Al Oxides}
37 & Al$_2$O$_3$           & 7.8--500 & porous & \cite{Begemann:1997}\\
38 & Al$_2$O$_3$           & 7.8--200 & compact & \cite{Begemann:1997}\\
%\cutinhead{Oxides}
\sidehead{Oxides}
39 & MgAl$_2$O$_4$         & 2--10000 & natural & \cite{Fabian:2001}\\
40 & MgAl$_2$O$_4$         & 2--6800 & natural, annealed & \cite{Fabian:2001}\\
41 & Mg$_{1.01}$Al$_{1.99}$O$_4$         & 1.67--6825 & nonstoich. spinel & \cite{Zeidler:2011}\\
42 & Mg$_{0.94}$Al$_{2.04}$O$_4$         & 1.67--6825 & nonstoich. spinel & \cite{Zeidler:2011}\\
43 & Mg$_{0.82}$Al$_{2.12}$O$_4$         & 1.67--6825 & nonstoich. spinel & \cite{Zeidler:2011}\\
44 & Mg$_{0.73}$Al$_{2.17}$O$_4$         & 1.67--6825 & nonstoich. spinel & \cite{Zeidler:2011}\\
45 & Mg$_{0.53}$Al$_{2.31}$O$_4$         & 1.67--6825 & nonstoich. spinel & \cite{Zeidler:2011}\\
46 & Mg$_{0.43}$Al$_{2.38}$O$_4$         & 2--10000   & nonstoich. spinel & \cite{Zeidler:2011}\\
47 & CaAl$_{12}$O$_{19}$                 & 2--10000 & hibonite   & \cite{Mutschke:2002}\\
48 & TiO$_{2}$ (1)                       & 2--5843  & anatase    & \cite{Posch:2003}; \cite{Zeidler:2011}\\
49 & TiO$_{2}$ (2)                       & 2--5843  & brookite   & \cite{Posch:2003}; \cite{Zeidler:2011}\\
50 & TiO$_{2}$ (3)                       & 0.47--36 & rutile     & \cite{Posch:2003}; \cite{Zeidler:2011}\\
51 & CaTiO$_{4}$                         & 2--5820  & perovskite & \cite{Posch:2003}; \cite{Zeidler:2011}\\
52 & Mg$_{0.6}$Fe$_{0.4}$O & 0.2--500 &  & \cite{Henning:1995}\\
53 & Mg$_{0.1}$Fe$_{0.9}$O & 0.2--500 &  & \cite{Henning:1995}\\
54 & FeO                   & 0.2--500 &  & \cite{Henning:1995}\\
55 & Fe$_3$O$_4$           & 0.1--100000 & & \cite{Mukai:1989}; Kozasa (priv. comm.)\\
%\cutinhead{Sulfides}
\sidehead{Sulfides}
56 & Mg$_{0.9}$Fe$_{0.1}$S & 10--500  &  & \cite{Begemann:1994}\\
57 & Mg$_{0.1}$Fe$_{0.9}$S & 10--500  &  & \cite{Begemann:1994}\\
58 & FeS                   & 0.1--100000 &  & \cite{Semenov:2003}; Kozasa (priv. comm.)\\
59 & SiS$_2$               & 13--60 &  & \cite{Begemann:1996}\\
\enddata
\label{tab:emiss}
\end{deluxetable}

%%%%%%%%%%%%%%%%%%%%%%%%%%%%%%%%%%%%%%%%%%%%%%%%%%%%%%%%%%%%%%%%%%%%%% 
\begin{deluxetable}{lll}
\rotate
\tabletypesize{\small}
\tablewidth{0pt}
\tablecaption{Best 2-Composition Fits of Cas A Dust}
\tablehead{
\colhead{Spectrum} &
\colhead{Compositions\tablenotemark{1}} &
\colhead{Total Dust Mass (M$_{\sun}$)\tablenotemark{2}}
}
\startdata
Ar {\sc ii} & {\bf Mg$_{0.7}$SiO$_{2.7}$} + [Graphite $\|$ C (ac) $\|$ FeS $\|$ C (glass) $\|$ Fe $\|$ C (be) $\|$ Al$_2$O$_3$ $\|$ Mg$_x$Al$_y$O$_4$ $\|$ Fe$_3$O$_4$] & $0.01 \pm 0.002$ \\
Ar {\sc iii} & {\bf Mg$_{0.7}$SiO$_{2.7}$} + [Mg$_x$Al$_y$O$_4$ $\|$ C (ac) $\|$ Al$_2$O$_3$ $\|$ C (glass) $\|$ Fe $\|$ C (be) $\|$ Graphite $\|$ CaAl$_{12}$O$_{19}$ $\|$ Fe$_3$O$_4$] & $0.004 \pm 0.001$\\
Ne {\sc ii} & {\bf Al$_2$O$_3$} + [C (be) $\|$ C (ac) $\|$ C (glass) $\|$ Fe $\|$ Graphite $\|$ FeS $\|$ Fe$_3$O$_4$ $\|$ Mg$_{0.1}$Fe$_{0.9}$S] & $0.004 \pm 0.001$\\
%            & ({\bf nonstoich. spinels} + [C (be) $\|$ C (glass) $\|$ Fe])\\
X-ray Fe & {\bf Mg$_{2.4}$SiO$_{4.4}$} + [C (be) $\|$ C (glass) $\|$ Fe $\|$ Mg$_{0.1}$Fe$_{0.9}$S $\|$ C (ac) $\|$ Graphite $\|$ FeS $\|$ Fe$_3$O$_4$] & $0.02 \pm 0.01$\\
         & {\bf MgFeSiO$_4$} + [C (be) $\|$ C (glass) $\|$ Fe $\|$ Mg$_{0.1}$Fe$_{0.9}$S $\|$ C (ac)]\\
%         & ({\bf other silicates} + [usual ``featureless'' components])\\
South Spot & {\bf MgFeSiO$_4$} + [C (glass) $\|$ C (ac) $\|$ C (be) $\|$ Fe] & $0.0001 \pm 3\times10^{-5}$\\
           & {\bf Mg$_{2.4}$SiO$_{4.4}$} + [C (ac) $\|$ Fe $\|$ C (be) $\|$ C (glass)]\\
%           & ({\bf other silicates} + [usual ``featureless'' components])\\
Radio & {\bf Mg$_{2.4}$SiO$_{4.4}$} + [Graphite $\|$ Mg$_x$Al$_y$O$_4$ $\|$ Mg$_{0.1}$Fe$_{0.9}$S $\|$ C (glass) $\|$ PAH$^+$ $\|$ Al$_2$O$_3$ $\|$ FeS $\|$ Fe $\|$ C (be)] & $0.0004 \pm 0.0004$\\
      & {\bf MgFeSiO$_4$} + [Graphite $\|$ C (glass) $\|$ Al$_2$O$_3$ $\|$ Fe $\|$ C (be) $\|$ FeS $\|$ Mg$_{0.1}$Fe$_{0.9}$S]\\
%      & ({\bf Mg$_2$SiO$_4$} + [C (glass) $\|$ Al$_2$O$_3$ $\|$ Graphite $\|$ Fe $\|$ C (be)])\\
Si {\sc ii} & Indeterminate & $\lesssim0.1$
\enddata
\tablenotetext{1}{The compositions in {\bf bold} need to be paired with any {\it one} of the 
generally more featureless components listed
after the ``+'' sign (i.e. $\|$ indicates ``or''). These components are listed in order of 
increasing $\chi^2$, but the differences are not significant. 
%Combinations listed in parentheses have $\chi^2$ values larger 
%than other combinations, but still within a factor of 2 of the minimum $\chi^2$.
}
\tablenotetext{2}{Mean and $\sigma$ for good models in which the bold components at left are the dominant mass.}
\label{tab:families}
\end{deluxetable}

%%%%%%%%%%%%%%%%%%%%%%%%%%%%%%%%%%%%%%%%%%%%%%%%%%%%%%%%%%%%%%%%%%%%%% 
\begin{deluxetable}{ll}
\tabletypesize{\small}
\tablewidth{0pt}
\tablecaption{Possible Compositions for the 12 $\micron$ Peak of \ion{Ar}{2}, \ion{Ar}{3} Dust}
\tablehead{
\colhead{Compositions} &
\colhead{Comment}
}
\startdata
SiO$_2$ & requires CDE calculation of $\kappa_{\lambda}$, also contributes to 9 and 21 $\micron$ features\\
SiC     & requires CDE calculation of $\kappa_{\lambda}$\\
Mg$_x$Al$_y$O$_4$ & requires extremely high temperatures ($>2000$ K) [$y=(8-2x)/3$]\\
CaAl$_{12}$O$_{19}$ & requires extremely high temperatures ($>2000$ K) 
\enddata
\label{tab:12um}
\end{deluxetable}

\end{document}